\documentclass[preprint,3p,sort&compress]{elsarticle}
\journal{Elsevier}
\usepackage[usenames,dvipsnames]{xcolor}
\usepackage{mathptmx}
\usepackage{graphicx}           
\usepackage{amsmath}
\usepackage{amsfonts}
\usepackage{amssymb}
\usepackage{epstopdf}
\usepackage{graphicx}
\usepackage{tikz}
\usepackage{xspace}
\usepackage{flowchart}
\usepackage{color}
\usepackage{subcaption}
\usepackage{booktabs}
\usepackage{bm}
\newcommand{\nc}{\newcommand}
\nc{\rnc}{\renewcommand}
\nc{\bs}{\boldsymbol}
\nc{\mrm}{\mathrm}
\nc{\mum}{$µ$\mrm{m}}
\rnc{\matrix}[2]{\left[\!\!\begin{array}{#1}
	#2\end{array}\!\!\right]}
\rnc{\vector}[1]{\matrix{c}{#1}}
\nc{\mm}[1]{\boldsymbol{#1}}
\nc{\mms}[1]{\boldsymbol{#1}}
\nc{\real}[1]{\mathrm{Re} \lbrace #1 \rbrace}
\nc{\imag}[1]{\mathrm{Im} \lbrace #1 \rbrace}
\nc{\dd}{\mathrm{d}}
\nc{\ii}{\mathrm{i}}
\nc{\ee}{\mathrm{e}}
\nc{\inv}{^{-1}} 
\nc{\herm}{^{\mathrm H}}
\nc{\tra}{^{\mathrm T}}
\nc{\conj}[1]{ \overline{#1} }
\nc{\normal}{\mathrm n}
\nc{\tangential}{\mathrm t}
\nc{\kn}{{k_{\normal}}}
\nc{\kt}{{k_{\tangential}}}
\nc{\COMMENT}[1]{\textcolor{red}{#1}}
\nc{\ie}{i.\,e.\xspace}
\nc{\eg}{e.\,g.\xspace}
\nc{\cf}{cf.\xspace}
\nc{\myquote}[1]{`#1'}
\nc{\etal}{et al.\xspace}
\nc{\fabstand}{\,}
\nc{\fp}{\fabstand.}
\nc{\fk}{\fabstand,}
\nc{\x}[1]{\mbox{#1}}
\nc{\g}[1]{\x{$#1$}}
\nc{\qb}{{\mm q}_{\mrm{b}}}
\nc{\N}{n_{\mrm{d}}}
\nc{\C}{n_{\mrm{c}}}
\nc{\nmod}{n_{\mrm{m}}}
\nc{\W}{\mm W}
\nc{\Wtra}{{\mm W}^{\mrm T}}
\nc{\Wb}{{\mm W}_{\mrm b}}
\nc{\Wbtra}{{\mm W}_{\mrm b}^{\mrm T}}
\nc{\qmod}{q_{\mrm{mod}}}
\nc{\Dmod}{D}
\nc{\ABAQUS}{\textsf{ABAQUS}\xspace}
\nc{\MATLAB}{\textsf{MATLAB}\xspace}
\nc{\epsAL}{\varepsilon}
\nc{\fig}[4][tbh]{
\begin{figure}[#1]
\centering
\includegraphics[width=#4\textwidth]{figures/#2}
\caption{#3\label{fig:#2}}
\end{figure}}
\nc{\e}[2]{\begin{equation} #1 \label {eq:#2} \end{equation}}
\nc{\est}[1]{\begin{equation*} #1 \end{equation*}}
\nc{\ea}[1]{
\begin{eqnarray}
#1 \end{eqnarray}}
\nc{\east}[1]{
\begin{eqnarray*}
#1 \end{eqnarray*}}
\nc{\fref}[1]{{Fig.~\ref{fig:#1}}}
\nc{\frefo}[1]{{\ref{fig:#1}}}
\nc{\frefs}[1]{{Figs.~\ref{fig:#1}}}
\nc{\tref}[1]{{Tab.~\ref{tab:#1}}}
\nc{\trefo}[1]{{\ref{tab:#1}}}
\nc{\trefs}[1]{{Tab.~\ref{tab:#1}}}
\nc{\eref}[1]{{Eq.~(\ref{eq:#1})}}
\nc{\erefo}[1]{(\ref{eq:#1})}
\nc{\erefs}[1]{{Eqs.~(\ref{eq:#1})}}
\nc{\sref}[1]{{Section~\ref{sec:#1}}}
\nc{\srefo}[1]{\ref{sec:#1}}
\nc{\srefs}[1]{{Sections~\ref{sec:#1}}}
\nc{\ssref}[1]{{Section~\ref{sec:#1}}}
\nc{\ssrefo}[1]{\ref{sec:#1}}
\nc{\ssrefs}[1]{{Sections~\ref{sec:#1}}}
\nc{\aref}[1]{{{\ref{asec:#1}}}}
\nc{\arefo}[1]{{\ref{asec:#1}}}
\nc{\arefs}[1]{{{Appendices~\ref{asec:#1}}}}

\makeindex

\begin{document}

\begin{frontmatter}
\title{A coupled FE-BE multi-scale method for the dynamics of jointed structures}
\author{
Hendrik D. Linder$^1$,
Johann Groß$^1$,
Malte Krack$^1$,
\\
}
\address{$^1$ University of Stuttgart, GERMANY}

\begin{abstract}
The damping of built-up structures stems largely from the microscopic dry frictional interactions in the contact interfaces.
The accurate prediction of friction damping has been an important scientific aim of the past several decades.
Recent research indicates that very good agreement with vibration measurements is to be expected if the actual contact surface topography is sufficiently well known and finely resolved, and frictional-unilateral interactions are modeled in terms of the Coulomb-Signorini conditions.
Resolving all relevant length scales in one finite element model leads to enormous or even prohibitive computation effort and regularization of the set-valued contact laws might be needed to ensure numerical stability.
In this work, we propose a multi-scale approach:
The stress and deformation field in the contact region is modeled using elastic half-space theory, implemented on a regular and fine grid of boundary elements (BE), so that the compliance matrix can be expressed in closed form.
The vibration behavior of the remaining region is described using a relatively coarse finite element (FE) model, which is further reduced via component mode synthesis.
The two models are coupled by enforcing compatibility and equilibrium conditions in the far field.
The set-valued Coulomb-Signorini conditions are enforced robustly and efficiently using a projected over-relaxation scheme in conjunction with an appropriate active-set strategy.
For the S4 beam benchmark, very good agreement with regard to the amplitude-dependent frequency and damping ratio of the first few modes is achieved, while the computation effort is reduced by several orders of magnitude compared to the full-FE reference.
The proposed multi-scale method permits a very fine resolution of the contact surface topography without suffering from numerical instability.
\end{abstract}


\end{frontmatter}

\section{Introduction}
To engineer mechanical systems against harmful vibrations, it is crucial to quantify their damping.
For most lightweight structures, dissipation within the material is negligible, and (dry) friction is the main cause of structural damping \cite{Brake.2018}.
Some interfaces are designed for gross sliding and/or liftoff.
This is the case, \eg, for friction dampers, like the under-platform dampers commonly used to mitigate vibrations of turbo-machinery blades \cite{srin1983}.
Other interfaces do not have damping but load transmission, leak tightness or alignment as primary purpose.
Common examples are contact interfaces at mechanical joints tightened by bolts or rivets.
Those contacts never reach gross sliding / liftoff during operation (before failure).
On the other hand, those contacts are not fully stuck either.
To understand this, it is important to emphasize that real technical surfaces feature form deviations, waviness and roughness \cite{raja2002,yas15}. 
Consequently, the real contact area may be a small fraction of the apparent (or nominal) contact area, even at high squeezing forces, and the contact pressure distribution is highly non-uniform, see \eg \cite{Campana.2011}.
Under vibratory loading, some parts of the contact area undergo microscopic relative motion, while the contact is still in a macroscopic stick phase.
This regime is called \emph{partial slip}, pre-sliding or microslip.
\\
When the contacts undergo gross slip / complete liftoff, reasonable predictions of the nonlinear vibration behavior can be obtained, see \eg \cite{vandeVorst.1996,vandeVorst.1998,Wagg.2002,Krishna.2012,firr2008,Claeys.2016b,Pesaresi.2017,MonjarazTec.2022,Schwarz.2023}.
In that regime, the surface irregularities (in particular waviness and roughness) play only a minor role, because their length scale is small compared to the relative contact displacements, and/or because they are worn away rapidly.
Consequently, the associated finite compliance (contact stiffness) can often be neglected, the contact model can be idealized to rigid unilateral behavior in the normal contact direction and rigid Coulomb friction in the tangential contact plane (\emph{Coulomb-Signorini conditions}), imposed between the nominal contact interfaces, and a relatively coarse contact mesh is sufficient.
The friction coefficient is then the only parameter of the contact model, and it is state of the art to identify this for the considered material pairs and ambient conditions from dedicated tests.
When the contacts undergo partial slip / partial liftoff only, in contrast, the predictive capabilities of today's state-of-the-art methods are quite limited \cite{Brake.2021}.
Here, it is common practice to calibrate the parameters of phenomenological models to vibration measurements, and it is known that the transfer of those parameters to other load scenarios or even other structures/systems is not valid.
Over the past decade, an important working hypothesis has been established:
\emph{If we know the as-built interface topography, reasonably accurate predictions of the vibration behavior can be obtained when imposing Coulomb-Signorini conditions.} 
This is supported, among others, by the theoretical and experimental works \cite{Willner.2004,Pesaresi.2017,Armand.2018,Brink.2020,Pinto.2022,Zare.2023,Porter.2023,Yuan.2023}.
Although there seems to be consensus on this hypothesis, it is still an open research question, down to what length scale the topography has to be resolved.
Some studies suggest that form and waviness length scales are more important than roughness \cite{Armand.2018,Yuan.2023}. 
The longer length scales have a strong influence on how the normal pressure is globally distributed within the contact interface, and this, of course, has a crucial effect on where and how much partial slip is induced by vibration loads.
It is important to note that vibrations are associated with length scales that are much longer than those of the roughness.
This is why one may assume that the effects of individual surface roughness asperities will average out when looking at integral quantities like the modal stiffness / damping provided by an entire contact interface.
\\
The development of efficient simulation methods is hampered by the multi-scale character of the problem:
The local relative displacements within a frictional clamping/joint may be in the sub-micrometer range, while the maximum (absolute) vibration level of the jointed structure may be much larger (away from the clamping/joint), \eg, in the order of several millimeters.
Similarly, the form deviations may be in the micrometer range, while the dimensions of the contact interfaces may be in the range of millimeters to centimeters, and the dimensions of the parts may be in the decimeter or meter range.
Resolving the solids and all relevant length scales in one finite element (FE) model is computationally infeasible \cite{hyun2004,pei2005,yastrebov2011}.
First, the resulting mathematical model order / number of degrees of freedom would lead to prohibitive computational effort.
Second, numerical stability and convergence problems are known to arise when imposing rigid contact constraints on sufficiently fine FE meshes, see \eg \cite{jewell20}.
\\
Due to the above described difficulties, it is the prevailing practice to consider the nominal, smooth geometry in the FE model, and to account for the neglected length scales of the surface topography via \emph{constitutive contact laws}.
An example for the latter are statistically averaged asperity models such as the classical Greenwood-Williamson model \cite{gree1966,john1989}.
Those provide an in general nonlinear relation between pressure/traction and relative displacement \cite{menq1986a,sanl1996a,popp2003a,Eriten.2010}.
Important limitations of those models are that they neglect the interaction among asperities and assume a rather specific shape (spherical or ellipsoidal) of the asperities (without good justification).
To overcome this, one may consider a pair of representative rough surface patches, simulate their contact interaction, and use the results to obtain constitutive laws \cite{Willner.2004,Willner.2008,Willner.2008b,Willner.2009,Zakharov.2020}.
The rough surface patches can either be obtained by scanning a real surface, or synthesizing one from an appropriate statistical model.
To describe the deformation of the underlying solids, the Boundary Element (BE) method is quite popular.
More specifically, elastic half-space theory is used, in accordance with Boussinesq and Cerruti \cite{love1906,john1989,barber2022}, and the surface is commonly discretized using a regular grid \cite{tian1996,Willner.2008b,putigano2012}.
This permits to obtain closed-form expressions for the influence coefficients / elements of the compliance matrix, relating the traction applied to one patch to the elastic displacement at another.
A downside of this method is its limitation to linear-elastic material behavior.
On surface roughness asperity level, stresses are so concentrated that plastic deformation is well-known to be relevant.
To approximate the inelastic asperity deformation, a yield limit can be imposed on the normal pressures \cite{Willner.2004}.
\\
A more integrated multi-scale BE/FE method has been pursued by Salles \etal \cite{armand2017,Armand.2018,Pesaresi.2018,Armand.2019,Yuan.2023}.
Those efforts use the method developed by Gallego \etal \cite{Gallego.2010}.
This method was initially intended for modeling fretting wear (by updating the gap according to an Archard-type law depending on the dissipated energy), but it has also been used for damping induced by partial slip.
On the FE level, contact is treated in terms of constitutive laws, whose parameters are informed by BE simulations.
More specifically, the BE domain receives rigid-body loads in accordance with the contact forces determined by the FE model.
With the BE method, pressure/gap and contact stiffness are determined and adopted in the FE model.
As the FE mesh is much coarser than the BE grid, the BE results have to be averaged in space.
Also, to obtain a single pressure/gap and contact stiffness parameter, the results are averaged over the vibration cycle, too.
Due to the restriction to rigid-body loads, and the cycle-averaging, the BE/FE coupling is to be characterized as rather loose.
It must be emphasized that no numerical validation against a scale resolving / monolithic FE model of this approach is known to date.
The method proposed by Bonari \etal \cite{Bonari.2020} implements a somewhat tighter coupling:
The relative displacement at each integration point of the FE model is fed into the BE simulation, which is directly used to obtain a contact pressure to be used further in the FE model.
It should be remarked that the method was restricted to the friction-less and static case.
Further, at each integration point, a \emph{separate BE model} is setup up, each of which considers an individual rough surface patch.
Consequently, there is no cross-coupling among FE integration points via the BE domain.
This seems justified, as the authors write, in the case of a strong separation of length scales captured by FE and BE domain, respectively, and implies that the FE domain accurately describes the local compliance related to form and waviness.
\\
Compared to the aforementioned efforts, the method proposed in the present work relies on a single BE model for the entire contact interface and a tight coupling with the FE model.
The latter is achieved by enforcing displacement compatibility and force equilibrium in the far field of the BE domain.
This way, rigid-body motion as well as elastic deformation of the interface, as obtained from the FE model, is consistently represented in the BE model.
Likewise, the actual contact stress distribution, as obtained from the BE model, is consistently accounted for in the FE model.
The proposed multi-scale method is described in \sref{method}.
The method is numerically validated against a scale resolving / monolithic FE model in \sref{results}.
This article ends with concluding remarks in \sref{conclusions}.

\section{Proposed multi-scale method\label{sec:method}}
As problem setting, a single body or a system of multiple bodies is considered.
Frictional-unilateral interactions are accounted for at the contacts.
In the present work, the focus is placed on the case of linear behavior everywhere outside the contact interface.
This is exploited in \ssrefs{ROM}-\ssrefo{condensation}.
The application to problems involving more sources of nonlinear behavior is feasible, and the example of friction-clamped (\eg cantilevered) structures undergoing large deflections, or the example of finite sliding seem particularly interesting perspectives of future work.
\begin{table}[b]
\small
\centering
\caption{Idea of proposed multi-scale method.}
\label{tab:idea}
\begin{tabular}{@{}lccc@{}}
\toprule
property & FE model & BE model\\
\midrule
purpose & structural dynamics & contact mechanics\\
mesh & coarse arbitrary 3D mesh & fine regular 2D grid\\
contact topography & nominal, smooth & real; form down to finite wavelength\\
\bottomrule
\end{tabular}
\end{table}
\\
\begin{figure}[h]
 \centering
 \def\svgwidth{1\textwidth}
\begingroup%
  \makeatletter%
  \providecommand\color[2][]{%
    \errmessage{(Inkscape) Color is used for the text in Inkscape, but the package 'color.sty' is not loaded}%
    \renewcommand\color[2][]{}%
  }%
  \providecommand\transparent[1]{%
    \errmessage{(Inkscape) Transparency is used (non-zero) for the text in Inkscape, but the package 'transparent.sty' is not loaded}%
    \renewcommand\transparent[1]{}%
  }%
  \providecommand\rotatebox[2]{#2}%
  \newcommand*\fsize{\dimexpr\f@size pt\relax}%
  \newcommand*\lineheight[1]{\fontsize{\fsize}{#1\fsize}\selectfont}%
  \ifx\svgwidth\undefined%
    \setlength{\unitlength}{598.22211906bp}%
    \ifx\svgscale\undefined%
      \relax%
    \else%
      \setlength{\unitlength}{\unitlength * \real{\svgscale}}%
    \fi%
  \else%
    \setlength{\unitlength}{\svgwidth}%
  \fi%
  \global\let\svgwidth\undefined%
  \global\let\svgscale\undefined%
  \makeatother%
  \begin{picture}(1,0.3966736)%
    \lineheight{1}%
    \setlength\tabcolsep{0pt}%
    \put(0,0){\includegraphics[width=\unitlength]{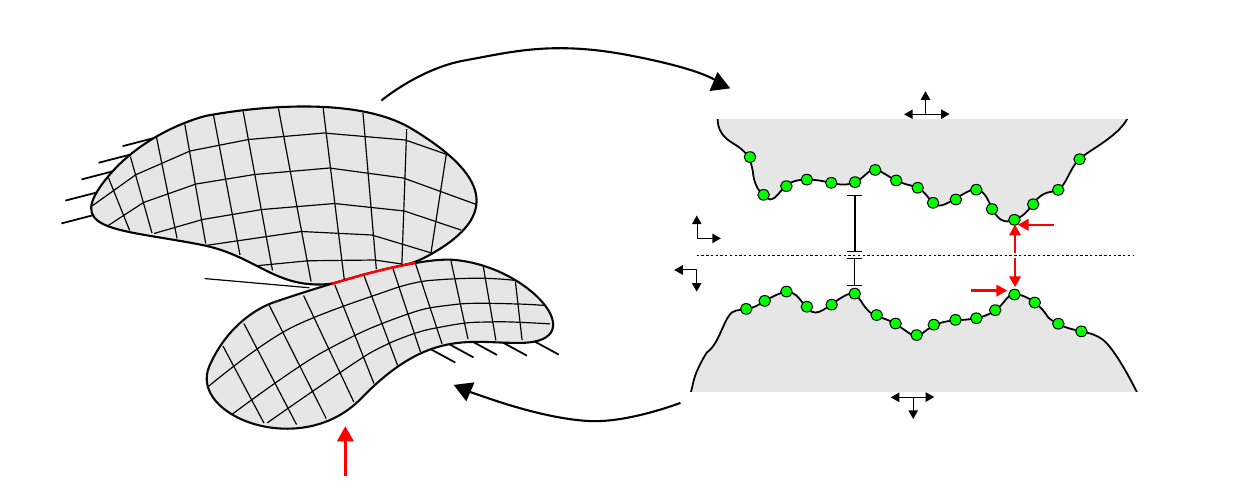}}%
    \put(0.55049245,0.23133888){\color[rgb]{0,0,0}\makebox(0,0)[lt]{\lineheight{1.25}\smash{\begin{tabular}[t]{l}$z^{(1)}$\end{tabular}}}}%
    \put(0.58111756,0.20265634){\color[rgb]{0,0,0}\makebox(0,0)[lt]{\lineheight{1.25}\smash{\begin{tabular}[t]{l}$x^{(1)}$\end{tabular}}}}%
    \put(0.55093868,0.14668294){\color[rgb]{0,0,0}\makebox(0,0)[lt]{\lineheight{1.25}\smash{\begin{tabular}[t]{l}$z^{(2)}$\end{tabular}}}}%
    \put(0.50181882,0.17658006){\color[rgb]{0,0,0}\makebox(0,0)[lt]{\lineheight{1.25}\smash{\begin{tabular}[t]{l}$x^{(2)}$\end{tabular}}}}%
    \put(0.77548119,0.20501359){\color[rgb]{0,0,0}\makebox(0,0)[lt]{\lineheight{1.25}\smash{\begin{tabular}[t]{l}$p_{\mrm n,j}$\end{tabular}}}}%
    \put(0.28768293,0.02006138){\color[rgb]{0,0,0}\makebox(0,0)[lt]{\lineheight{1.25}\smash{\begin{tabular}[t]{l}$\mm f_{\mrm{ex}}$\end{tabular}}}}%
    \put(0.21165545,0.23761564){\color[rgb]{0,0,0}\makebox(0,0)[lt]{\lineheight{1.25}\smash{\begin{tabular}[t]{l}$\mm M$\end{tabular}}}}%
    \put(0.24224608,0.24093742){\color[rgb]{0,0,0}\makebox(0,0)[lt]{\lineheight{1.25}\smash{\begin{tabular}[t]{l}$\mm K$\end{tabular}}}}%
    \put(0.01640204,0.17032325){\color[rgb]{0,0,0}\makebox(0,0)[lt]{\lineheight{1.25}\smash{\begin{tabular}[t]{l}boundary with \\nominal geometry\end{tabular}}}}%
    \put(0.40828694,0.37725155){\color[rgb]{0,0,0}\makebox(0,0)[lt]{\lineheight{1.25}\smash{\begin{tabular}[t]{l}$\Wtra\mm q$\end{tabular}}}}%
    \put(0.45550558,0.03019591){\color[rgb]{0,0,0}\makebox(0,0)[lt]{\lineheight{1.25}\smash{\begin{tabular}[t]{l}$\W\mm\lambda$\end{tabular}}}}%
    \put(0.92020646,0.19348404){\color[rgb]{0,0,0}\makebox(0,0)[lt]{\lineheight{1.25}\smash{\begin{tabular}[t]{l}reference\\level\end{tabular}}}}%
    \put(0.85136727,0.21326486){\color[rgb]{0,0,0}\makebox(0,0)[lt]{\lineheight{1.25}\smash{\begin{tabular}[t]{l}$p_{\mrm t1,j}$\end{tabular}}}}%
    \put(0.64364364,0.21395315){\color[rgb]{0,0,0}\makebox(0,0)[lt]{\lineheight{1.25}\smash{\begin{tabular}[t]{l}$h_l^{(1)}$\end{tabular}}}}%
    \put(0.64537056,0.16833808){\color[rgb]{0,0,0}\makebox(0,0)[lt]{\lineheight{1.25}\smash{\begin{tabular}[t]{l}$h_l^{(2)}$\end{tabular}}}}%
    \put(0.64206751,0.27400208){\color[rgb]{0,0,0}\makebox(0,0)[lt]{\lineheight{1.25}\smash{\begin{tabular}[t]{l}$C^{(1)}$\end{tabular}}}}%
    \put(0.66097642,0.096696){\color[rgb]{0,0,0}\makebox(0,0)[lt]{\lineheight{1.25}\smash{\begin{tabular}[t]{l}$C^{(2)}$\end{tabular}}}}%
    \put(0.73793971,0.16485958){\color[rgb]{0,0,0}\makebox(0,0)[lt]{\lineheight{1.25}\smash{\begin{tabular}[t]{l}$p_{\mrm t1,j}$\end{tabular}}}}%
    \put(0.82099962,0.17515435){\color[rgb]{0,0,0}\makebox(0,0)[lt]{\lineheight{1.25}\smash{\begin{tabular}[t]{l}$p_{\mrm n,j}$\end{tabular}}}}%
    \put(0.73383548,0.32680078){\color[rgb]{0,0,0}\makebox(0,0)[lt]{\lineheight{1.25}\smash{\begin{tabular}[t]{l}$\infty$\end{tabular}}}}%
    \put(0.76715314,0.30133865){\color[rgb]{0,0,0}\makebox(0,0)[lt]{\lineheight{1.25}\smash{\begin{tabular}[t]{l}$\infty$\end{tabular}}}}%
    \put(0.72203758,0.04862606){\color[rgb]{0,0,0}\makebox(0,0)[lt]{\lineheight{1.25}\smash{\begin{tabular}[t]{l}$\infty$\end{tabular}}}}%
    \put(0.75295456,0.07366998){\color[rgb]{0,0,0}\makebox(0,0)[lt]{\lineheight{1.25}\smash{\begin{tabular}[t]{l}$\infty$\end{tabular}}}}%
    \put(0.69477732,0.07245098){\color[rgb]{0,0,0}\makebox(0,0)[lt]{\lineheight{1.25}\smash{\begin{tabular}[t]{l}$\infty$\end{tabular}}}}%
    \put(0.70013642,0.3020318){\color[rgb]{0,0,0}\makebox(0,0)[lt]{\lineheight{1.25}\smash{\begin{tabular}[t]{l}$\infty$\end{tabular}}}}%
  \end{picture}%
\endgroup%
 \caption{Schematic illustration of the coupled FE-BE model.}
 \label{fig:FE-BE-MSM}
\end{figure}
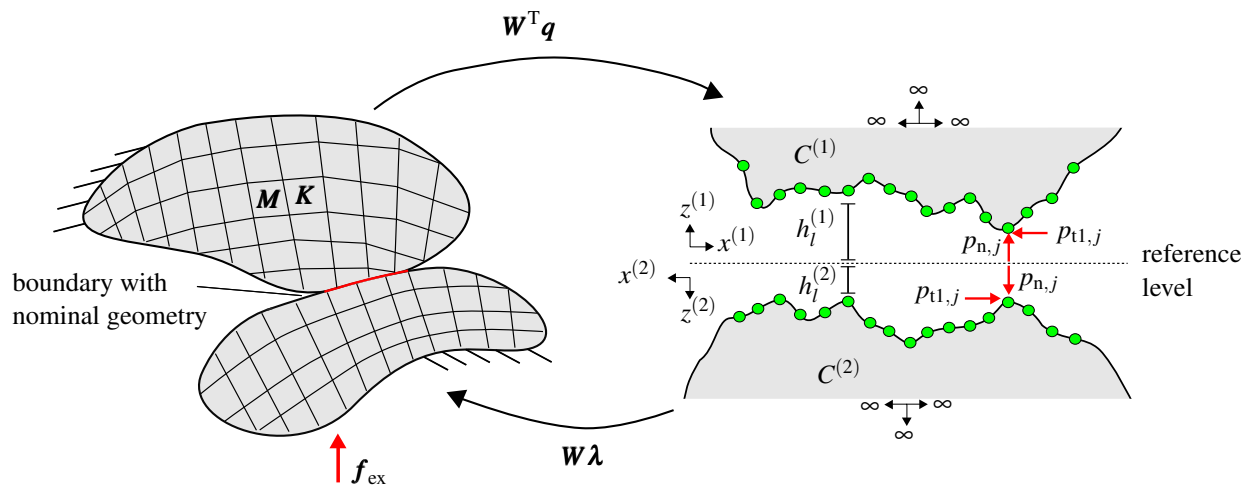
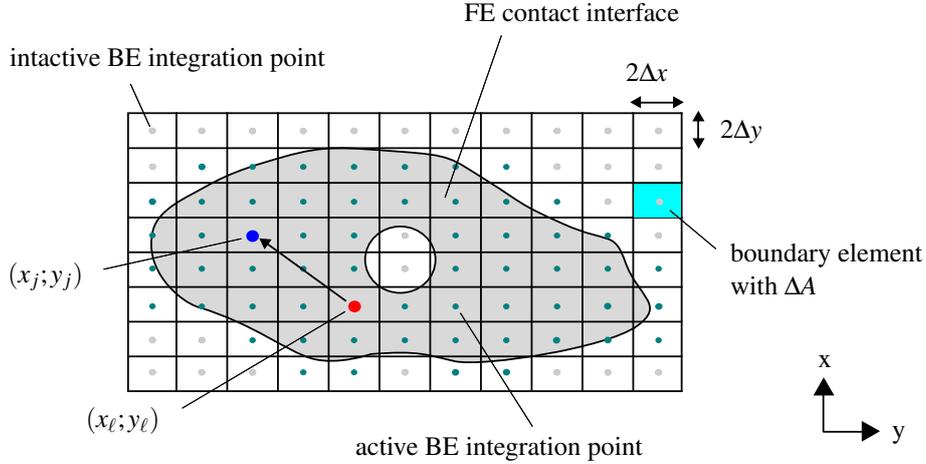
\begin{figure}[h]
 \centering
 \def\svgwidth{0.8\textwidth}
\begingroup%
  \makeatletter%
  \providecommand\color[2][]{%
    \errmessage{(Inkscape) Color is used for the text in Inkscape, but the package 'color.sty' is not loaded}%
    \renewcommand\color[2][]{}%
  }%
  \providecommand\transparent[1]{%
    \errmessage{(Inkscape) Transparency is used (non-zero) for the text in Inkscape, but the package 'transparent.sty' is not loaded}%
    \renewcommand\transparent[1]{}%
  }%
  \providecommand\rotatebox[2]{#2}%
  \newcommand*\fsize{\dimexpr\f@size pt\relax}%
  \newcommand*\lineheight[1]{\fontsize{\fsize}{#1\fsize}\selectfont}%
  \ifx\svgwidth\undefined%
    \setlength{\unitlength}{314.08039527bp}%
    \ifx\svgscale\undefined%
      \relax%
    \else%
      \setlength{\unitlength}{\unitlength * \real{\svgscale}}%
    \fi%
  \else%
    \setlength{\unitlength}{\svgwidth}%
  \fi%
  \global\let\svgwidth\undefined%
  \global\let\svgscale\undefined%
  \makeatother%
  \begin{picture}(1,0.4930344)%
    \lineheight{1}%
    \setlength\tabcolsep{0pt}%
    \put(0,0){\includegraphics[width=\unitlength]{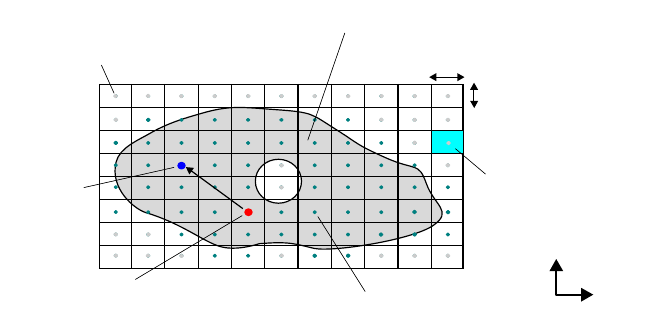}}%
    \put(0.91606655,0.03593776){\color[rgb]{0,0,0}\makebox(0,0)[lt]{\lineheight{1.25}\smash{\begin{tabular}[t]{l}y\end{tabular}}}}%
    \put(0.84214953,0.10712218){\color[rgb]{0,0,0}\makebox(0,0)[lt]{\lineheight{1.25}\smash{\begin{tabular}[t]{l}x\end{tabular}}}}%
    \put(0.03380739,0.41108392){\color[rgb]{0,0,0}\makebox(0,0)[lt]{\lineheight{1.25}\smash{\begin{tabular}[t]{l}intactive BE integration point  \end{tabular}}}}%
    \put(0.48832198,0.45654255){\color[rgb]{0,0,0}\makebox(0,0)[lt]{\lineheight{1.25}\smash{\begin{tabular}[t]{l}FE contact interface\end{tabular}}}}%
    \put(0.37955086,0.01866502){\color[rgb]{0,0,0}\makebox(0,0)[lt]{\lineheight{1.25}\smash{\begin{tabular}[t]{l}active BE integration point  \end{tabular}}}}%
    \put(0.75431556,0.21419676){\color[rgb]{0,0,0}\makebox(0,0)[lt]{\lineheight{1.25}\smash{\begin{tabular}[t]{l}boundary element\\with $\Delta A$\end{tabular}}}}%
    \put(0.64981317,0.39636748){\color[rgb]{0,0,0}\makebox(0,0)[lt]{\lineheight{1.25}\smash{\begin{tabular}[t]{l}$2 \Delta x$\end{tabular}}}}%
    \put(0.74435884,0.33824393){\color[rgb]{0,0,0}\makebox(0,0)[lt]{\lineheight{1.25}\smash{\begin{tabular}[t]{l}$2 \Delta y$\end{tabular}}}}%
    \put(0.10981879,0.04694599){\color[rgb]{0,0,0}\makebox(0,0)[lt]{\lineheight{1.25}\smash{\begin{tabular}[t]{l}$(x_{\ell};y_{\ell})$\end{tabular}}}}%
    \put(0.03178381,0.1925065){\color[rgb]{0,0,0}\makebox(0,0)[lt]{\lineheight{1.25}\smash{\begin{tabular}[t]{l}$(x_{j};y_{j})$\end{tabular}}}}%
  \end{picture}%
\endgroup%
 \caption{Schematic illustration of the BE grid. The entire apparent contact area is depicted in gray.}
 \label{fig:BE_grid}
\end{figure}
%
Key features of the proposed multi-scale method are summarized in \tref{idea}.
The structural dynamics is described with an FE model, the contact mechanics with a BE model.
Thus, a relatively coarse mesh in the FE domain is sufficient, both at the contact interface and within the bulk, and only the nominal contact topography is accounted for.
In contrast, a fine mesh is needed in the BE domain, where he actual contact topography is resolved.
The nominal contact interface of the FE model is coupled via displacement compatibility and force equilibrium conditions with the BE model.
Displacement compatibility is weakly enforced by imposing it in the far field of the BE domain.
\\
Within the BE domain, elastic half-space theory will be used.
As stated in the introduction, this has the important advantage that the elements of the compliance matrix can be expressed in closed form.
More specifically, this is the case for a regular grid, which is considered in the present work.
On the other hand, relying on elastic half-space theory induces two fundamental limitations:
\begin{enumerate}
    \item One is limited to \emph{linear elasticity}.
    \item Valid results can only be expected if the \emph{actual contact dimensions are much smaller than the dimensions of the underlying solids}.
\end{enumerate}
It is useful to recall that approximations are available for plastic deformation of asperities and wear, as described in the introduction, alleviating the first limitation.
But an extension of the proposed multi-scale method towards nonlinear elasticity or plasticity within the sub-surface part of the contact region does not seem straight-forward.
The \emph{actual contact dimensions} depend on the resolution.
The finer the real topography is resolved, the smaller is the actual contact area, and hence the better justified is the proposed multi-scale method.
On the other hand, the method cannot be expected to perform well if the contact topography is considered as nominal, smooth on the microscopic scale.
This implies that an appropriate model of the surface topography is needed, which has to be obtained by scanning or by simulation.
Reduced accuracy is to be expected if the actual contact area occurs near an edge of the underlying bodies.
In fact, this happens for the benchmark considered in \sref{results}, where the resulting error produced by \emph{edge effects} typical for Boussinesq-Cerruti theory will be inspected.
\\
The FE model captures the \emph{global structural compliance} linked to the geometry, material and boundary conditions of the considered system.
The BE model captures the \emph{local compliance of the contact region}  linked to the underlying bulk material and features of the real contact topography.
In a certain sense, the contact region is described in both the FE and in the BE model.
The inertia forces within the contact region are captured in the FE model.\footnote{
Since only the nominal geometry is considered in the FE model, the effect of the topography deviations on the distribution of the inertia forces is neglected.
This seems reasonable thanks to the microscopic scale of the topography deviations.
}
A coarse FE model does not capture the local compliance of the contact region.
If one were to increase the mesh density in the FE model until the results stabilize, however, one would have considered the compliance of the contact region twice, namely both in the FE and the BE model.
In this sense, the proposed multi-scale method is inconsistent with respect to the spatially continuous reference.
However, the avoidance of a fine resolution of the FE model is precisely the motivation for the proposed multi-scale method.
In other words, using an extremely fine FE model defies the purpose of the proposed multi-scale method, and is expected to lead to higher computation effort without improved accuracy.
\\
The \emph{benefits} of the proposed multi-scale method are as follows:
First, only the contact surface must be finely meshed.
In fact, the problem will be further restricted to the active parts of the contact interfaces.
In contrast to FE modeling, the fine contact mesh does not propagate into the volume.
Second, the \emph{contact problem is quasi-static}; \ie, no inertia forces are explicitly accounted for within the BE domain.
This facilitates the numerically robust and efficient solution of the contact problem (reduced index of the differential-algebraic equation system; no spurious oscillations) \cite{Ascher.1998,Leine.2004}.
Note that the quasi-static treatment is well-justified because the inertia forces of the entire domain are already captured in the FE model.
Third, if the topography is replaced, \eg, in the course of uncertainty quantification or parameter studies, the FE model (and also its reduced version proposed later) remains unchanged.
It is useful to emphasize that the contact conditions are only imposed in the BE model, whereas the nominal contact interface in the FE model merely serves for the coupling of the two domains.
Consequently, no tough requirements are placed on the FE model at the interface.
For instance, incompatible/non-matching meshes and different element types can be quite easily handled.
That is the fourth and last benefit of the proposed multi-scale method.
\\
In the following, the proposed multi-scale method is mathematically formulated.
The coupled FE-BE problem is described in \ssref{math}.
A reduction of the FE model is proposed in \ssref{ROM}.
It is shown how the coupled problem can be restricted to the BE domain in \ssref{condensation}.
Finally, details on the solution of the contact problem, including an active set strategy are presented in \ssref{algorithm}.

\subsection{Formulation of the coupled FE-BE model\label{sec:math}}
The coupled FE-BE model is schematically illustrated in \fref{FE-BE-MSM}.
The dynamic force balance within the FE model is described by the ordinary differential equation system \erefo{FEM}, while the algebraic equation system \erefo{BEM} describes the quasi-static behavior of the contacting elastic half spaces.
\ea{
\mm f(\mm q,\dot{\mm q}, \ddot{\mm q},t) &=& \W\mm\lambda\fk \label{eq:FEM} \\
\mm g &=& -\mm h + \mm C\mm\lambda + \Wtra\mm q \fp \label{eq:BEM}
}
%
Herein, $\mm q\in\mathbb R^{\N\times 1}$ is the vector of $\N$ nodal coordinates of the FE model, overdot denotes derivative with respect to time $t$.
$\mm\lambda\in\mathbb R^{3\C\times 1}$ is the vector of Lagrange multipliers which can be interpreted as three-dimensional contact forces acting on $\C$ BE integration points.
The matrix $\W\in\mathbb R^{\N\times 3\C}$ describes how the contact forces act on the FE model. 
$\mm f\left(\mm q,\dot{\mm q},\ddot{\mm q},t\right)$ is the vector of all remaining (non-contact) forces such as inertia, restoring, damping, and excitation forces.
Linear behavior of the FE model is not yet required, as it becomes relevant only for the simplifications proposed in \ssref{ROM}-\ssrefo{condensation}.
$\mm g\in\mathbb R^{\C\times 1}$ is the vector of relative contact displacements at the respective BE integration point, $\mm h\in\mathbb R^{\C\times 1}$ describes the contact topography (height profile), and $\mm C\in\mathbb R^{3\C\times3\C}$ is the compliance matrix of the BE model.
$\mm g$ will be referred to as \emph{gap} in the present work; in the BE context, $-\mm g$ (interference) is commonly used as variable instead.
Note that the larger $\mm h$, the smaller $\mm g$, which is why there is a minus sign in \eref{BEM}.
Finally, $\Wtra\mm q$ describes that part of the relative displacement at the contact interface which is imposed by the displacement $\mm q$ of the FE model.
Note that the virtual work done by the contact forces must satisfy $\delta W_{\mrm c}=\mm\lambda^{\mrm T}\delta\mm g = \mm f_{\mrm{c}}^{\mrm T}\delta\mm q$;
using \eref{BEM}, we find that $\delta\mm g = \Wtra\delta \mm q$, so that we can follow that the contact forces acting on the FE model are indeed $\mm f_{\mrm c}= \W\mm\lambda$.
It is important to emphasize that the term $\Wtra\mm q$ in \eref{BEM} ensures that the interface displacement of the FE model is compatible with the far-field displacement of the BE model.
This way, effectively, the compliance of the pair of elastic half spaces described by the BE model is aligned in series with that of the underlying FE model.
Likewise, the contact forces are distributed to the FE model in a way consistent with the principle of virtual work.
\\
The BE model is implemented on a regular grid (\fref{BE_grid}).
It is important to emphasize that the BE grid spans the entire apparent contact interface; integration points outside the interface, \eg, within bore holes, are associated with a $-\infty$ height, so that they are never activated.
For ease of presentation, only one contact interface is shown in the schematic diagram, and treated in the following.
In practice, it may be useful to split the problem into multiple contacts regions, in particular if the interfaces are disjoint.
Each integration point of the BE model is associated with the same area $\Delta A=2\Delta x\cdot 2\Delta y$.
Consequently, we simply have $\mm \lambda = \mm p\Delta A$, where $\mm p$ is the contact stress vector.
The gap $\mm g$ is sorted as
\ea{
\mm g = \vector{\mm g_1 \\ \vdots \\ \mm g_{\C}}\fk \quad \mm g_j = \vector{g_{\mrm{n},j}\\ g_{\mrm{t1},j}\\ g_{\mrm{t2},j}}\fk \label{eq:gap}
}
where $g_{\mrm{n},j}$ is the normal and $g_{\mrm{t1},j}$, $g_{\mrm{t2},j}$ are the two orthogonal gaps in the tangential plane.
$\mm \lambda$, $\mm p$ and $\mm h$ are sorted analogous to $\mm g$.
The contact topography $\mm h=\mm h^{(1)}+\mm h^{(2)}$ is composed of the height profiles $\mm h^{(1)}$, $\mm h^{(2)}$ of the two paired surfaces, where $\mm h=\left[h_1;0;0;h_2;0;0;\ldots\right]$ contains the composite height values $h_j$ at each integration point $j$ (semicolon denotes vertical concatenation).
The individual height profile is counted positive in the respective negative $z$-direction, so that surface hills have greater height than valleys.
To ensure that the gap in the non-deformed ($\Wtra\mm q=\mm 0$), unstressed ($\mm\lambda = \mm 0$) configuration is $\mm g=-\mm h$ (\cf \eref{BEM}), the height is counted from the same reference level as indicated in \fref{FE-BE-MSM}-right.
This configuration must be compatible with contact constraints; \ie, $g_{\mrm{n},j}\geq$ must hold for all integration points $j$ (no initial interference allowed).
It is an open research question whether the oblique (shoulder-to-shoulder) contact among non-aligned peaks on either side of the contact has a significant effect on the structural dynamics.
By simply adding the height profiles, such effects are ignored in the present work for simplicity.
To justify this, it was verified that the topographies considered in \sref{results} are indeed sufficiently flat.
The matrix $\mm C$ is the sum of the compliance matrices $\mm C^{(1)}$ and $\mm C^{(2)}$ of the two half spaces in contact (for appropriate orientation of the coordinate systems, \cf \fref{FE-BE-MSM}-right),
\ea{
\mm C = \mm C^{(1)} + \mm C^{(2)} = \left[{\mm C}_{j\ell}\right] \fk \label{eq:C}
}
$\mm C$ is composed of $\C\times\C$ blocks of the $3\times 3$ sub-matrices ${\mm C}_{j\ell}$.
The sub-matrices ${\mm C}_{j\ell}$ are defined in \aref{C}, based on the well-known Boussinesq and Cerruti potential theory, for which closed-form expressions are available in the considered case of a regular grid.
\\
The contact laws in \eref{claws} are expressed as inclusions into the normal cone, $\mathcal N_{\mathcal C}$, to the admissible set, $\mathcal C$, of the contact stress $\mm p$,
\ea{
&-\mm{\gamma} \in \mathcal N_{\mathcal C}\left(\mm p\right)\fk \label{eq:claws} \\
&	\mathcal C = \mathcal C_1 \times \ldots \times \mathcal C_{\C}\fk \quad \mathcal C_j = \mathbb R_0^+ \times \mathcal D\left(\mu p_{\mathrm{n},j}\right)\fp \label{eq:admissibleset}
}
To treat friction, the contact conditions are formulated on velocity level, where $\mm\gamma=\dot{\mm g}$, and apply to all active contacts (closed normal gap, $g_{\mrm{n},j}=0$).
The admissible set $\mathcal C$ is given by \eref{admissibleset}.
In the normal direction, unilateral interaction is considered (Signorini conditions).
Coulomb's law of dry friction is considered in the tangential contact plane.
Herein, $\mathcal D\left(r\right)$ denotes the planar disk of radius $r$.
\\
\erefs{FEM}-\erefo{BEM} represent a system of differential-algebraic equations subjected to the set-valued contact conditions defined in \eref{claws}.
The solution of this coupled problem requires dedicated numerical techniques.
Before the contact conditions are further resolved in \ssref{algorithm}, the FE model is reduced (\ssref{ROM}) and the problem is restricted to the BE domain (\ssref{condensation}).

\subsection{Reduction of the FE model via component mode synthesis\label{sec:ROM}}
In the present work, we focus on linear-elastic behavior within the FE domain.
We thus set
\ea{
\mm f(\mm q,\dot{\mm q},\ddot{\mm q},t)=\mm M\ddot{\mm q} + \mm K\mm q - \mm f_{\mrm{ex}}\fk \label{eq:flin}
}
where $\mm M\in\mathbb R^{\N\times\N}$ is the symmetric and positive definite FE mass matrix, $\mm K\in\mathbb R^{\N\times\N}$ is the symmetric and positive semi-definite FE stiffness matrix, and $\mm f_{\mrm{ex}}\in\mathbb R^{\N\times 1}$ is the vector of imposed forces with known explicit time dependence.
The latter generally contains static forces which are responsible for the initial load applied to the joints, and it contains dynamic loads as well.
Other linear forces, in particular, viscous damping could be added easily to the following derivation, but we prefer to omit this for reducing the writing effort.
In linear structural dynamics, component mode synthesis is quite popular.
In the present context, it is useful to reduce the model order and to facilitate the restriction to the contact problem proposed in \ssref{condensation}.
The Craig-Bampton method is by far the most popular component mode synthesis method.
It is the natural choice for jointed structures, as the contacts should remain in a macroscopic stick phase, so that fixed-interface methods are expected to converge better than free-interface methods.
To keep this article self-contained, the Craig-Bampton method is briefly described in the following and applied to the considered problem setting.
\\
The method relies on splitting the vector of coordinates $\mm q$ in boundary coordinates $\qb$ and inner coordinates ${\mm q}_{\mrm i}$.
For convenience, we assume that $\mm q$ has been sorted accordingly as $\mm q = [\qb;\mm q_{\mrm i}]$.
Only the boundary coordinates (at the nominal contact interface) are involved into the coupling to the BE model, so that  $\Wtra\mm q = \Wbtra\qb$, and $\W\mm\lambda = \left[\Wb;\mm 0\right]\mm \lambda$.
The coordinates $\mm q$ are approximated as
\ea{
\mm q = \vector{\qb \\ \mm q_{\mrm i}}  \simeq \underbrace{\matrix{cc}{\mm I & \mm 0 \\ \mm\Psi & \mm\Theta}}_{\mm R} \underbrace{\vector{\qb\\ \mm\eta}}_{\tilde{\mm q}}\fk \label{eq:CB}
}
in terms of component modes (columns of matrix $\mm R$) and associated coordinates (elements of vector $\tilde{\mm q}$).
To facilitate the coupling of FE and BE models, the boundary coordinates $\qb$ are retained.
In contrast, the inner coordinates are replaced by a reduced set of generalized coordinates $\mm\eta$.
In \eref{CB}, $\mm \Theta= \left[\mm\theta_1,\ldots,\mm\theta_{\nmod}\right]$ denotes the set of $\nmod$ lowest-frequency fixed-interface normal modes (restricted to the inner coordinates), and the first columns correspond to the static constraint modes with $\mm\Psi = -\mm K\inv_{\mathrm{ii}}\mm K_{\mathrm{ib}}$.
The $j$-th column of $\mm\Psi$ represents the static deflection of $\mm q_{\mrm i}$ due to unit displacement imposed at the $j$-th element of $\mm{q}_{\mrm b}$ (with the other boundary coordinates fixed).
The modes $\mm\theta_m$ and associated angular frequencies $\omega_m$ are obtained from the eigenvalue problem $\left({\mm K}_{\mrm{ii}} - \omega_m^2{\mm M}_{\mrm{ii}} \right)\mm\theta_m = \mm 0$, where ${\mm K}_{\mrm{ii}}$, ${\mm M}_{\mrm{ii}}$ are the restrictions of $\mm K$ respectively $\mm M$ to the square block associated with ${\mm q}_{\mrm i}$.
The modes are mass-normalized; \ie, $\mm\theta\tra_m{\mm M}_{\mrm{ii}}\mm\theta_m = 1$.
By requiring that the residual produced by substituting \eref{flin} and \eref{CB} into \eref{FEM} is orthogonal with respect to $\mm R$, one obtains the reduced-order model
\ea{
\tilde{\mm M}\ddot{\tilde{\mm q}} + \tilde{\mm K}\tilde{\mm q} = \mm R^{\mrm T}\W\mm\lambda + \tilde{\mm f}_{\mrm{ex}} \fp \label{eq:ROM}
}
Herein,
\ea{
\tilde{\mm M} &=& \mm R^{\mrm T} \mm M\mm R = \matrix{cc}{\tilde{\mm M}_{\mrm{bb}} & \tilde{\mm M}_{\mrm{bi}}\\ \text{sym.} & \mm I} \fk \label{eq:Mtil} \\
\tilde{\mm K} &=& \mm R^{\mrm T} \mm K\mm R = \matrix{cc}{\tilde{\mm K}_{\mrm{bb}} & \mm 0\\ \text{sym.} & \operatorname{diag}\left(\omega_m^2\right)} \fk \label{eq:Ktil} \\
\tilde{\mm f}_{\mrm{ex}} &=& \mm R^{\mrm T}{\mm f}_{\mrm{ex}}\fk \quad \mm R^{\mrm T}\W = \vector{\Wb \\ \mm 0}\fp \label{eq:ftil}
}
It is an important property of the Craig-Bampton method that the reduced model captures the static compliance with respect to loads applied at the boundary as accurately as the parent FE model.
The retained normal modes should cover the relevant frequency range of the dynamic response, and for most vibration problems a very small number of normal modes is sufficient.

\subsection{Block condensation to the BE domain\label{sec:condensation}}
Different techniques are available for casting the ordinary differential equation system \erefo{ROM} into an algebraic equation system.
We refer to these collectively as \emph{time discretization} methods.
Examples are numerical time step integration, or any technique belonging to the wide class of methods of mean weighted residuals, including collocation and Harmonic Balance.
In any case, the resulting algebraic equation system is linear in $\tilde{\mm q}$ and $\mm\lambda$, or more generally the parameters associated with their time discrete approximation, because the differential equation system \erefo{ROM} is linear.
This can be exploited by applying a block condensation.
More specifically, one solves for $\tilde{\mm q}$ as function of $\mm\lambda$, and substitutes this into \eref{BEM}.
This way, the problem is restricted to the BE domain.
\\
The numerical illustrations in the present work are limited to quasi-static analysis.
Hence, the above described restriction is shown for the simple case of a static force balance; \ie, $\dot{\mm q}=\mm 0$, $\ddot{\mm q}=\mm 0$, which readily casts \eref{ROM} into an algebraic equation system.
More specifically, the inertia term $\tilde{\mm M}\ddot{\tilde{\mm q}}$ in \eref{ROM} vanishes.
Thanks to the block diagonal form of $\tilde{\mm K}$ in \eref{Ktil}, we can solve the first hyper-row of \eref{ROM} for $\qb$,
\ea{
\qb &=& \tilde{\mm K}_{\mrm{bb}}^{-1}\left(\Wb\mm\lambda + \tilde{\mm f}_{\mrm{ex,b}}\right) \fp \label{eq:qbconds}
}
Substituting this into \eref{BEM}, we obtain
\ea{
\mm g &=& \mm C_*\mm\lambda + \mm g_{\mrm{ex}}\fk \label{eq:Gconds}\\
\mm C_* &=& \mm C + \Wbtra~\tilde{\mm K}_{\mrm{bb}}^{-1}~\Wb\fk \label{eq:Cstar}\\
\mm g_{\mrm{ex}} &=& \Wbtra~\tilde{\mm K}_{\mrm{bb}}^{-1}~\tilde{\mm f}_{\mrm{ex,b}} - \mm h\fp \label{eq:gex}
}
The setup of the matrix $\mm C_*$ reflects that the compliance of the pair of elastic half spaces is aligned in series with the compliance of the underlying structures.
The variable $\mm g_{\mrm{ex}}$ can be interpreted as the superposition of the initial height profile, and the gap caused by the structural far field displacement due to the imposed loading.
\\
In the quasi-static case, the velocity $\mm\gamma = \dot{\mm g}$ in \eref{claws} is not well-defined.
Instead, we treat a sequence of finite load increments, $\mm f_{\mrm{ex}}^k = \mm f_{\mrm{ex}}^{k-1}+\Delta\mm f_{\mrm{ex}}$, with $k=1,\ldots$ and $\mm f_{\mrm{ex}}^0 = \mm 0$, and determine the resulting gap and contact load incrementally:
\ea{
\mm g^{k} &=& \mm g^{k-1} + \Delta\mm g\fk \quad \mm \lambda^{k} = \mm \lambda^{k-1} + \Delta\mm \lambda\fk \label{eq:incr} \\
\Delta\mm g &=& \mm C_*\Delta\mm\lambda + \Delta \mm g_{\mrm{ex}}\fk \label{eq:BEMincr} \\
-\Delta\mm g &\in & \mathcal N_{\mathcal C}\left(\mm p\right) \fp \label{eq:clawsincr}
}
Recall here that we simply have $\mm p = \mm\lambda/\Delta A$ thanks to the considered regular grid (\fref{BE_grid}).

\subsection{Solution of the contact problem\label{sec:algorithm}}
An important downside of the BE method is that it generally leads to fully populated matrices, in contrast to the FE method which leads to sparse matrices.
More specifically, the compliance matrix $\mm C_*$ in \eref{Cstar} is fully populated.
In fact, this is the case not only due to the matrix $\mm C$ in \eref{C}, but also due to the proposed reduction of the FE model based on component mode synthesis, which also yields a fully populated sub-matrix $\tilde{\mm K}_{\mrm{bb}}$.
To reduce the computational effort, and appropriate \emph{active set strategy} is proposed.
This strategy is described in the following, along with specifics of the algorithm used to solve the restricted contact problem.
\\
The proposed active set strategy consists of multiple steps.
The first step is a geometric restriction similar to \cite{bemporad2015} which takes into account the actual contact topography.
The idea is to consider only integration points on the BE grid that lie on hills of the composite surface.
No error is made if the neglected integration points never close, \ie, if the valleys never touch.
In the simplest implementation, one retains all integration points down to some distance from the highest peak.
The set of retained integration points should be enlarged if contact occurs at any of the points located on a boundary of the retained sets.
Finding an appropriate restriction becomes more complicated if the structural dynamics induce a deformation which exceeds the initial height profile.
This is the case for the out-of-phase bending mode of the considered benchmark system in \sref{results}, which induces a rolling-type deformation.
The effective composite surface, with the deformation superimposed, changes severely in that case, and the active contact region moves from one edge of the interface to the opposing one.
Aside from such extreme cases, a considerable reduction of the retained integration points can be achieved by geometric restriction.
\\
The second step of the proposed active set strategy is to distinguish open (separated) and closed integration points.
Here and in the following, only the geometrically restricted subset of $\C$ points of the complete BE grid is taken into account.
The set of open points, $\mathcal I_{\mrm{sep}}$, is the set of points with open gap and, at the same time, zero pressure, at the previous load level,
\ea{
\mathcal I_{\mrm{sep}} &=& \lbrace j \vert g_{\mrm{n},j}^{k-1} > 0 \land p_{\mrm{n},j}^{k-1} = 0  \rbrace\fk \label{eq:Isep}\\
\mathcal I_{\mrm{cl}} &=& \lbrace 1,\ldots,\C\rbrace \backslash \mathcal I_{\mrm{sep}}\fk \label{eq:Icl}
}
and the remaining points form the set of closed points $\mathcal I_{\mrm{cl}}$.
\\
The third step of the proposed active set strategy is to further differentiate between presumably sticking and active points within the set of closed points.
This step is in line with the strategy presented in \cite{hueber16}.
A prediction is made assuming that all closed points are sticking, $\Delta \mm g^{\mrm{pre}~(\mrm{cl})}=\mm 0$.
Here and in the following, $\square^{(\mrm{cl})}$ denotes the partition of the vector $\square$ associated to the set $\mathcal I_{\mrm{cl}}$, and analogously to other sets.
Using that open contacts have $\Delta\mm\lambda^{\mrm{pre}~(\mrm{sep})}=\mm 0$, we can predict the contact forces from \eref{BEMincr} as
\ea{
\mm\lambda^{\mrm{pre}} &=& \mm\lambda^{k-1} + \Delta\mm\lambda^{\mrm{pre}}\fk \label{eq:lampre} \\
\Delta\mm\lambda^{\mrm{pre}~(\mrm{cl})} &=& - \left(\mm C_*^{(\mrm{cl},\mrm{cl})}\right)^{-1}~\Delta\mm g_{\mrm{ex}}^{(\mrm{cl})}\fp \label{eq:dlamprecl}
}
Herein, $\mm C_*^{(\mrm{cl},\mrm{cl})}$ is the partition of $\mm C_*$ associated to the set $\mathcal I_{\mrm{cl}}$, and analogously for other sets.
The set of presumably sticking points, $\mathcal I_{\mrm{st}}$, contains the points which have been sticking at the previous load level, and have a predicted contact force within the Coulomb cone,
\ea{
\mathcal I_{\mrm{st}} &=& \lbrace j\in \mathcal I_{\mrm{cl}} \vert \Delta \mm g^{k-1}_{\mrm{t},j}=\mm 0 \land \|\mm\lambda^{\mrm{pre}}_{\mrm{t},j}\| < \mu \lambda^{\mrm{pre}}_{\mrm{n},j} \rbrace\fk \label{eq:Ist}\\
\mathcal I_{\mrm{a}} &=& \mathcal I_{\mrm{cl}} \backslash \mathcal I_{\mrm{st}}\fp \label{eq:Ia}
}
The remaining closed points form the set of active points $\mathcal I_{\mrm{a}}$.
\\
Under the conditions that the open points are free from contact forces, $\Delta\mm\lambda^{(\mrm{sep})}=\mm 0$, and the presumably sticking points have $\Delta \mm g^{(\mrm{st})}=\mm 0$, we can derive from \eref{BEMincr}:
\ea{
\Delta\mm g^{(\mrm{a})} &=& \mm G \Delta\mm \lambda^{(\mrm a)} + \mm c\fk \label{eq:Glamc}\\
\mm G &=& \mm C_*^{(\mrm{a,a})}  -  \mm C_*^{(\mrm{a,st})} \left(\mm C_*^{(\mrm{st,st})}\right)^{-1}\mm C_*^{(\mrm{st,a})}\fk \label{eq:G}\\
\mm c &=& \Delta\mm g_{\mrm{ex}}^{(\mrm a)} - \mm C_*^{(\mrm{a,st})} \left(\mm C_*^{(\mrm{st,st})}\right)^{-1} \Delta\mm g_{\mrm{ex}}^{(\mrm{st})}\fp \label{eq:c}
}
This corresponds to a block condensation to the active set; it is exact under the condition that all contacts that are actually sliding are contained in the active set.
The contact problem in \eref{clawsincr} can be restricted accordingly, $-\Delta\mm g^{(\mrm{a})} \in \mathcal N_{\mathcal C}\left(\mm p^{(\mrm a)}\right)$, where the gap and contact force increments of only the active set have to be iterated.
The solution of such implicit algebraic inclusions is a standard task of non-smooth analysis.
A projected Jacobi over-relaxation algorithm was used in the present work, as described in \aref{iai}.

\subsection{Overview of the proposed multi-scale method and discussion of its computational effort\label{sec:overview}}
An overview on the implemented algorithm is given in \fref{flowChart}.
Before the actual simulation, the FE model is set up, the nodes are split into boundary and internal ones, and the model is reduced following the technique described in \ssref{ROM}.
This yields, in particular, the reduced mass and stiffness matrices $\tilde{\mm M}$, $\tilde{\mm K}$, and the reduced load vector $\tilde{\mm f}_{\mrm{ex}}$.
Also, the BE grid is defined over the entire apparent contact interface, and an associated height profile must be specified.
Besides the grid and the height profile, the input parameters of the BE model are the elastic material properties (in the case of a pair of identical isotropic materials, the parameters are the Young's modulus $E$ and the Poisson ratio $\nu$).
Subsequently, the proposed geometric restriction of the grid points, as described in \ssref{algorithm}, is carried out.
Based on the locations of the grid points within the FE faces, the matrix $\Wb$ is determined, taking into account the shape functions of the respective surface elements.
The compliance matrix $\mm C$ is set up for the restricted set of points (\eref{C}, \aref{C}), and so is the effective compliance matrix $\mm C_*$ (\eref{Cstar}).
\\
\definecolor{myBlue}{rgb}{0 0.4470 0.7410}
\definecolor{myWhite}{rgb}{1 1 1}
\tikzset{ input/.style={
        draw,
        fill=myWhite!30,
        trapezium,
        trapezium left angle=60,
        trapezium right angle=120,
        align=left
    }
}
\tikzset{
preparation/.style={
        draw,
        fill=myWhite!30,
        chamfered rectangle,
        chamfered rectangle xsep=2cm,
        align=left
    }
}
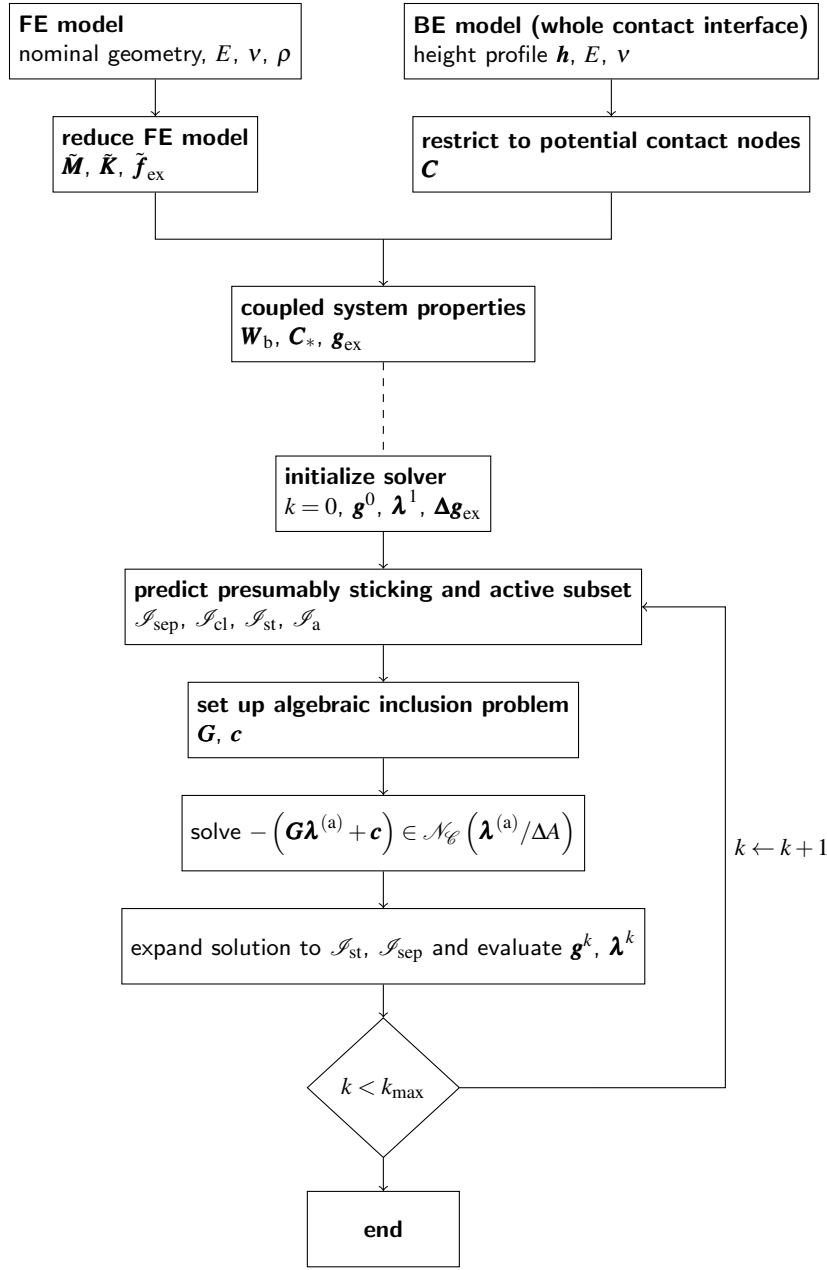
\begin{figure}[!]
\begin{tikzpicture}[font ={\sf \small}, scale=0.75]
\node (start_FE) at (1,0)  [draw, process,align=left,
minimum width=2cm,
minimum height=1cm,fill=myWhite!30] {\textbf{FE model} \\ nominal geometry, $E$, $\nu$, $\rho$};
\node (ROM_FE) at (1,-2)  [draw, process,align=left,
minimum width=2cm,
minimum height=1cm,fill=myWhite!30] {\textbf{reduce FE model} \\ $\tilde{\mm M}$, $\tilde{\mm K}$, $\tilde{\mm f}_{\mrm{ex}}$};
\node (start_BE) at (9,0)  [draw, process,align=left,
minimum width=2cm,
minimum height=1cm,fill=myWhite!30] {\textbf{BE model (whole contact interface)} \\ height profile $\mm h$, $E$, $\nu$};
\node (reduce_BE) at (9,-2)  [draw, process,align=left,
minimum width=2cm,
minimum height=1cm,fill=myWhite!30] {\textbf{restrict to potential contact nodes} \\ $\mm C$};
\node (coupling) at (5,-5)  [draw, process,align=left,
minimum width=2cm,
minimum height=1cm,fill=myWhite!30] {\textbf{coupled system properties} \\$\Wb$, $\mm C_*$, $\mm g_{\mrm{ex}}$};
\node (init) at (5,-8)  [draw, process,align=left,
minimum width=2cm,
minimum height=1cm,fill=myWhite!30] {\textbf{initialize solver} \\ $k = 0$, $\mm g^{0}$, $\mm\lambda^1$, $\mm \Delta \mm g_{\mrm{ex}}$};
\node (activeSet) at (5,-10)  [draw, process,align=left,
minimum width=2cm,
minimum height=1cm,fill=myWhite!30] {\textbf{predict presumably sticking and active subset} \\ $\mathcal I_{\mrm{sep}}$, $\mathcal I_{\mrm{cl}}$, $\mathcal I_{\mrm{st}}$, $\mathcal I_{\mrm{a}}$};
\node (Delassus) at (5,-12)  [draw, process,align=left,
minimum width=2cm,
minimum height=1cm,fill=myWhite!30] {\textbf{set up algebraic inclusion problem} \\ $\mm G$, $\mm c$};
\node (solve) at (5,-14)  [draw, process,align=left,
minimum width=2cm,
minimum height=1cm,fill=myWhite!30] {solve $-\left(\mm G\mm\lambda^{(\mrm a)} + \mm c\right) \in \mathcal N_{\mathcal C}\left(\mm \lambda^{(\mrm a)}/\Delta A\right)$};
\node (reconstruct) at (5,-16)  [draw, process,align=left,
minimum width=2cm,
minimum height=1cm,fill=myWhite!30] {expand solution to $\mathcal I_{\mrm{st}}$, $\mathcal I_{\mrm{sep}}$ and evaluate $\mm g^{k}$, $\mm\lambda^k$};
\node (check) at (5,-18.5) [draw, decision,align=left,
minimum width=2cm,
minimum height=1cm,fill=myWhite!30] {$k < k_{\mathrm{max}}$};
\node (bulk) at (5,-21) [draw, process,align=center,
minimum width=2cm,
minimum height=1cm,fill=myWhite!30] {\textbf{end}};
\coordinate (point1) at (1,-3.5) {};
\coordinate (point2) at (5,-3.5) {};
\coordinate (point3) at (9,-3.5) {};
\coordinate (point4) at (9,-6.5) {};
\coordinate (point5) at (11,-18.5) {};
\coordinate (point6) at (11,-10) {};
\draw[->] (start_FE) -- (ROM_FE);
\draw[->] (start_BE) -- (reduce_BE);
\draw[-] (ROM_FE) -- (point1) -- (point3) -- (reduce_BE);
\draw[->] (point2) -- (coupling);
\draw[dashed] (coupling) -- (init);
\draw[->] (init) -- (activeSet);
\draw[->] (activeSet) -- (Delassus);
\draw[->] (Delassus) -- (solve);
\draw[->] (solve) -- (reconstruct);
\draw[->] (reconstruct) -- (check);
\draw[->] (check) -- (bulk);
\draw[->] (check) -- (point5) -- node[right]{$k\leftarrow k+1$} (point6) -- (activeSet);
\end{tikzpicture}
\caption{Flowchart of the simulation algorithm using the coupled FE/BE multi-scale method (quasi-static case).}
\label{fig:flowChart}
\end{figure}
The simulation starts, as in reality, from a configuration where initially the contacts are open (or touching without interference) and unstressed.
The simulation is split into a preload and a dynamic load step.
The assembly process is modeled, for instance, by imposing static forces.
If imposed displacements seem more appropriate, the associated degrees of freedom can be removed, and equivalent imposed forces can be determined.
In any case, the imposed forces are stepped from zero to their final magnitude in an incremental way.
Subsequently, the dynamic loading is imposed.
In the present work, the dynamic load step takes the form of a quasi-static analysis as well.
Hence the treatment is analogous to that of the preload simulation.
\\
At each load level, the separated, the closed, the presumably sticking and the active set are estimated using \erefs{Isep}-\erefo{Ia}.
Subsequently, the matrix $\mm G$ and vector $\mm c$ defined in \erefs{G}-\erefo{c} are evaluated considering $\mm g_{\mrm{ex}}$ defined in \eref{gex}.
Finally, starting from the initial guess adopted from the previous load level, the contact problem is resolved using the projected Jacobi over-relaxation algorithm described in \aref{iai}.
\\
It is useful to discuss how the computational effort is affected by the proposed active set strategy, and how it scales with the BE grid density / resolution of the contact topography.
The dimension of the contact problem, $3\C$, equals to the apparent contact area, times the BE grid density, times the geometric restriction factor.
This dimension drives the memory required for storing the generally fully-populated, symmetric matrix $\mm C_*$.
It also drives the computation effort needed to obtain $\mm G$ and $\mm c$; note that the block condensation to the active set involves the inverse of $\mm C_*^{(\mrm{st,st})}$ (\erefs{G}-\erefo{c}).
The dimension of the active set, \ie, the number of points that are closed but not presumably sticking, $\left|\mathcal I_{\mrm{a}}\right|$, depends on the contact pressure distribution, as well as the distribution of the dynamic deformation.
For increased level of dynamic deformation, it is expected that the number of presumably sticking points diminishes, and the contact interface tends towards a gross slip / liftoff state.
The dimension $\left|\mathcal I_{\mrm{a}}\right|$ determines the number of non-smooth equations to be solved, and one may expect fewer iterations for fewer equations, and hence shorter simulation times.
As explained in the beginning of \sref{method}, the actual contact area decreases when the resolution of the contact topography is increased, \ie, more hills and valleys are considered within the same apparent contact area.
Consequently, $\C$ does not necessarily increase upon refinement of the surface topography.
This is in complete contrast to the situation where the contact region is modeled using finite elements.

\section{Numerical results\label{sec:results}}
The purpose of the present section is to validate the proposed multi-scale method.
For reference, a full-FE analysis was carried out in which both the contact topography and the underlying solids were modeled using a single (monolithic) FE model.
To this end, the commercial, general-purpose FE tool \ABAQUS was used.
As benchmark, the well-known \emph{S4 beam} (also known as C-beam) was considered, as described in \ssref{benchmark}.
The preload was simulated, and the amplitude-dependent frequency and damping ratio of the three lowest-order modes were computed.
For the latter, a quasi-static modal analysis was carried out; this was preferred over a fully-dynamic/transient alternative in order to obtain full-FE reference results with reasonable effort.
Besides the validation in \ssref{validation}, the relevance of different length scales of the contact topography was studied (\ssref{rough}).
Finally, the computational effort of the proposed multi-scale method was compared against the state of the art (\ssref{effort}).

\subsection{The benchmark system and its modeling\label{sec:benchmark}}
The S4 beam benchmark was initially proposed in \cite{Singh.2019}.
It consists of two C-shaped beams bolted at the ends (\fref{S4Beam}).
Linear-elastic, isotropic material behavior was considered as in \cite{Zare.2023}, with a mass density of $\rho=7861~\mrm{kg}/\mrm{m}^3$, a Young's modulus of $E=194~\mrm{GPa}$, and a Poisson ratio $\nu=0.2854$.
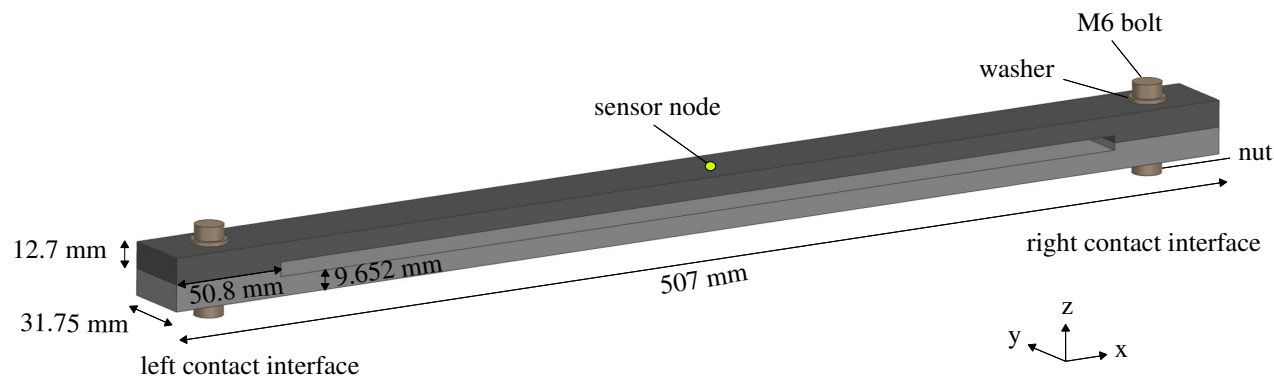
\begin{figure}[b]
 \centering
 \def\svgwidth{1.0\textwidth}
\begingroup%
  \makeatletter%
  \providecommand\color[2][]{%
    \errmessage{(Inkscape) Color is used for the text in Inkscape, but the package 'color.sty' is not loaded}%
    \renewcommand\color[2][]{}%
  }%
  \providecommand\transparent[1]{%
    \errmessage{(Inkscape) Transparency is used (non-zero) for the text in Inkscape, but the package 'transparent.sty' is not loaded}%
    \renewcommand\transparent[1]{}%
  }%
  \providecommand\rotatebox[2]{#2}%
  \newcommand*\fsize{\dimexpr\f@size pt\relax}%
  \newcommand*\lineheight[1]{\fontsize{\fsize}{#1\fsize}\selectfont}%
  \ifx\svgwidth\undefined%
    \setlength{\unitlength}{1015.49928474bp}%
    \ifx\svgscale\undefined%
      \relax%
    \else%
      \setlength{\unitlength}{\unitlength * \real{\svgscale}}%
    \fi%
  \else%
    \setlength{\unitlength}{\svgwidth}%
  \fi%
  \global\let\svgwidth\undefined%
  \global\let\svgscale\undefined%
  \makeatother%
  \begin{picture}(1,0.29538966)%
    \lineheight{1}%
    \setlength\tabcolsep{0pt}%
    \put(0,0){\includegraphics[width=\unitlength]{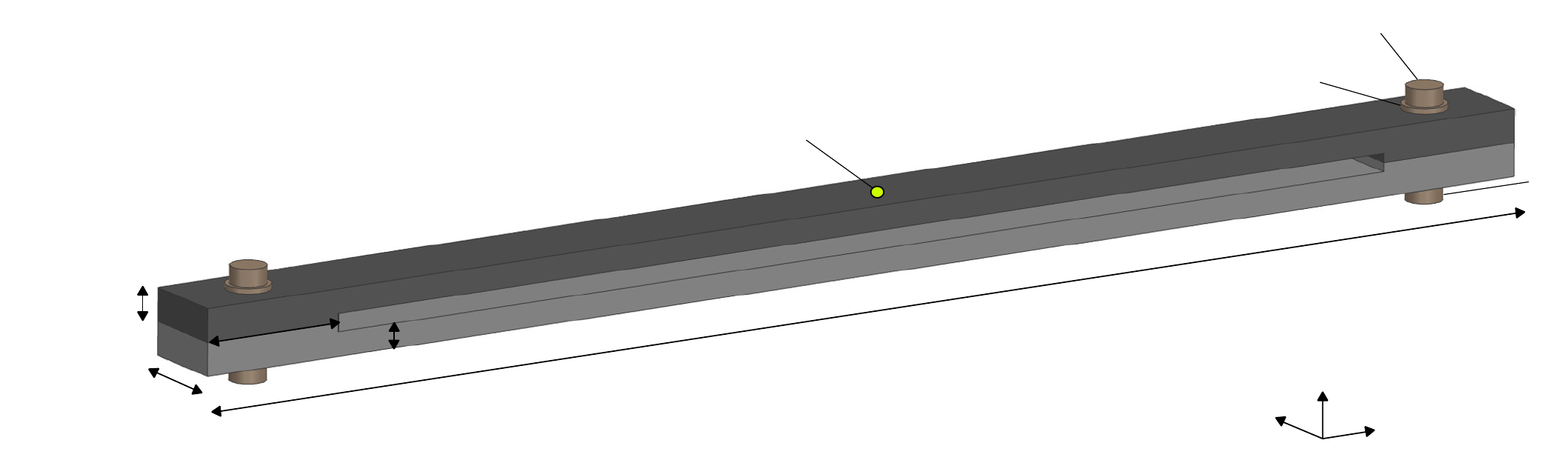}}%
    \put(0.46657153,0.21424625){\color[rgb]{0,0,0}\makebox(0,0)[lt]{\lineheight{1.25}\smash{\begin{tabular}[t]{l}sensor node\end{tabular}}}}%
    \put(0.10352665,0.00538695){\color[rgb]{0,0,0}\makebox(0,0)[lt]{\lineheight{1.25}\smash{\begin{tabular}[t]{l}left contact interface\end{tabular}}}}%
    \put(0.81275281,0.10447896){\color[rgb]{0,0,0}\makebox(0,0)[lt]{\lineheight{1.25}\smash{\begin{tabular}[t]{l}right contact interface\end{tabular}}}}%
    \put(0.85292969,0.28200873){\color[rgb]{0,0,0}\makebox(0,0)[lt]{\lineheight{1.25}\smash{\begin{tabular}[t]{l}M6 bolt\end{tabular}}}}%
    \put(0.98206421,0.17816962){\color[rgb]{0,0,0}\makebox(0,0)[lt]{\lineheight{1.25}\smash{\begin{tabular}[t]{l}nut\end{tabular}}}}%
    \put(0.77439389,0.24467308){\color[rgb]{0,0,0}\makebox(0,0)[lt]{\lineheight{1.25}\smash{\begin{tabular}[t]{l}washer\end{tabular}}}}%
    \put(0.52044042,0.06972506){\color[rgb]{0,0,0}\rotatebox{9.077958}{\makebox(0,0)[lt]{\lineheight{1.25}\smash{\begin{tabular}[t]{l}507 mm\end{tabular}}}}}%
    \put(0.00741259,0.04005146){\color[rgb]{0,0,0}\rotatebox{-1.0252444}{\makebox(0,0)[lt]{\lineheight{1.25}\smash{\begin{tabular}[t]{l}31.75 mm\end{tabular}}}}}%
    \put(0.14345338,0.06132318){\color[rgb]{0,0,0}\rotatebox{6.9632936}{\makebox(0,0)[lt]{\lineheight{1.25}\smash{\begin{tabular}[t]{l}50.8 mm\end{tabular}}}}}%
    \put(0.00083215,0.0986772){\color[rgb]{0,0,0}\makebox(0,0)[lt]{\lineheight{1.25}\smash{\begin{tabular}[t]{l}12.7 mm\end{tabular}}}}%
    \put(0.2600954,0.07640557){\color[rgb]{0,0,0}\rotatebox{7.9965242}{\makebox(0,0)[lt]{\lineheight{1.25}\smash{\begin{tabular}[t]{l}9.652 mm\end{tabular}}}}}%
    \put(0.84038747,0.05334771){\color[rgb]{0,0,0}\makebox(0,0)[lt]{\lineheight{1.25}\smash{\begin{tabular}[t]{l}z\end{tabular}}}}%
    \put(0.88313748,0.0190588){\color[rgb]{0,0,0}\makebox(0,0)[lt]{\lineheight{1.25}\smash{\begin{tabular}[t]{l}x\end{tabular}}}}%
    \put(0.7975993,0.03018769){\color[rgb]{0,0,0}\makebox(0,0)[lt]{\lineheight{1.25}\smash{\begin{tabular}[t]{l}y\end{tabular}}}}%
  \end{picture}%
\endgroup%
 \caption{Considered benchmark system: S4 beam.}
 \label{fig:S4Beam}
\end{figure}
\\
The contact topography was adopted from \cite{Wall.2022}.
The height profile was obtained from a coordinate measurement machine.
Due to the relatively low precision, the height profile was smoothed as proposed in \cite{Zare.2023}. 
This way, only the form deviation was retained (\fref{surfaceTopography}).
To estimate the effects of smaller wave lengths of the topography, a synthetic height profile is superimposed in \ssref{rough}.
Essentially, the form is characterized by a hill near the bore hole.
The maximum peak-to-peak deviation is $12.14~\mum$, which should be seen in relation to the dimensions of the apparent contact area of $50.8~\mrm{mm}~\times~ 31.75~\mrm{mm}$.
\begin{figure}[t]%
  \centering
  \subfloat[][]{\includegraphics[width=0.5\linewidth]{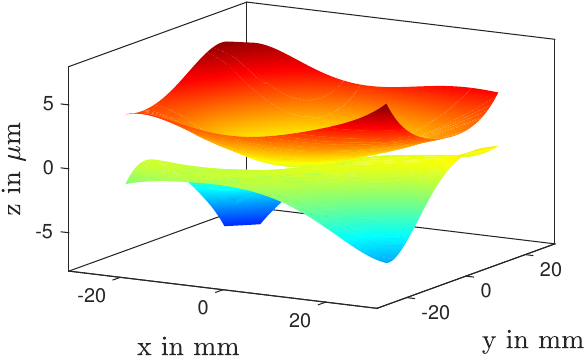}}%
  \subfloat[][]{\includegraphics[width=0.5\linewidth]{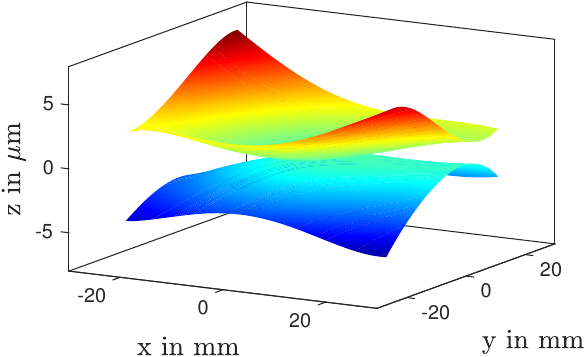}}%
  \caption{Contact topography (form deviation): (a) left, (b) right interface.}%
  \label{fig:surfaceTopography}
\end{figure}
\\
Frictional-unilateral interactions were accounted for between the two beams, with a friction coefficient of $0.6$.
The apparent contact interfaces between bolt head and washer, between washer and beam surface, between bolt and nut, and between nut beam surface were modeled as tied for simplicity.
To suppress rigid body movements, some nodes in each bolt's center plane (orthogonal to the $z$-axis in \fref{S4Beam}) were constrained in all three directions.
This simplification seems appropriate here, since the goal is a numerical validation, rather than the agreement with a particular experiment.
In any case, it has been verified that these artificial constraints do not distort severely the depicted results.
To introduce the tightening load of the bolts, a uniform area load was applied to the nodes in the planes adjacent to the aforementioned bolt center plane; those at one element row in the positive $z$-direction receive a load in the negative $z$-direction, and vice versa with those nodes one element row in the negative $z$-direction from the center plane.
The load was scaled in such a way that an integral preload of $5.71~\mrm{kN}$ was achieved per contact zone.
The magnitude of the load is consistent with the norm for the given M6 bolts assuming a medium strength class (ISO 898/1).
\\
The system was initially meshed using brick elements with quadratic shape functions (C3D20R elements).
Poor convergence behavior of \ABAQUS was observed during the nonlinear contact simulation.
To alleviate this, bricks with linear shape functions were instead used in the bolted regions (C3D8 elements).
At the apparent contact interfaces, matching nodes on the opposing surfaces were enforced.
\\
Four different FE meshes are shown in \fref{S4beam_contactMeshes}, restricted to one apparent contact interface.
All four contact interfaces were meshed in the same way.
The coarse, fine and super fine meshes have about $500$, $4400$ and $24000$ nodes per interface, respectively.
The coarse mesh was only used for the FE domain within the proposed multi-scale method.
The fine and super fine meshes were used for the full-FE reference.
To obtain a comparable resolution with the proposed multi-scale method, fine and super fine BE grids were introduced which have about $4300$ and $25000$ integration points, respectively (\fref{BEM_mesh}).
For the full-FE model, the node locations at the interface were moved in accordance with the specified height profile (mesh morphing).
In accordance with the proposed multi-scale method, the node locations of the considered coarse FE model were not moved; instead the height profile was only considered in the BE model.
The fine FE mesh is of similar density as that used in \cite{Zare.2023}, where good agreement was achieved with experimentally obtained amplitude-dependent properties of the S4 beam's first symmetric bending mode and the first torsional mode.
The super-fine FE mesh is comparable to that used in \cite{Brink.2020}.
%
\begin{figure}[t]%
  \centering
  \includegraphics[width=1.0\linewidth]{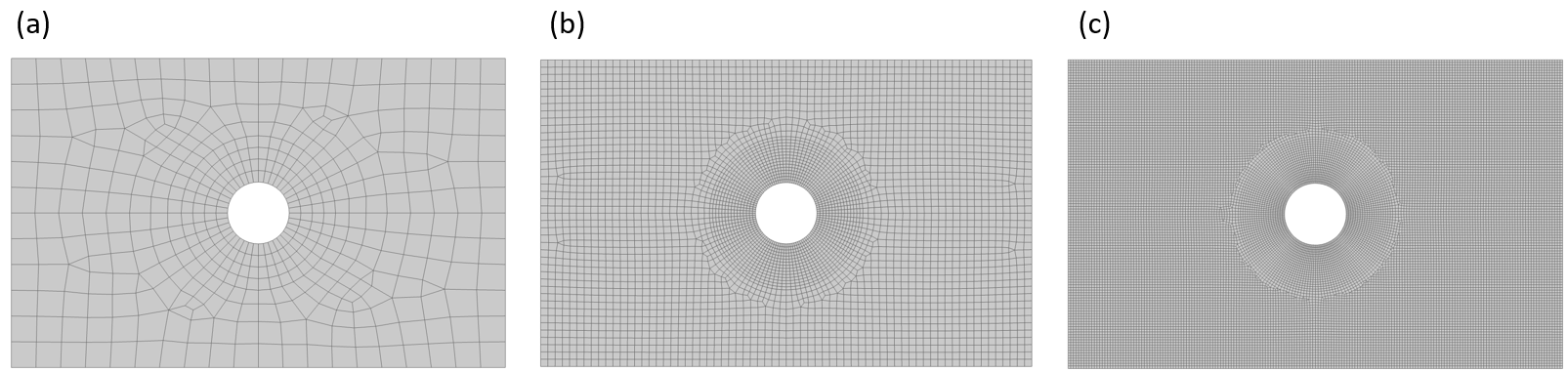}%
  \caption{FE meshes of the interfaces: (a) coarse mesh used in proposed multi-scale method; (b) fine and (c) super fine contact mesh used in full-FE method. The meshes have 500, 4400 and 24000 nodes, respectively.
  }
  \label{fig:S4beam_contactMeshes}
\end{figure}
\begin{figure}[t]
 \centering
 \def\svgwidth{1.0\textwidth}
\begingroup%
  \makeatletter%
  \providecommand\color[2][]{%
    \errmessage{(Inkscape) Color is used for the text in Inkscape, but the package 'color.sty' is not loaded}%
    \renewcommand\color[2][]{}%
  }%
  \providecommand\transparent[1]{%
    \errmessage{(Inkscape) Transparency is used (non-zero) for the text in Inkscape, but the package 'transparent.sty' is not loaded}%
    \renewcommand\transparent[1]{}%
  }%
  \providecommand\rotatebox[2]{#2}%
  \newcommand*\fsize{\dimexpr\f@size pt\relax}%
  \newcommand*\lineheight[1]{\fontsize{\fsize}{#1\fsize}\selectfont}%
  \ifx\svgwidth\undefined%
    \setlength{\unitlength}{530.76129317bp}%
    \ifx\svgscale\undefined%
      \relax%
    \else%
      \setlength{\unitlength}{\unitlength * \real{\svgscale}}%
    \fi%
  \else%
    \setlength{\unitlength}{\svgwidth}%
  \fi%
  \global\let\svgwidth\undefined%
  \global\let\svgscale\undefined%
  \makeatother%
  \begin{picture}(1,0.29815663)%
    \lineheight{1}%
    \setlength\tabcolsep{0pt}%
    \put(0,0){\includegraphics[width=\unitlength]{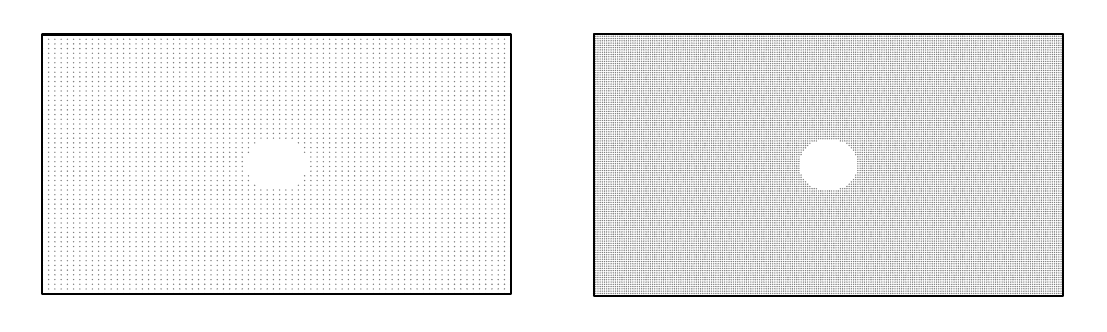}}%
    \put(0.03357217,0.28426982){\color[rgb]{0,0,0}\makebox(0,0)[lt]{\lineheight{1.25}\smash{\begin{tabular}[t]{l}(a)\end{tabular}}}}%
    \put(0.53305225,0.28475483){\color[rgb]{0,0,0}\makebox(0,0)[lt]{\lineheight{1.25}\smash{\begin{tabular}[t]{l}(b)\end{tabular}}}}%
  \end{picture}%
\endgroup%
 \caption{BE grids: (a) fine and (b) super fine grid. The grids have 4300 and 25000 integration points, respectively.}
 \label{fig:BEM_mesh}
\end{figure}
\\
For the full-FE analysis, contact was treated using the Lagrange multiplier technique, unless stated otherwise.
Finite sliding was enabled, as this was found to improve convergence, even though it is not expected to be relevant from a physical perspective for realistic vibration levels (\cf \aref{verification}).
\\
The proposed multi-scale method was implemented in \MATLAB.
The mass and stiffness matrices of the coarse FE model, along with the contact mesh, were adopted from \ABAQUS.
A crucial question is how fine the FE model has to be for the proposed multi-scale method.
It is postulated that the FE model should be fine enough to ensure that the frequencies of the modes of interest are accurately described for sticking contacts.
This was verified for the selected coarse FE model in \fref{S4beam_contactMeshes}.
Before application of the Craig-Bampton method, a transform was applied from absolute coordinates at the matching nodes on either side of the interface to one set of relative and one set of absolute displacements.
This reduces the number of required static constraint modes by factor two, without loss of accuracy, and improves the convergence with respect to the number of retained normal modes.
With the about 500 FE nodes per interface, the three relative degrees of freedom per node, the two interfaces are associated with about $3,000$ static constraint modes.
In addition, the $25$ lowest-frequency fixed-interface normal modes were retained in the reduction basis.
With this, the frequencies of the eight lowest-order free-interface normal modes are described as accurate as in the parent (coarse) FE model.
Of course, the $25$ fixed-interface normal modes are a much smaller burden than the $3000$ static constraint modes.
\\
Given that a commercial FE tool was used for reference, we do not have access to the implementation of the contact solver.
In principle, this could lead to a systematic error even for the same model.
To exclude this source of uncertainty, the contact solver was compared separately.
To this end, the compliance of the BE model was removed and the FE nodes were directly used as integration points.
Exemplary results of this verification are presented in \aref{verification}.

\subsection{Validation of the proposed multi-scale method against the state-of-the-art\label{sec:validation}}
\begin{figure}[h!]
 \centering
 \def\svgwidth{1.0\textwidth}
\begingroup%
  \makeatletter%
  \providecommand\color[2][]{%
    \errmessage{(Inkscape) Color is used for the text in Inkscape, but the package 'color.sty' is not loaded}%
    \renewcommand\color[2][]{}%
  }%
  \providecommand\transparent[1]{%
    \errmessage{(Inkscape) Transparency is used (non-zero) for the text in Inkscape, but the package 'transparent.sty' is not loaded}%
    \renewcommand\transparent[1]{}%
  }%
  \providecommand\rotatebox[2]{#2}%
  \newcommand*\fsize{\dimexpr\f@size pt\relax}%
  \newcommand*\lineheight[1]{\fontsize{\fsize}{#1\fsize}\selectfont}%
  \ifx\svgwidth\undefined%
    \setlength{\unitlength}{1030.82909094bp}%
    \ifx\svgscale\undefined%
      \relax%
    \else%
      \setlength{\unitlength}{\unitlength * \real{\svgscale}}%
    \fi%
  \else%
    \setlength{\unitlength}{\svgwidth}%
  \fi%
  \global\let\svgwidth\undefined%
  \global\let\svgscale\undefined%
  \makeatother%
  \begin{picture}(1,0.14710611)%
    \lineheight{1}%
    \setlength\tabcolsep{0pt}%
    \put(0,0){\includegraphics[width=\unitlength]{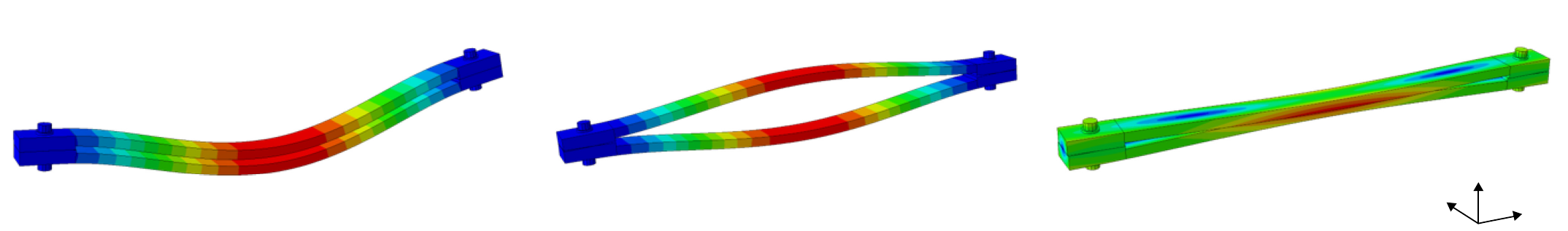}}%
    \put(0.0100113,0.13860501){\color[rgb]{0,0,0}\makebox(0,0)[lt]{\lineheight{1.25}\smash{\begin{tabular}[t]{l}(1)\end{tabular}}}}%
    \put(0.35039946,0.1385041){\color[rgb]{0,0,0}\makebox(0,0)[lt]{\lineheight{1.25}\smash{\begin{tabular}[t]{l}(2)\end{tabular}}}}%
    \put(0.68307881,0.13886656){\color[rgb]{0,0,0}\makebox(0,0)[lt]{\lineheight{1.25}\smash{\begin{tabular}[t]{l}(3)\end{tabular}}}}%
    \put(0.97497989,0.00813908){\color[rgb]{0,0,0}\makebox(0,0)[lt]{\lineheight{1.25}\smash{\begin{tabular}[t]{l}x\end{tabular}}}}%
    \put(0.9389544,0.03810707){\color[rgb]{0,0,0}\makebox(0,0)[lt]{\lineheight{1.25}\smash{\begin{tabular}[t]{l}z\end{tabular}}}}%
    \put(0.9085813,0.01864513){\color[rgb]{0,0,0}\makebox(0,0)[lt]{\lineheight{1.25}\smash{\begin{tabular}[t]{l}y\end{tabular}}}}%
  \end{picture}%
\endgroup%
 \caption{Lowest-order mode shapes for tied apparent contact areas: (1) first in-phase bending mode, (2) first out-of-phase bending mode, (3) first torsion mode.
  }
  \label{fig:modeShapes}
\end{figure}
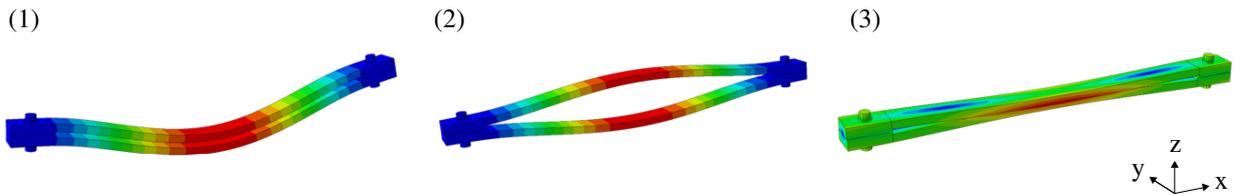
The simulation was split into a preload step, where the bolts were tightened, and a nonlinear modal analysis step, where the amplitude-dependent frequency and damping ratio were determined for each mode of interest.
In the latter, the three lowest-frequency elastic modes were considered:
(1) the first in-phase bending, (2) the first out-of-phase bending, and (3) the first torsion mode.
The corresponding deflection shapes are illustrated in \fref{modeShapes}, for the case of \emph{tied apparent contact area}.
The associated modal frequencies are
$\omega_{\mrm{lin},1}/(2\pi)=217.29~\mrm{Hz}$,
$\omega_{\mrm{lin},2}/(2\pi)=271.63~\mrm{Hz}$
$\omega_{\mrm{lin},3}/(2\pi)=976.65~\mrm{Hz}$.
Those linear modal frequencies will be used throughout the present work for normalization of the amplitude-dependent ones.
Those three modes induce quite different, yet typical kinematics in the contact interfaces:
The first in-phase bending mode provokes slip in the $x$-direction, the first out-of-phase bending mode provokes a rolling around the $y$-axis associated with opening-closing interactions, and the first torsion mode provokes rotational slip around the bolt axis ($z$-direction).

\subsubsection{Preload simulation\label{sec:assembly}}
The contact pressure distribution obtained after tightening the bolts is shown in \fref{PreLoadComparison}.
Overall, the agreement is good.
The contact pressure is distributed more uniformly, and over a larger area according to the proposed multi-scale method.
A reason for this may be the spurious compliance of the contact region obtained by superimposing the contributions of both the pair of elastic half spaces and the (coarse) FE model.
Additionally, edge effects are observed around the bore hole:
While the contact pressure decreases towards the bore hole in the full-FE reference, artificial stress peaks occur there with the proposed multi-scale method.
This is attributed to the half-space theory, which models the bore hole as if it was filled with elastic material, only with zero contact stress at the interface.
Upon close inspection one may note that the pressure distribution in \fref{PreLoadComparison}(b) shows some \myquote{fanning out}, which is attributed to the underlying FE mesh.
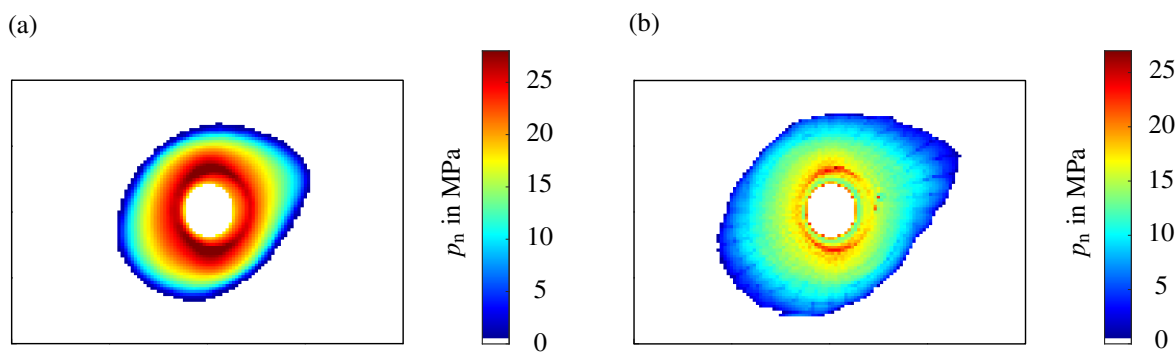
\begin{figure}[h]
 \centering
 \def\svgwidth{1.0\textwidth}
\begingroup%
  \makeatletter%
  \providecommand\color[2][]{%
    \errmessage{(Inkscape) Color is used for the text in Inkscape, but the package 'color.sty' is not loaded}%
    \renewcommand\color[2][]{}%
  }%
  \providecommand\transparent[1]{%
    \errmessage{(Inkscape) Transparency is used (non-zero) for the text in Inkscape, but the package 'transparent.sty' is not loaded}%
    \renewcommand\transparent[1]{}%
  }%
  \providecommand\rotatebox[2]{#2}%
  \newcommand*\fsize{\dimexpr\f@size pt\relax}%
  \newcommand*\lineheight[1]{\fontsize{\fsize}{#1\fsize}\selectfont}%
  \ifx\svgwidth\undefined%
    \setlength{\unitlength}{736.3426586bp}%
    \ifx\svgscale\undefined%
      \relax%
    \else%
      \setlength{\unitlength}{\unitlength * \real{\svgscale}}%
    \fi%
  \else%
    \setlength{\unitlength}{\svgwidth}%
  \fi%
  \global\let\svgwidth\undefined%
  \global\let\svgscale\undefined%
  \makeatother%
  \begin{picture}(1,0.3278757)%
    \lineheight{1}%
    \setlength\tabcolsep{0pt}%
    \put(0,0){\includegraphics[width=\unitlength]{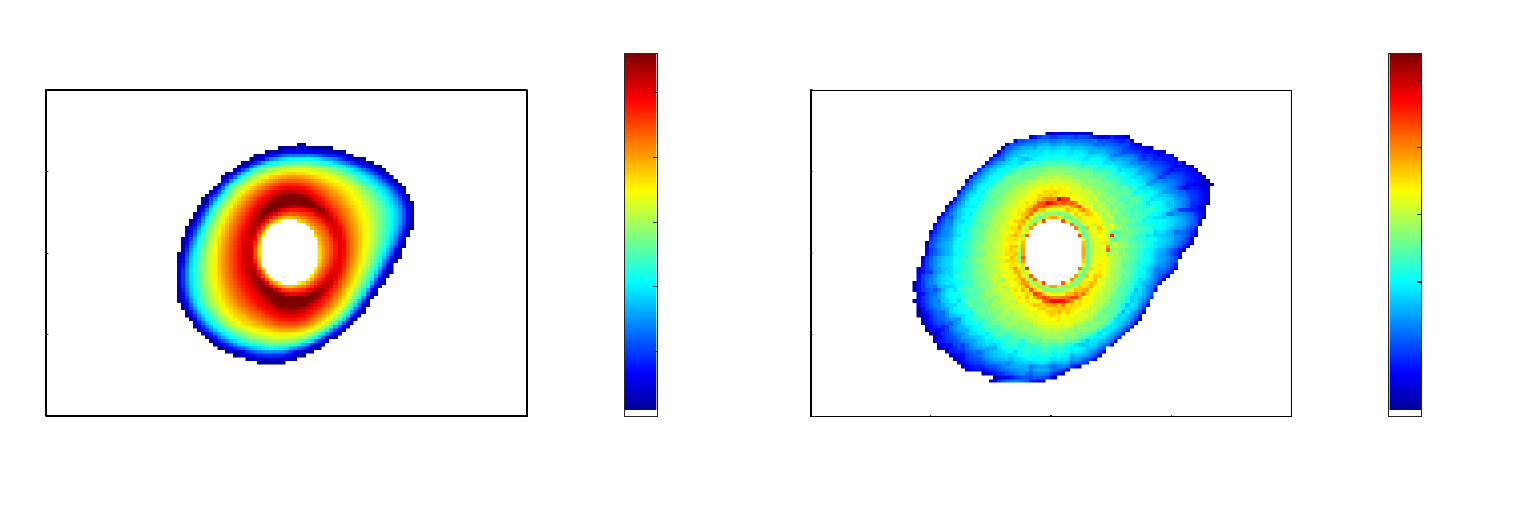}}%
    \put(0.44800093,0.05167058){\color[rgb]{0,0,0}\makebox(0,0)[lt]{\lineheight{1.25}\smash{\begin{tabular}[t]{l}0\end{tabular}}}}%
    \put(0.44800093,0.09380356){\color[rgb]{0,0,0}\makebox(0,0)[lt]{\lineheight{1.25}\smash{\begin{tabular}[t]{l}5\end{tabular}}}}%
    \put(0.4420591,0.13665679){\color[rgb]{0,0,0}\makebox(0,0)[lt]{\lineheight{1.25}\smash{\begin{tabular}[t]{l}10\end{tabular}}}}%
    \put(0.44151895,0.17950998){\color[rgb]{0,0,0}\makebox(0,0)[lt]{\lineheight{1.25}\smash{\begin{tabular}[t]{l}15\end{tabular}}}}%
    \put(0.44205909,0.22074269){\color[rgb]{0,0,0}\makebox(0,0)[lt]{\lineheight{1.25}\smash{\begin{tabular}[t]{l}20\end{tabular}}}}%
    \put(0.44187906,0.26359589){\color[rgb]{0,0,0}\makebox(0,0)[lt]{\lineheight{1.25}\smash{\begin{tabular}[t]{l}25\end{tabular}}}}%
    \put(0.94472451,0.05374959){\color[rgb]{0,0,0}\makebox(0,0)[lt]{\lineheight{1.25}\smash{\begin{tabular}[t]{l}0\end{tabular}}}}%
    \put(0.94472459,0.097143){\color[rgb]{0,0,0}\makebox(0,0)[lt]{\lineheight{1.25}\smash{\begin{tabular}[t]{l}5\end{tabular}}}}%
    \put(0.93896281,0.13981615){\color[rgb]{0,0,0}\makebox(0,0)[lt]{\lineheight{1.25}\smash{\begin{tabular}[t]{l}10\end{tabular}}}}%
    \put(0.93914294,0.18266935){\color[rgb]{0,0,0}\makebox(0,0)[lt]{\lineheight{1.25}\smash{\begin{tabular}[t]{l}15\end{tabular}}}}%
    \put(0.9395558,0.22695769){\color[rgb]{0,0,0}\makebox(0,0)[lt]{\lineheight{1.25}\smash{\begin{tabular}[t]{l}20\end{tabular}}}}%
    \put(0.94001212,0.27146603){\color[rgb]{0,0,0}\makebox(0,0)[lt]{\lineheight{1.25}\smash{\begin{tabular}[t]{l}25\end{tabular}}}}%
    \put(0.38857471,0.12609181){\color[rgb]{0,0,0}\rotatebox{90}{\makebox(0,0)[lt]{\lineheight{1.25}\smash{\begin{tabular}[t]{l}$p_{\mrm n}$ in MPa\end{tabular}}}}}%
    \put(0.88877381,0.12616194){\color[rgb]{0,0,0}\rotatebox{90}{\makebox(0,0)[lt]{\lineheight{1.25}\smash{\begin{tabular}[t]{l}$p_{\mrm n}$ in MPa\end{tabular}}}}}%
    \put(0.02727588,0.30792665){\color[rgb]{0,0,0}\makebox(0,0)[lt]{\lineheight{1.25}\smash{\begin{tabular}[t]{l}(a)\end{tabular}}}}%
    \put(0.52380156,0.30909321){\color[rgb]{0,0,0}\makebox(0,0)[lt]{\lineheight{1.25}\smash{\begin{tabular}[t]{l}(b)\end{tabular}}}}%
  \end{picture}%
\endgroup%
\caption{Contact pressure distribution obtained at the left interface after tightening the bolts: (a) full-FE reference, (b) proposed multi-scale method.
  }%
  \label{fig:PreLoadComparison}
\end{figure}

\subsubsection{Modal analysis\label{sec:modalAnalysis}}
To obtain the amplitude-dependent frequency and damping ratio of the modes of interest, quasi-static modal analysis was employed.
The particular implementation of this technique is described in \aref{QSMA}.
As mentioned before, a quasi-static analysis was preferred over a fully-dynamic/transient alternative in order to obtain full-FE reference results with reasonable effort.
At least in the weakly-linear partial slip regime, no severe change of the overall modal deflection shape, nor a severe interaction with another mode is expected.
Under those conditions, the results of a quasi-static modal analysis should be consistent with a fully-dynamic modal analysis.
Further, it should be recalled that the purpose of this section is a numerical validation.
For this, it is of prime importance that the same type of analysis is applied to the proposed multi-scale and the full-FE model.
In contrast, it is only of secondary relevance that the compared frequency and damping measures accurately characterize the vibration behavior (as long as those measures capture the effect of the contact stresses on the structural dynamics).
\begin{figure}[t!]
 \centering
 \def\svgwidth{1.0\textwidth}
\begingroup%
  \makeatletter%
  \providecommand\color[2][]{%
    \errmessage{(Inkscape) Color is used for the text in Inkscape, but the package 'color.sty' is not loaded}%
    \renewcommand\color[2][]{}%
  }%
  \providecommand\transparent[1]{%
    \errmessage{(Inkscape) Transparency is used (non-zero) for the text in Inkscape, but the package 'transparent.sty' is not loaded}%
    \renewcommand\transparent[1]{}%
  }%
  \providecommand\rotatebox[2]{#2}%
  \newcommand*\fsize{\dimexpr\f@size pt\relax}%
  \newcommand*\lineheight[1]{\fontsize{\fsize}{#1\fsize}\selectfont}%
  \ifx\svgwidth\undefined%
    \setlength{\unitlength}{966.74879696bp}%
    \ifx\svgscale\undefined%
      \relax%
    \else%
      \setlength{\unitlength}{\unitlength * \real{\svgscale}}%
    \fi%
  \else%
    \setlength{\unitlength}{\svgwidth}%
  \fi%
  \global\let\svgwidth\undefined%
  \global\let\svgscale\undefined%
  \makeatother%
  \begin{picture}(1,0.32822763)%
    \lineheight{1}%
    \setlength\tabcolsep{0pt}%
    \put(0,0){\includegraphics[width=\unitlength]{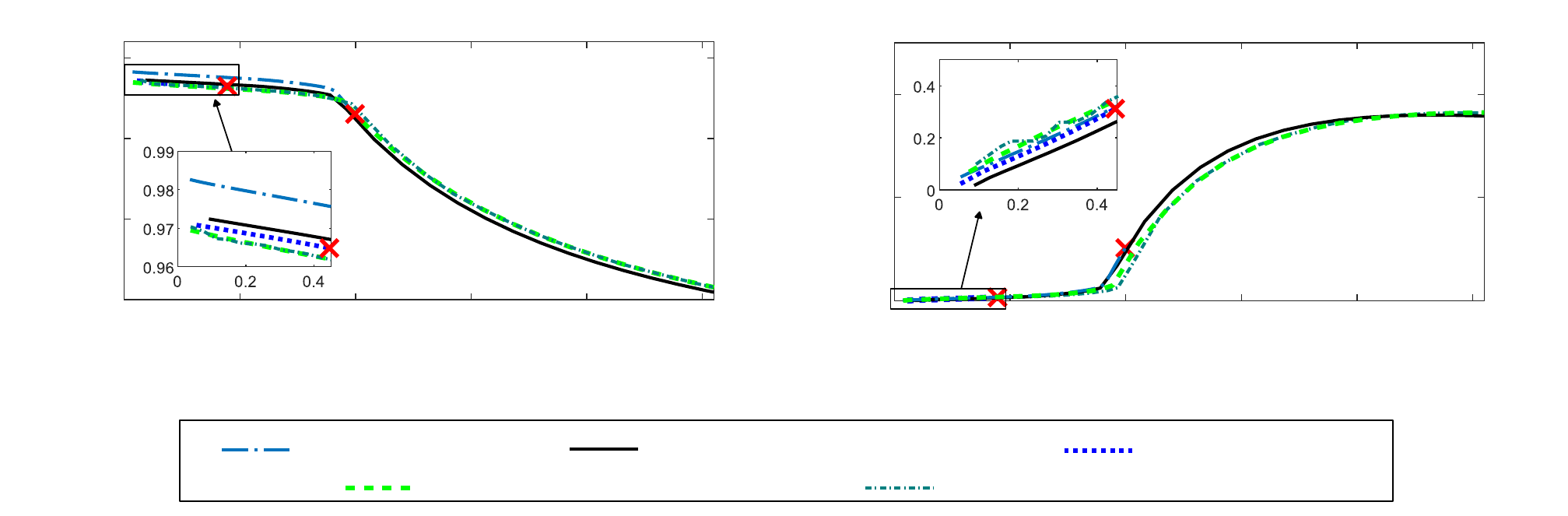}}%
    \put(0.07468758,0.11264093){\color[rgb]{0,0,0}\makebox(0,0)[lt]{\lineheight{1.25}\smash{\begin{tabular}[t]{l}0\end{tabular}}}}%
    \put(0.1446201,0.11249808){\color[rgb]{0,0,0}\makebox(0,0)[lt]{\lineheight{1.25}\smash{\begin{tabular}[t]{l}0.5\end{tabular}}}}%
    \put(0.22195512,0.11379069){\color[rgb]{0,0,0}\makebox(0,0)[lt]{\lineheight{1.25}\smash{\begin{tabular}[t]{l}1\end{tabular}}}}%
    \put(0.29157837,0.1119757){\color[rgb]{0,0,0}\makebox(0,0)[lt]{\lineheight{1.25}\smash{\begin{tabular}[t]{l}1.5\end{tabular}}}}%
    \put(0.37063105,0.11207639){\color[rgb]{0,0,0}\makebox(0,0)[lt]{\lineheight{1.25}\smash{\begin{tabular}[t]{l}2\end{tabular}}}}%
    \put(0.43841954,0.11191521){\color[rgb]{0,0,0}\makebox(0,0)[lt]{\lineheight{1.25}\smash{\begin{tabular}[t]{l}2.5\end{tabular}}}}%
    \put(0.56652505,0.11224082){\color[rgb]{0,0,0}\makebox(0,0)[lt]{\lineheight{1.25}\smash{\begin{tabular}[t]{l}0\end{tabular}}}}%
    \put(0.63581469,0.11124083){\color[rgb]{0,0,0}\makebox(0,0)[lt]{\lineheight{1.25}\smash{\begin{tabular}[t]{l}0.5\end{tabular}}}}%
    \put(0.71422115,0.11146199){\color[rgb]{0,0,0}\makebox(0,0)[lt]{\lineheight{1.25}\smash{\begin{tabular}[t]{l}1\end{tabular}}}}%
    \put(0.78491583,0.11093275){\color[rgb]{0,0,0}\makebox(0,0)[lt]{\lineheight{1.25}\smash{\begin{tabular}[t]{l}1.5\end{tabular}}}}%
    \put(0.86032572,0.11081915){\color[rgb]{0,0,0}\makebox(0,0)[lt]{\lineheight{1.25}\smash{\begin{tabular}[t]{l}2\end{tabular}}}}%
    \put(0.9338998,0.11108654){\color[rgb]{0,0,0}\makebox(0,0)[lt]{\lineheight{1.25}\smash{\begin{tabular}[t]{l}2.5\end{tabular}}}}%
    \put(0.04580073,0.13470411){\color[rgb]{0,0,0}\makebox(0,0)[lt]{\lineheight{1.25}\smash{\begin{tabular}[t]{l}0.7\end{tabular}}}}%
    \put(0.04538931,0.18530973){\color[rgb]{0,0,0}\makebox(0,0)[lt]{\lineheight{1.25}\smash{\begin{tabular}[t]{l}0.8\end{tabular}}}}%
    \put(0.04546646,0.23825539){\color[rgb]{0,0,0}\makebox(0,0)[lt]{\lineheight{1.25}\smash{\begin{tabular}[t]{l}0.9\end{tabular}}}}%
    \put(0.05305215,0.28900676){\color[rgb]{0,0,0}\makebox(0,0)[lt]{\lineheight{1.25}\smash{\begin{tabular}[t]{l}1\end{tabular}}}}%
    \put(0.54247849,0.2681723){\color[rgb]{0,0,0}\makebox(0,0)[lt]{\lineheight{1.25}\smash{\begin{tabular}[t]{l}20\end{tabular}}}}%
    \put(0.54181485,0.19977991){\color[rgb]{0,0,0}\makebox(0,0)[lt]{\lineheight{1.25}\smash{\begin{tabular}[t]{l}10\end{tabular}}}}%
    \put(0.54580095,0.13385334){\color[rgb]{0,0,0}\makebox(0,0)[lt]{\lineheight{1.25}\smash{\begin{tabular}[t]{l}0\end{tabular}}}}%
    \put(0.68819632,0.08043497){\color[rgb]{0,0,0}\makebox(0,0)[lt]{\lineheight{1.25}\smash{\begin{tabular}[t]{l}amplitude in mm\end{tabular}}}}%
    \put(0.21782482,0.08155984){\color[rgb]{0,0,0}\makebox(0,0)[lt]{\lineheight{1.25}\smash{\begin{tabular}[t]{l}amplitude in mm\end{tabular}}}}%
    \put(0.51668869,0.10595205){\color[rgb]{0,0,0}\rotatebox{90}{\makebox(0,0)[lt]{\lineheight{1.25}\smash{\begin{tabular}[t]{l}damping ratio $D_1$ in \%\end{tabular}}}}}%
    \put(0.4990693,0.31812312){\color[rgb]{0,0,0}\makebox(0,0)[lt]{\lineheight{1.25}\smash{\begin{tabular}[t]{l}(b)\end{tabular}}}}%
    \put(0.01981413,0.13050228){\color[rgb]{0,0,0}\rotatebox{90}{\makebox(0,0)[lt]{\lineheight{1.25}\smash{\begin{tabular}[t]{l}frequency $\omega_1$/$\omega_{\mrm{lin},1}$\end{tabular}}}}}%
    \put(0.00055922,0.31773832){\color[rgb]{0,0,0}\makebox(0,0)[lt]{\lineheight{1.25}\smash{\begin{tabular}[t]{l}(a)\end{tabular}}}}%
    \put(0.19306705,0.03864126){\color[rgb]{0,0,0}\makebox(0,0)[lt]{\lineheight{1.25}\smash{\begin{tabular}[t]{l}\small{full-FE, fine}\end{tabular}}}}%
    \put(0.7313168,0.0381709){\color[rgb]{0,0,0}\makebox(0,0)[lt]{\lineheight{1.25}\smash{\begin{tabular}[t]{l}\small{full-FE super-fine}\end{tabular}}}}%
    \put(0.4146646,0.03878226){\color[rgb]{0,0,0}\makebox(0,0)[lt]{\lineheight{1.25}\smash{\begin{tabular}[t]{l}\small{full-FE, fine, regularized}\end{tabular}}}}%
    \put(0.28603929,0.01458548){\color[rgb]{0,0,0}\makebox(0,0)[lt]{\lineheight{1.25}\smash{\begin{tabular}[t]{l}\small{multi-scale, fine}\end{tabular}}}}%
    \put(0.61940944,0.01457904){\color[rgb]{0,0,0}\makebox(0,0)[lt]{\lineheight{1.25}\smash{\begin{tabular}[t]{l}\small{multi-scale, refined}\end{tabular}}}}%
  \end{picture}%
\endgroup%
  \caption{Amplitude-dependent properties of the first in-phase bending mode for the contact topography in \fref{surfaceTopography} (form, smooth): (a) frequency, (b) damping ratio. Crosses indicate where the full-FE solver failed to converge.
  }
  \label{fig:MeshConvergenceModal}
\end{figure}
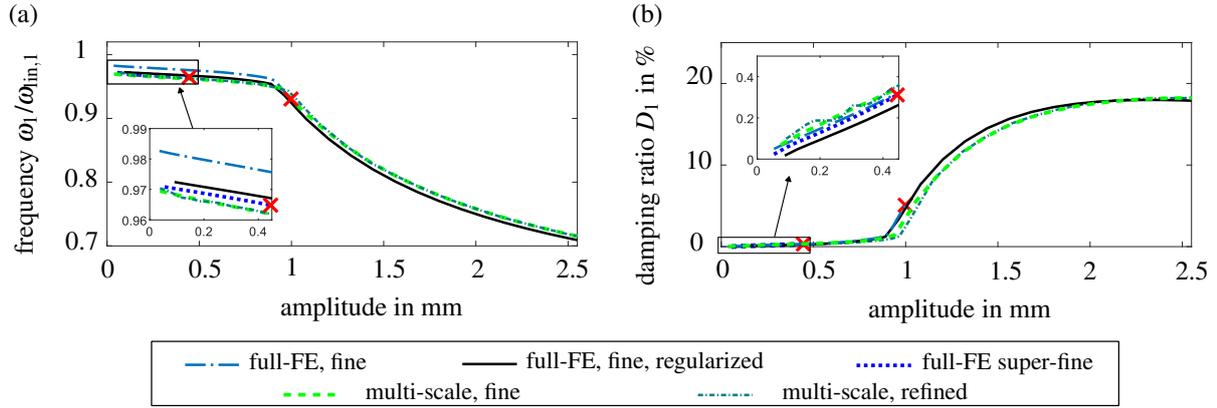
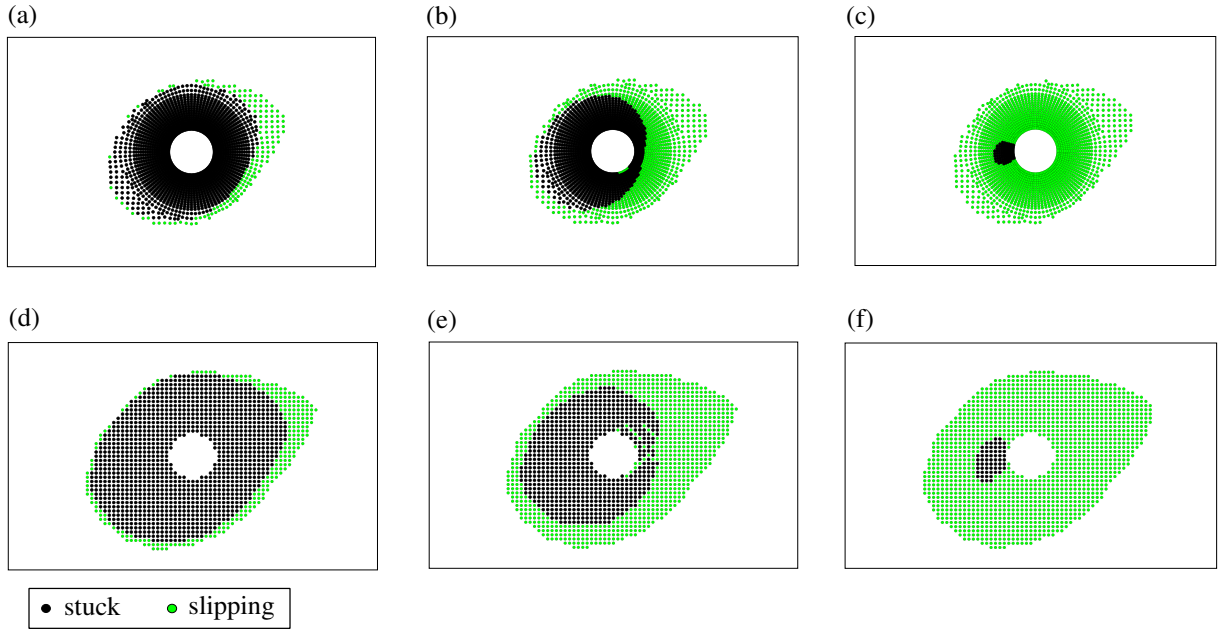
\begin{figure}[h!]
 \centering
 \begin{subfigure}{\textwidth}
  \centering
  \def\svgwidth{1.0\textwidth}
\begingroup%
  \makeatletter%
  \providecommand\color[2][]{%
    \errmessage{(Inkscape) Color is used for the text in Inkscape, but the package 'color.sty' is not loaded}%
    \renewcommand\color[2][]{}%
  }%
  \providecommand\transparent[1]{%
    \errmessage{(Inkscape) Transparency is used (non-zero) for the text in Inkscape, but the package 'transparent.sty' is not loaded}%
    \renewcommand\transparent[1]{}%
  }%
  \providecommand\rotatebox[2]{#2}%
  \newcommand*\fsize{\dimexpr\f@size pt\relax}%
  \newcommand*\lineheight[1]{\fontsize{\fsize}{#1\fsize}\selectfont}%
  \ifx\svgwidth\undefined%
    \setlength{\unitlength}{945.589055bp}%
    \ifx\svgscale\undefined%
      \relax%
    \else%
      \setlength{\unitlength}{\unitlength * \real{\svgscale}}%
    \fi%
  \else%
    \setlength{\unitlength}{\svgwidth}%
  \fi%
  \global\let\svgwidth\undefined%
  \global\let\svgscale\undefined%
  \makeatother%
  \begin{picture}(1,0.23906562)%
    \lineheight{1}%
    \setlength\tabcolsep{0pt}%
    \put(0,0){\includegraphics[width=\unitlength]{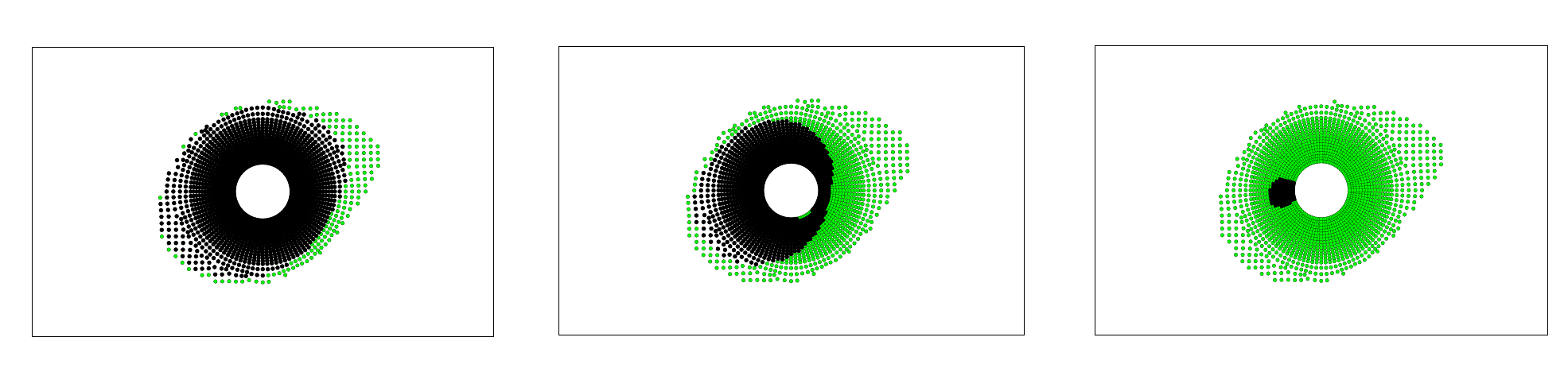}}%
    \put(0.0207157,0.22176889){\color[rgb]{0,0,0}\makebox(0,0)[lt]{\lineheight{1.25}\smash{\begin{tabular}[t]{l}(a)\end{tabular}}}}%
    \put(0.35542048,0.22089107){\color[rgb]{0,0,0}\makebox(0,0)[lt]{\lineheight{1.25}\smash{\begin{tabular}[t]{l}(b)\end{tabular}}}}%
    \put(0.69136607,0.2209738){\color[rgb]{0,0,0}\makebox(0,0)[lt]{\lineheight{1.25}\smash{\begin{tabular}[t]{l}(c)\end{tabular}}}}%
  \end{picture}%
\endgroup%
 \end{subfigure}
 \begin{subfigure}{\textwidth}
  \centering
  \def\svgwidth{1.0\textwidth}
\begingroup%
  \makeatletter%
  \providecommand\color[2][]{%
    \errmessage{(Inkscape) Color is used for the text in Inkscape, but the package 'color.sty' is not loaded}%
    \renewcommand\color[2][]{}%
  }%
  \providecommand\transparent[1]{%
    \errmessage{(Inkscape) Transparency is used (non-zero) for the text in Inkscape, but the package 'transparent.sty' is not loaded}%
    \renewcommand\transparent[1]{}%
  }%
  \providecommand\rotatebox[2]{#2}%
  \newcommand*\fsize{\dimexpr\f@size pt\relax}%
  \newcommand*\lineheight[1]{\fontsize{\fsize}{#1\fsize}\selectfont}%
  \ifx\svgwidth\undefined%
    \setlength{\unitlength}{945.5890915bp}%
    \ifx\svgscale\undefined%
      \relax%
    \else%
      \setlength{\unitlength}{\unitlength * \real{\svgscale}}%
    \fi%
  \else%
    \setlength{\unitlength}{\svgwidth}%
  \fi%
  \global\let\svgwidth\undefined%
  \global\let\svgscale\undefined%
  \makeatother%
  \begin{picture}(1,0.27996275)%
    \lineheight{1}%
    \setlength\tabcolsep{0pt}%
    \put(0,0){\includegraphics[width=\unitlength]{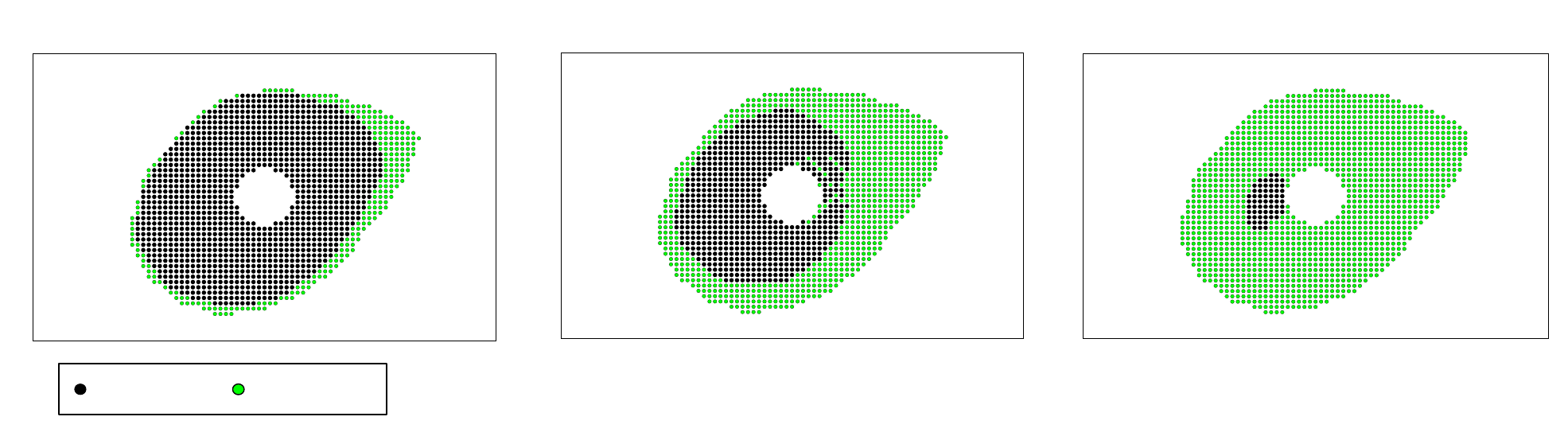}}%
    \put(0.02155972,0.25921159){\color[rgb]{0,0,0}\makebox(0,0)[lt]{\lineheight{1.25}\smash{\begin{tabular}[t]{l}(d)\end{tabular}}}}%
    \put(0.3562645,0.25833375){\color[rgb]{0,0,0}\makebox(0,0)[lt]{\lineheight{1.25}\smash{\begin{tabular}[t]{l}(e)\end{tabular}}}}%
    \put(0.69221008,0.2584165){\color[rgb]{0,0,0}\makebox(0,0)[lt]{\lineheight{1.25}\smash{\begin{tabular}[t]{l}(f)\end{tabular}}}}%
    \put(0.16531015,0.02738247){\color[rgb]{0,0,0}\makebox(0,0)[lt]{\lineheight{1.25}\smash{\begin{tabular}[t]{l}slipping\end{tabular}}}}%
    \put(0.06582167,0.02671628){\color[rgb]{0,0,0}\makebox(0,0)[lt]{\lineheight{1.25}\smash{\begin{tabular}[t]{l}stuck\end{tabular}}}}%
  \end{picture}%
\endgroup%
 \end{subfigure}
 \caption{Evolution of contact states for increasing vibration amplitude for the contact topography in \fref{surfaceTopography} (form, smooth). From left to right: $0.15~\mrm{mm}$, $0.68~\mrm{mm}$, $0.97~\mrm{mm}$ amplitude. Top row: full-FE reference, fine mesh. Bottom row: proposed multi-scale method, fine grid.}
 \label{fig:CSTATE}
\end{figure}
\\
\fref{MeshConvergenceModal} shows the results obtained for the first \emph{in-phase bending mode}.
Throughout the present work, the \emph{amplitude} is measured in terms of the Euclidean norm of the displacement at the sensor location indicated in \fref{S4Beam}, $\|\mm q_{\mrm{sensor}}\|_2$.
Recall that the frequencies are normalized by the respective linear one for tied apparent contact interface.
The softening trend and the damping increase are typical for jointed structures, and the results are in good quantitative agreement with the numerical results from \cite{Zare.2023}, and the experimental results from \cite{Wall.2022}.
To gain further insight into the contact behavior, the contact state (stuck vs. slipping) of the closed contacts is shown in \fref{CSTATE}.
The amplitude increases from $0.15~\mrm{mm}$ in the left column via $0.68~\mrm{mm}$ in the middle to $0.97~\mrm{mm}$ in the right column of the figure.
Those are associated with a maximum slip distance $0.93~\mum$, $3.8~\mum$, and $13.3~\mum$, respectively.
As expected, the larger the amplitude, the larger the slipping and the smaller the sticking part of the contact area.
\\
Severe convergence problems were encountered with \ABAQUS.
The amplitude at which the respective simulation stopped due to failed convergence is indicated by a cross in \fref{MeshConvergenceModal}.
The finer the mesh in the full-FE model, the earlier does the solver fail to converge.
This is accordance with the observations in \cite{jewell20}.
\\
When comparing the full-FE results obtained with the fine versus the super fine model, the natural frequency drops by an appreciable extent ($1.2 \%$).
On the other hand, the damping results are deemed stabilized.
Those convergence problems encountered with \ABAQUS were the reason why a penalty regularization was used instead of a Lagrange multiplier technique in \cite{jewell20}.
When adopting the regularization parameters from \cite{jewell20}, and simulating the fine full-FE model again, one is able to obtain results even for high amplitudes.
However, the regularization does not only improve the convergence behavior, but obviously distorts the results.
In particular, the modal behavior is noteworthy softer, and the damping ratio is much lower.
The latter is explained by the elastic sticking phase introduced by the regularization.
\\
In contrast to the full-FE reference, no convergence problems were encountered with the proposed multi-scale method.
Besides the fine mesh, results of the proposed method are also shown for a refined mesh (6000 instead of 4300 integration points).
The excellent agreement shows that the fine mesh has already converged.
The converged results obtained with the proposed multi-scale method agree well with the super-fine full-FE model (in the range where \ABAQUS converged), and with the fine full-FE model at higher vibration levels.
One may notice a slight shift of the gross slip kink to higher vibration levels.
This is attributed to the aforementioned edge effects near the bore hole:
For increasing vibration level, the contact transitions from stick to slip, from the outside towards the bore hole.
Directly around the bore hole, artificial stress peaks occur (\fref{PreLoadComparison}b), which delay the transition to gross slip.
It should be emphasized that the gross slip phase has limited relevance from a technical perspective.
Before the onset of gross slip in a bolted joint, contact is likely to occur between bolt and bore hole in reality (pinning), or the vibration stresses cause fatigue failure.
\\
\\
\fref{valiStudi} shows the results obtained for the first \emph{out-of-phase bending} and the \emph{first torsion} mode.
Recall that the first out-of-phase bending mode provokes a rolling around the $y$-axis (\cf \fref{modeShapes}) associated with opening-closing interactions.
Compared to the other two considered modes, the amplitude-dependence is rather weak.
This is in line with the experimental findings in \cite{Wall.2022}.
Apparently, the active contact area increases due to rolling, which leads to a slight hardening trend.
At the same time, the contact tends to stick more, leading to a slight damping decrease.
For higher vibration levels, the typical frictional softening and damping increase is predominant.
For a vibration amplitude of $>2.1~ \mrm{mm}$, the rolling extends towards the outer edges of the apparent contact area.
This scenario can be regarded as a worst case in terms of the edge effects expected from half-space theory.
In view of this, the agreement of the proposed multi-scale method with the full-FE reference is surprisingly good.
\\
Recall that the first torsion mode provokes rotational slip around the bolt axis ($z$-direction in \fref{modeShapes}).
The qualitative evolution of the modal properties with the amplitude is in line with the first in-phase bending mode, and the results are in good agreement with the reported numerical data from \cite{Zare.2023}.
Again, the agreement of the proposed multi-scale method with the full-FE reference is deemed very good.
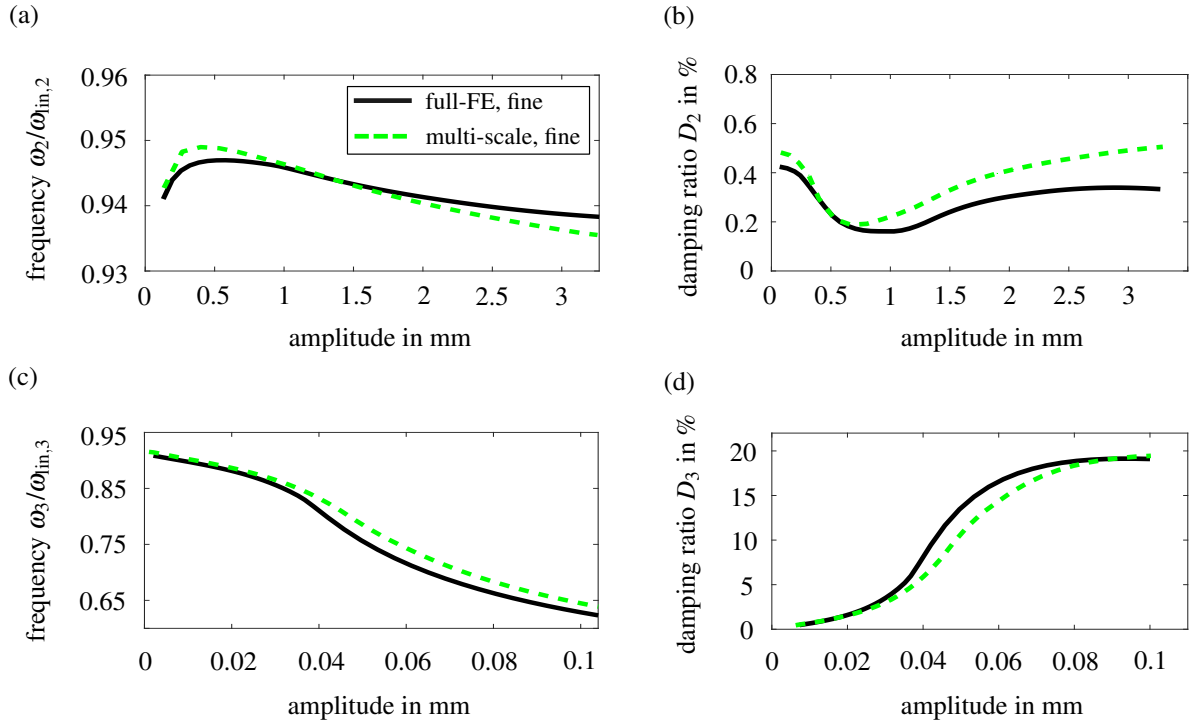
\begin{figure}[h]
 \centering
 \def\svgwidth{1.0\textwidth}
\begingroup%
  \makeatletter%
  \providecommand\color[2][]{%
    \errmessage{(Inkscape) Color is used for the text in Inkscape, but the package 'color.sty' is not loaded}%
    \renewcommand\color[2][]{}%
  }%
  \providecommand\transparent[1]{%
    \errmessage{(Inkscape) Transparency is used (non-zero) for the text in Inkscape, but the package 'transparent.sty' is not loaded}%
    \renewcommand\transparent[1]{}%
  }%
  \providecommand\rotatebox[2]{#2}%
  \newcommand*\fsize{\dimexpr\f@size pt\relax}%
  \newcommand*\lineheight[1]{\fontsize{\fsize}{#1\fsize}\selectfont}%
  \ifx\svgwidth\undefined%
    \setlength{\unitlength}{519.88186395bp}%
    \ifx\svgscale\undefined%
      \relax%
    \else%
      \setlength{\unitlength}{\unitlength * \real{\svgscale}}%
    \fi%
  \else%
    \setlength{\unitlength}{\svgwidth}%
  \fi%
  \global\let\svgwidth\undefined%
  \global\let\svgscale\undefined%
  \makeatother%
  \begin{picture}(1,0.59714736)%
    \lineheight{1}%
    \setlength\tabcolsep{0pt}%
    \put(0,0){\includegraphics[width=\unitlength]{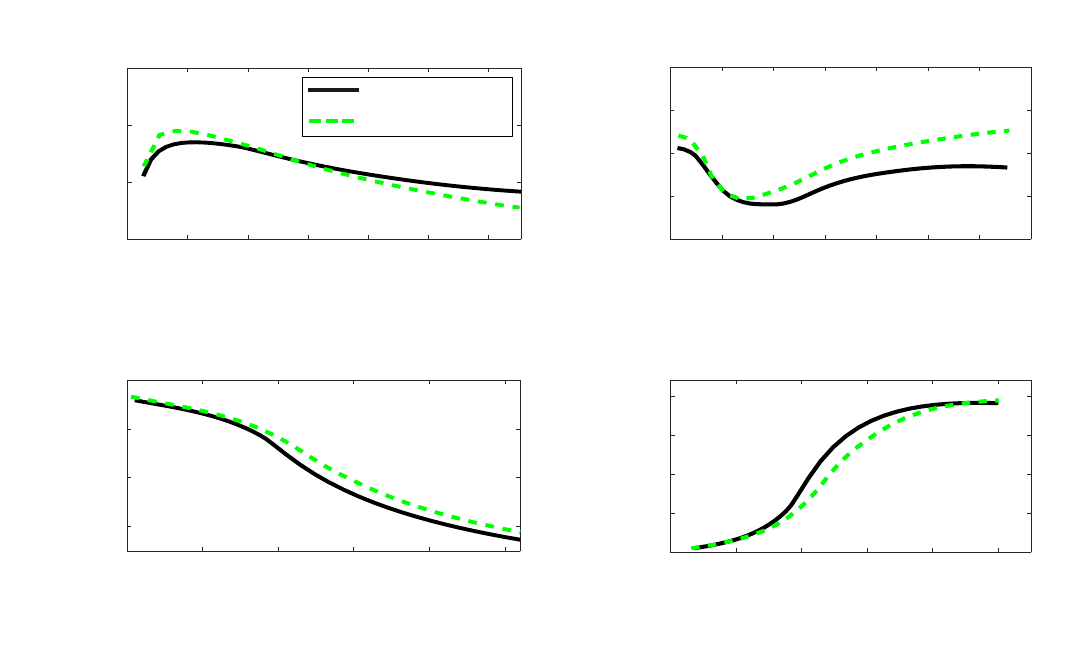}}%
    \put(0.23182983,0.02009629){\color[rgb]{0,0,0}\makebox(0,0)[lt]{\lineheight{1.25}\smash{\begin{tabular}[t]{l}amplitude in mm\end{tabular}}}}%
    \put(0.11264334,0.0559962){\color[rgb]{0,0,0}\makebox(0,0)[lt]{\lineheight{1.25}\smash{\begin{tabular}[t]{l}0\end{tabular}}}}%
    \put(0.16997377,0.0559962){\color[rgb]{0,0,0}\makebox(0,0)[lt]{\lineheight{1.25}\smash{\begin{tabular}[t]{l}0.02\end{tabular}}}}%
    \put(0.24041581,0.05599603){\color[rgb]{0,0,0}\makebox(0,0)[lt]{\lineheight{1.25}\smash{\begin{tabular}[t]{l}0.04\end{tabular}}}}%
    \put(0.30957269,0.05599603){\color[rgb]{0,0,0}\makebox(0,0)[lt]{\lineheight{1.25}\smash{\begin{tabular}[t]{l}0.06\end{tabular}}}}%
    \put(0.378821,0.05621118){\color[rgb]{0,0,0}\makebox(0,0)[lt]{\lineheight{1.25}\smash{\begin{tabular}[t]{l}0.08\end{tabular}}}}%
    \put(0.45232765,0.05648806){\color[rgb]{0,0,0}\makebox(0,0)[lt]{\lineheight{1.25}\smash{\begin{tabular}[t]{l}0.1\end{tabular}}}}%
    \put(0.61392353,0.05619637){\color[rgb]{0,0,0}\makebox(0,0)[lt]{\lineheight{1.25}\smash{\begin{tabular}[t]{l}0\end{tabular}}}}%
    \put(0.66082866,0.05619637){\color[rgb]{0,0,0}\makebox(0,0)[lt]{\lineheight{1.25}\smash{\begin{tabular}[t]{l}0.02\end{tabular}}}}%
    \put(0.72169072,0.05647808){\color[rgb]{0,0,0}\makebox(0,0)[lt]{\lineheight{1.25}\smash{\begin{tabular}[t]{l}0.04\end{tabular}}}}%
    \put(0.78183115,0.05619637){\color[rgb]{0,0,0}\makebox(0,0)[lt]{\lineheight{1.25}\smash{\begin{tabular}[t]{l}0.06\end{tabular}}}}%
    \put(0.84347174,0.0561294){\color[rgb]{0,0,0}\makebox(0,0)[lt]{\lineheight{1.25}\smash{\begin{tabular}[t]{l}0.08\end{tabular}}}}%
    \put(0.90880724,0.05640636){\color[rgb]{0,0,0}\makebox(0,0)[lt]{\lineheight{1.25}\smash{\begin{tabular}[t]{l}0.1\end{tabular}}}}%
    \put(0.59377433,0.08096258){\color[rgb]{0,0,0}\makebox(0,0)[lt]{\lineheight{1.25}\smash{\begin{tabular}[t]{l}0\end{tabular}}}}%
    \put(0.5940037,0.11628606){\color[rgb]{0,0,0}\makebox(0,0)[lt]{\lineheight{1.25}\smash{\begin{tabular}[t]{l}5\end{tabular}}}}%
    \put(0.5862624,0.15281433){\color[rgb]{0,0,0}\makebox(0,0)[lt]{\lineheight{1.25}\smash{\begin{tabular}[t]{l}10\end{tabular}}}}%
    \put(0.58579125,0.18798048){\color[rgb]{0,0,0}\makebox(0,0)[lt]{\lineheight{1.25}\smash{\begin{tabular}[t]{l}15\end{tabular}}}}%
    \put(0.58539219,0.22411914){\color[rgb]{0,0,0}\makebox(0,0)[lt]{\lineheight{1.25}\smash{\begin{tabular}[t]{l}20\end{tabular}}}}%
    \put(0.72230674,0.02047803){\color[rgb]{0,0,0}\makebox(0,0)[lt]{\lineheight{1.25}\smash{\begin{tabular}[t]{l}amplitude in mm\end{tabular}}}}%
    \put(0.55747013,0.06846685){\color[rgb]{0,0,0}\rotatebox{90}{\makebox(0,0)[lt]{\lineheight{1.25}\smash{\begin{tabular}[t]{l}damping ratio $D_3$ in \%\end{tabular}}}}}%
    \put(0.06529663,0.10404443){\color[rgb]{0,0,0}\makebox(0,0)[lt]{\lineheight{1.25}\smash{\begin{tabular}[t]{l}0.65\end{tabular}}}}%
    \put(0.06569109,0.15007544){\color[rgb]{0,0,0}\makebox(0,0)[lt]{\lineheight{1.25}\smash{\begin{tabular}[t]{l}0.75\end{tabular}}}}%
    \put(0.06608956,0.1948381){\color[rgb]{0,0,0}\makebox(0,0)[lt]{\lineheight{1.25}\smash{\begin{tabular}[t]{l}0.85\end{tabular}}}}%
    \put(0.0662835,0.23715106){\color[rgb]{0,0,0}\makebox(0,0)[lt]{\lineheight{1.25}\smash{\begin{tabular}[t]{l}0.95\end{tabular}}}}%
    \put(0.03730719,0.3697098){\color[rgb]{0,0,0}\rotatebox{90}{\makebox(0,0)[lt]{\lineheight{1.25}\smash{\begin{tabular}[t]{l}frequency $\omega_2$/$\omega_{\mrm{lin},2}$\end{tabular}}}}}%
    \put(0.03732926,0.07844392){\color[rgb]{0,0,0}\rotatebox{90}{\makebox(0,0)[lt]{\lineheight{1.25}\smash{\begin{tabular}[t]{l}frequency $\omega_3$/$\omega_{\mrm{lin},3}$\end{tabular}}}}}%
    \put(0.23265463,0.31733045){\color[rgb]{0,0,0}\makebox(0,0)[lt]{\lineheight{1.25}\smash{\begin{tabular}[t]{l}amplitude in mm\end{tabular}}}}%
    \put(0.72223145,0.3173495){\color[rgb]{0,0,0}\makebox(0,0)[lt]{\lineheight{1.25}\smash{\begin{tabular}[t]{l}amplitude in mm\end{tabular}}}}%
    \put(0.61226123,0.35063206){\color[rgb]{0,0,0}\makebox(0,0)[lt]{\lineheight{1.25}\smash{\begin{tabular}[t]{l}0\end{tabular}}}}%
    \put(0.65330404,0.35070378){\color[rgb]{0,0,0}\makebox(0,0)[lt]{\lineheight{1.25}\smash{\begin{tabular}[t]{l}0.5\end{tabular}}}}%
    \put(0.70870397,0.35021191){\color[rgb]{0,0,0}\makebox(0,0)[lt]{\lineheight{1.25}\smash{\begin{tabular}[t]{l}1\end{tabular}}}}%
    \put(0.74864803,0.35070378){\color[rgb]{0,0,0}\makebox(0,0)[lt]{\lineheight{1.25}\smash{\begin{tabular}[t]{l}1.5\end{tabular}}}}%
    \put(0.80401463,0.35077545){\color[rgb]{0,0,0}\makebox(0,0)[lt]{\lineheight{1.25}\smash{\begin{tabular}[t]{l}2\end{tabular}}}}%
    \put(0.84399198,0.35059874){\color[rgb]{0,0,0}\makebox(0,0)[lt]{\lineheight{1.25}\smash{\begin{tabular}[t]{l}2.5\end{tabular}}}}%
    \put(0.89932522,0.35080881){\color[rgb]{0,0,0}\makebox(0,0)[lt]{\lineheight{1.25}\smash{\begin{tabular}[t]{l}3\end{tabular}}}}%
    \put(0.11190945,0.35018404){\color[rgb]{0,0,0}\makebox(0,0)[lt]{\lineheight{1.25}\smash{\begin{tabular}[t]{l}0\end{tabular}}}}%
    \put(0.15915107,0.35081925){\color[rgb]{0,0,0}\makebox(0,0)[lt]{\lineheight{1.25}\smash{\begin{tabular}[t]{l}0.5\end{tabular}}}}%
    \put(0.22384925,0.35117268){\color[rgb]{0,0,0}\makebox(0,0)[lt]{\lineheight{1.25}\smash{\begin{tabular}[t]{l}1\end{tabular}}}}%
    \put(0.27337325,0.35081921){\color[rgb]{0,0,0}\makebox(0,0)[lt]{\lineheight{1.25}\smash{\begin{tabular}[t]{l}1.5\end{tabular}}}}%
    \put(0.33550221,0.35060918){\color[rgb]{0,0,0}\makebox(0,0)[lt]{\lineheight{1.25}\smash{\begin{tabular}[t]{l}2\end{tabular}}}}%
    \put(0.3819602,0.35043242){\color[rgb]{0,0,0}\makebox(0,0)[lt]{\lineheight{1.25}\smash{\begin{tabular}[t]{l}2.5\end{tabular}}}}%
    \put(0.44574634,0.35007896){\color[rgb]{0,0,0}\makebox(0,0)[lt]{\lineheight{1.25}\smash{\begin{tabular}[t]{l}3\end{tabular}}}}%
    \put(0.5587861,0.35651691){\color[rgb]{0,0,0}\rotatebox{90}{\makebox(0,0)[lt]{\lineheight{1.25}\smash{\begin{tabular}[t]{l}damping ratio $D_2$ in \%\end{tabular}}}}}%
    \put(0.59382549,0.37067032){\color[rgb]{0,0,0}\makebox(0,0)[lt]{\lineheight{1.25}\smash{\begin{tabular}[t]{l}0\end{tabular}}}}%
    \put(0.58164435,0.40929348){\color[rgb]{0,0,0}\makebox(0,0)[lt]{\lineheight{1.25}\smash{\begin{tabular}[t]{l}0.2\end{tabular}}}}%
    \put(0.58060454,0.44747099){\color[rgb]{0,0,0}\makebox(0,0)[lt]{\lineheight{1.25}\smash{\begin{tabular}[t]{l}0.4\end{tabular}}}}%
    \put(0.58075301,0.48847105){\color[rgb]{0,0,0}\makebox(0,0)[lt]{\lineheight{1.25}\smash{\begin{tabular}[t]{l}0.6\end{tabular}}}}%
    \put(0.58119866,0.52872831){\color[rgb]{0,0,0}\makebox(0,0)[lt]{\lineheight{1.25}\smash{\begin{tabular}[t]{l}0.8\end{tabular}}}}%
    \put(0.06694223,0.36814978){\color[rgb]{0,0,0}\makebox(0,0)[lt]{\lineheight{1.25}\smash{\begin{tabular}[t]{l}0.93\end{tabular}}}}%
    \put(0.06661534,0.424574){\color[rgb]{0,0,0}\makebox(0,0)[lt]{\lineheight{1.25}\smash{\begin{tabular}[t]{l}0.94\end{tabular}}}}%
    \put(0.06602125,0.47683274){\color[rgb]{0,0,0}\makebox(0,0)[lt]{\lineheight{1.25}\smash{\begin{tabular}[t]{l}0.95\end{tabular}}}}%
    \put(0.06643377,0.52650488){\color[rgb]{0,0,0}\makebox(0,0)[lt]{\lineheight{1.25}\smash{\begin{tabular}[t]{l}0.96\end{tabular}}}}%
    \put(0.0090782,0.57730429){\color[rgb]{0,0,0}\makebox(0,0)[lt]{\lineheight{1.25}\smash{\begin{tabular}[t]{l}(a)\end{tabular}}}}%
    \put(0.5323283,0.57682913){\color[rgb]{0,0,0}\makebox(0,0)[lt]{\lineheight{1.25}\smash{\begin{tabular}[t]{l}(b)\end{tabular}}}}%
    \put(0.00828125,0.28539881){\color[rgb]{0,0,0}\makebox(0,0)[lt]{\lineheight{1.25}\smash{\begin{tabular}[t]{l}(c)\end{tabular}}}}%
    \put(0.53267499,0.28181267){\color[rgb]{0,0,0}\makebox(0,0)[lt]{\lineheight{1.25}\smash{\begin{tabular}[t]{l}(d)\end{tabular}}}}%
    \put(0.34208615,0.50787359){\color[rgb]{0,0,0}\makebox(0,0)[lt]{\lineheight{1.25}\smash{\begin{tabular}[t]{l} \small{full-FE, fine}\end{tabular}}}}%
    \put(0.34215768,0.47994552){\color[rgb]{0,0,0}\makebox(0,0)[lt]{\lineheight{1.25}\smash{\begin{tabular}[t]{l} \small{multi-scale, fine}\end{tabular}}}}%
  \end{picture}%
\endgroup%
  \caption{Amplitude-dependent properties of the first out-of-phase bending mode (top) and the first torsion mode (bottom) for the contact topography in \fref{surfaceTopography} (form, smooth): (left) frequency, (right) damping ratio.
  }%
  \label{fig:valiStudi}
\end{figure}

\subsection{On the relevance of form deviations and roughness\label{sec:rough}}
\begin{figure}[h]%
  \centering
  \subfloat[][]{\includegraphics[width=0.5\linewidth]{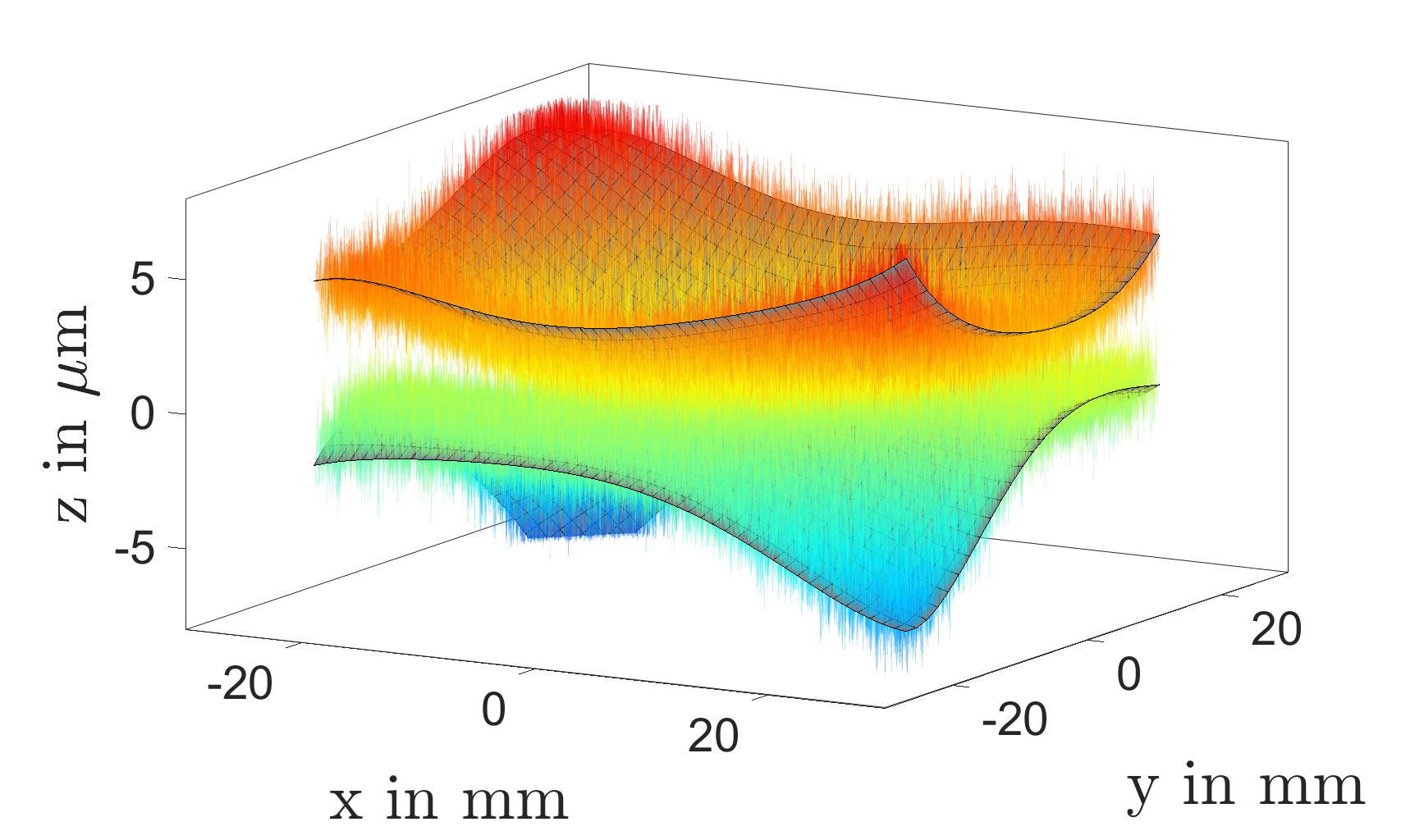}}%
  \subfloat[][]{\includegraphics[width=0.5\linewidth]{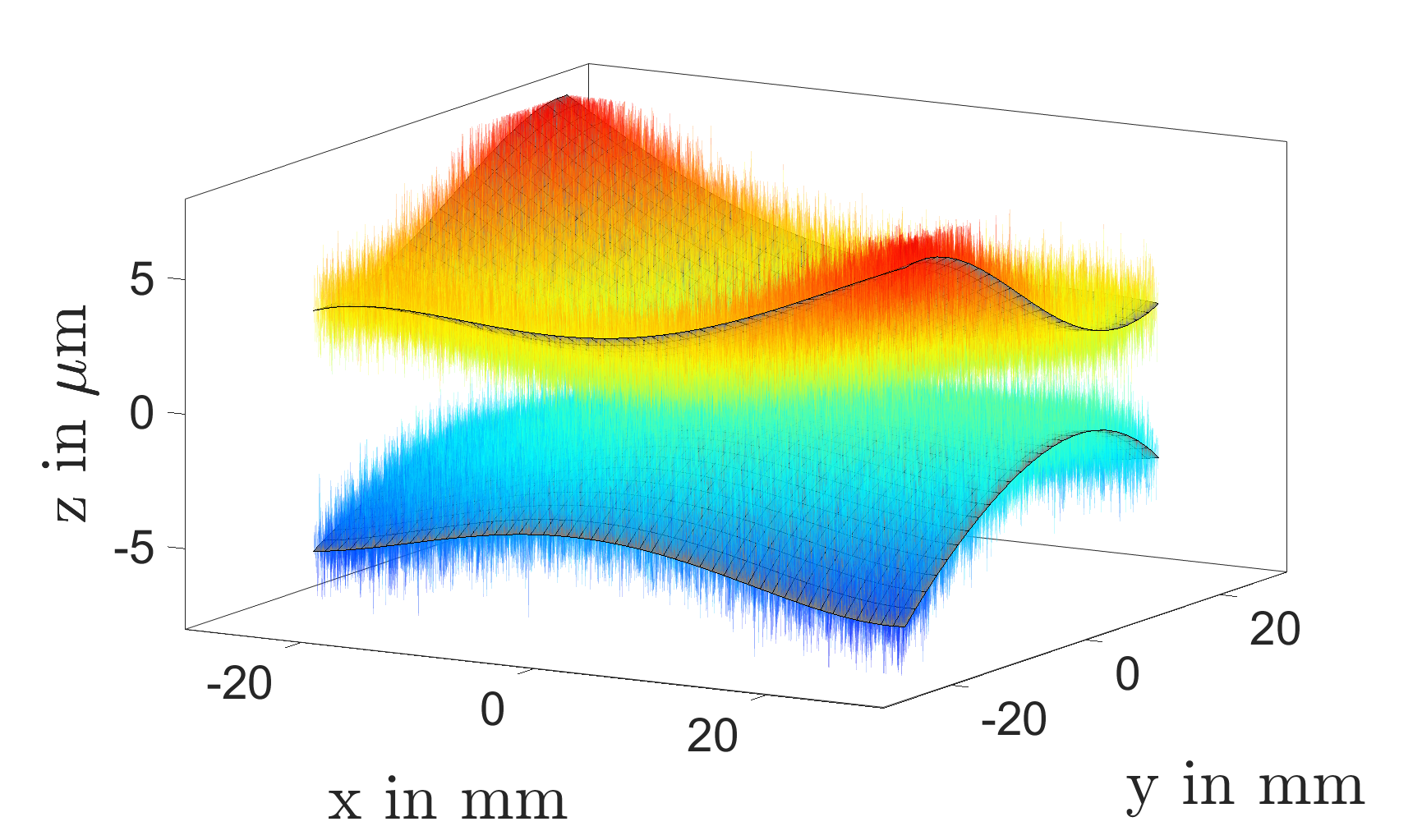}}%
  \caption{Contact topography as in \fref{surfaceTopography} but with synthetic roughness added, $\sigma=1~\mum$: (a) left, (b) right interface.}%
  \label{fig:surfaceTopography_rough}
\end{figure}
%
Recall that it is still an open research question, down to what length scale the topography has to be resolved for an accurate damping prediction.
The proposed multi-scale method opens up new opportunities to address this question.
In lack of sufficiently accurate measurement data, a synthetic height profile was generated, as detailed below, and superimposed onto the experimentally determined form (\fref{surfaceTopography}).
In this section, the focus is placed on the first in-phase bending mode.
\\
To generate the rough surface, a constant auto-correlation of $\sigma^2$ was considered, the corresponding discrete power spectral density was computed, and truncated to the band of wave lengths from $0.5~\mathrm{mm}$ to $5~\mathrm{mm}$.
The height profile was then obtained as the inverse discrete Fourier transform, where the magnitude is given by the power spectral density, and a pseudo-random phase was drawn for each spectral line from the uniform distribution between $0$ and $2\pi$.
Different standard height deviations were considered, $\sigma \in \lbrace 0.1~\mum,\, 1~\mum,\, 5~\mum\rbrace$, which should be viewed in perspective to the maximum peak-to-peak form deviation of $12.14~ \mum$.
Note that the upper bound of the wave length band, $5~\mrm{mm}$, is small compared to the wave length of the form deviation of about $120 ~\mrm{mm}$, and the lower bound, $0.5~\mathrm{mm}$, is sufficiently large so that it can be resolved with the super fine BE grid ($\Delta x=\Delta y = 0.25 ~\mrm{mm}$).
The resulting topography is illustrated in \fref{surfaceTopography_rough} for $\sigma=1~\mum$.
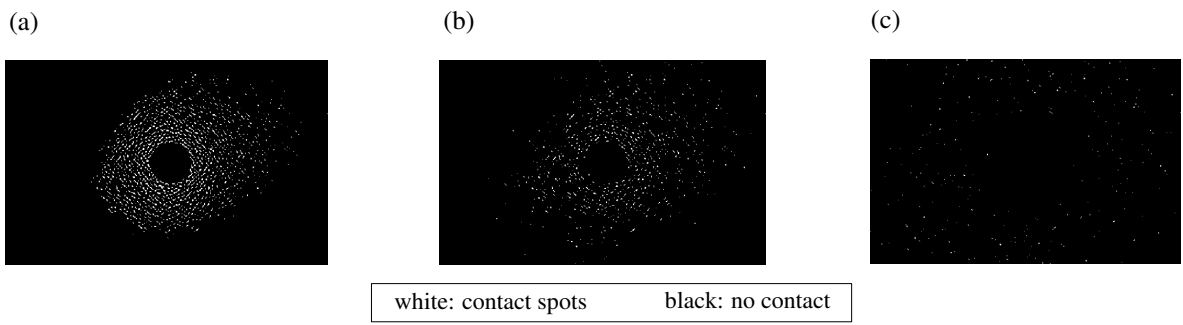
\begin{figure}[b]
 \centering
 \def\svgwidth{1.0\textwidth}
\begingroup%
  \makeatletter%
  \providecommand\color[2][]{%
    \errmessage{(Inkscape) Color is used for the text in Inkscape, but the package 'color.sty' is not loaded}%
    \renewcommand\color[2][]{}%
  }%
  \providecommand\transparent[1]{%
    \errmessage{(Inkscape) Transparency is used (non-zero) for the text in Inkscape, but the package 'transparent.sty' is not loaded}%
    \renewcommand\transparent[1]{}%
  }%
  \providecommand\rotatebox[2]{#2}%
  \newcommand*\fsize{\dimexpr\f@size pt\relax}%
  \newcommand*\lineheight[1]{\fontsize{\fsize}{#1\fsize}\selectfont}%
  \ifx\svgwidth\undefined%
    \setlength{\unitlength}{1022.54270543bp}%
    \ifx\svgscale\undefined%
      \relax%
    \else%
      \setlength{\unitlength}{\unitlength * \real{\svgscale}}%
    \fi%
  \else%
    \setlength{\unitlength}{\svgwidth}%
  \fi%
  \global\let\svgwidth\undefined%
  \global\let\svgscale\undefined%
  \makeatother%
  \begin{picture}(1,0.26751586)%
    \lineheight{1}%
    \setlength\tabcolsep{0pt}%
    \put(0,0){\includegraphics[width=\unitlength]{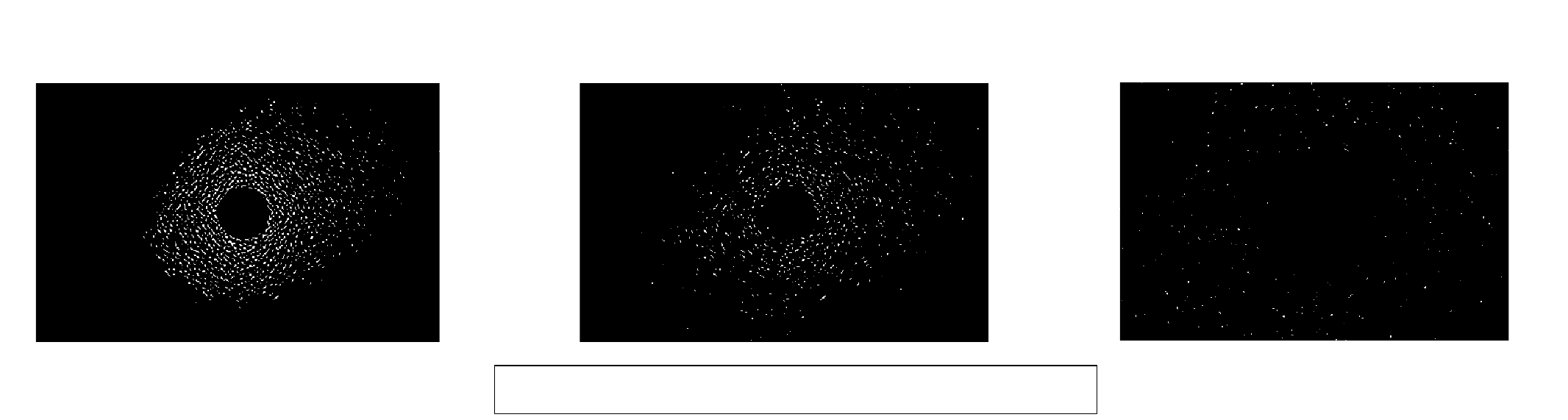}}%
    \put(0.02568092,0.2393928){\color[rgb]{0,0,0}\makebox(0,0)[lt]{\lineheight{1.25}\smash{\begin{tabular}[t]{l}(a)\end{tabular}}}}%
    \put(0.37333394,0.24010187){\color[rgb]{0,0,0}\makebox(0,0)[lt]{\lineheight{1.25}\smash{\begin{tabular}[t]{l}(b)\end{tabular}}}}%
    \put(0.71518079,0.24042715){\color[rgb]{0,0,0}\makebox(0,0)[lt]{\lineheight{1.25}\smash{\begin{tabular}[t]{l}(c)\end{tabular}}}}%
    \put(0.33411759,0.01444759){\color[rgb]{0,0,0}\makebox(0,0)[lt]{\lineheight{1.25}\smash{\begin{tabular}[t]{l}\small{white: contact spots}\end{tabular}}}}%
    \put(0.55038649,0.01498575){\color[rgb]{0,0,0}\makebox(0,0)[lt]{\lineheight{1.25}\smash{\begin{tabular}[t]{l}\small{black: no contact}\end{tabular}}}}%
  \end{picture}%
\endgroup%
  \caption{Contact area after preload application according to proposed multi-scale method with a super-fine grid: (a) $\sigma=0.1~\mum$, (b) $\sigma=1~\mum$ and (c) $\sigma=5~\mum$.}%
  \label{fig:contactArea}
\end{figure}
\\
The resulting contact area obtained after the preload step is shown in \fref{contactArea}.
For $\sigma=0.1~\mum$, the overall spatial distribution of the active contacts is clearly dominated by the form (\cf \fref{PreLoadComparison}, $\sigma=0$).
Yet, the contact is localized to individual peaks.
For larger $\sigma$, the effect of the form is much less prominent and practically lost for sufficiently high $\sigma$.
The higher $\sigma$, the smaller the actual contact area, and the higher the maximum stresses.
When the topography is resolved down to the roughness length scale, it is expected that the stresses are so concentrated that plastic deformation is inevitable.
The highest computed stresses, considering those obtained with the BE model and with the full-FE model, were $<2100~\mrm{MPa}$.
This is deemed sufficiently small to justify restricting the present work to linear elasticity.
\begin{figure}[t]
 \centering
 \def\svgwidth{1.0\textwidth}
\begingroup%
  \makeatletter%
  \providecommand\color[2][]{%
    \errmessage{(Inkscape) Color is used for the text in Inkscape, but the package 'color.sty' is not loaded}%
    \renewcommand\color[2][]{}%
  }%
  \providecommand\transparent[1]{%
    \errmessage{(Inkscape) Transparency is used (non-zero) for the text in Inkscape, but the package 'transparent.sty' is not loaded}%
    \renewcommand\transparent[1]{}%
  }%
  \providecommand\rotatebox[2]{#2}%
  \newcommand*\fsize{\dimexpr\f@size pt\relax}%
  \newcommand*\lineheight[1]{\fontsize{\fsize}{#1\fsize}\selectfont}%
  \ifx\svgwidth\undefined%
    \setlength{\unitlength}{829.05462113bp}%
    \ifx\svgscale\undefined%
      \relax%
    \else%
      \setlength{\unitlength}{\unitlength * \real{\svgscale}}%
    \fi%
  \else%
    \setlength{\unitlength}{\svgwidth}%
  \fi%
  \global\let\svgwidth\undefined%
  \global\let\svgscale\undefined%
  \makeatother%
  \begin{picture}(1,0.28729346)%
    \lineheight{1}%
    \setlength\tabcolsep{0pt}%
    \put(0,0){\includegraphics[width=\unitlength]{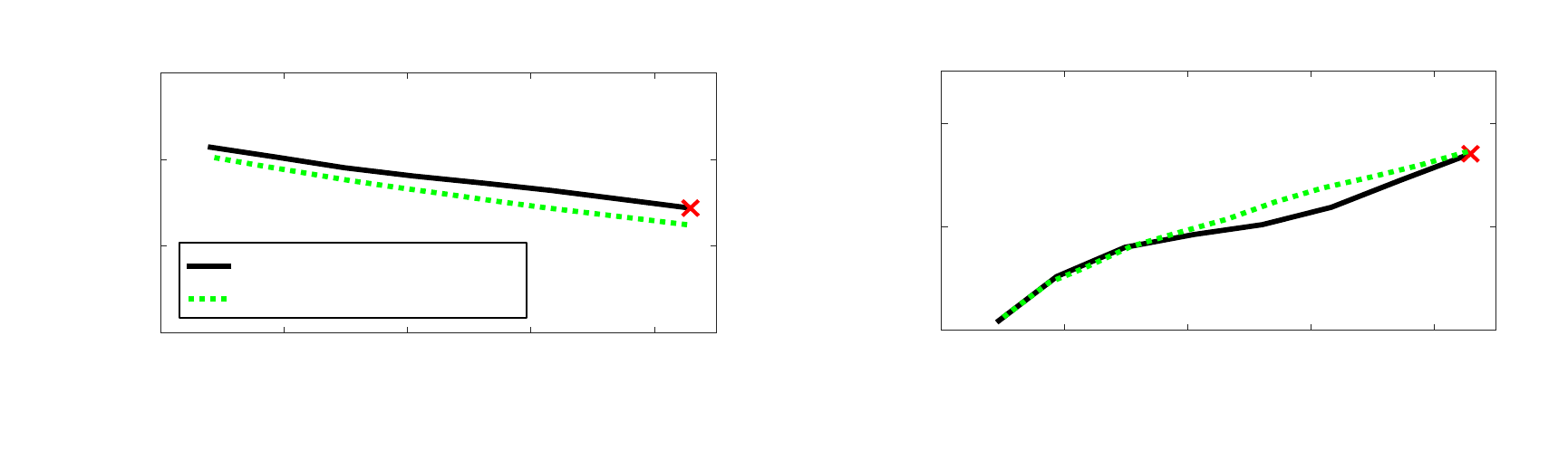}}%
    \put(0.05207886,0.1269928){\color[rgb]{0,0,0}\makebox(0,0)[lt]{\lineheight{1.25}\smash{\begin{tabular}[t]{l}0.96\end{tabular}}}}%
    \put(0.05218491,0.07294489){\color[rgb]{0,0,0}\makebox(0,0)[lt]{\lineheight{1.25}\smash{\begin{tabular}[t]{l}0.95\end{tabular}}}}%
    \put(0.0524903,0.18078112){\color[rgb]{0,0,0}\makebox(0,0)[lt]{\lineheight{1.25}\smash{\begin{tabular}[t]{l}0.97\end{tabular}}}}%
    \put(0.05190441,0.2373405){\color[rgb]{0,0,0}\makebox(0,0)[lt]{\lineheight{1.25}\smash{\begin{tabular}[t]{l}0.98\end{tabular}}}}%
    \put(0.67171419,0.047523){\color[rgb]{0,0,0}\makebox(0,0)[lt]{\lineheight{1.25}\smash{\begin{tabular}[t]{l}0.1\end{tabular}}}}%
    \put(0.74924131,0.04763942){\color[rgb]{0,0,0}\makebox(0,0)[lt]{\lineheight{1.25}\smash{\begin{tabular}[t]{l}0.2\end{tabular}}}}%
    \put(0.82758643,0.04675634){\color[rgb]{0,0,0}\makebox(0,0)[lt]{\lineheight{1.25}\smash{\begin{tabular}[t]{l}0.3\end{tabular}}}}%
    \put(0.90747093,0.04700718){\color[rgb]{0,0,0}\makebox(0,0)[lt]{\lineheight{1.25}\smash{\begin{tabular}[t]{l}0.4\end{tabular}}}}%
    \put(0.59620753,0.0475193){\color[rgb]{0,0,0}\makebox(0,0)[lt]{\lineheight{1.25}\smash{\begin{tabular}[t]{l}0\end{tabular}}}}%
    \put(0.17211063,0.04675532){\color[rgb]{0,0,0}\makebox(0,0)[lt]{\lineheight{1.25}\smash{\begin{tabular}[t]{l}0.1\end{tabular}}}}%
    \put(0.25138693,0.04787122){\color[rgb]{0,0,0}\makebox(0,0)[lt]{\lineheight{1.25}\smash{\begin{tabular}[t]{l}0.2\end{tabular}}}}%
    \put(0.32973199,0.04748789){\color[rgb]{0,0,0}\makebox(0,0)[lt]{\lineheight{1.25}\smash{\begin{tabular}[t]{l}0.3\end{tabular}}}}%
    \put(0.40911677,0.047239){\color[rgb]{0,0,0}\makebox(0,0)[lt]{\lineheight{1.25}\smash{\begin{tabular}[t]{l}0.4\end{tabular}}}}%
    \put(0.09810321,0.04750122){\color[rgb]{0,0,0}\makebox(0,0)[lt]{\lineheight{1.25}\smash{\begin{tabular}[t]{l}0\end{tabular}}}}%
    \put(0.57174465,0.07320883){\color[rgb]{0,0,0}\makebox(0,0)[lt]{\lineheight{1.25}\smash{\begin{tabular}[t]{l}0\end{tabular}}}}%
    \put(0.56396344,0.14089609){\color[rgb]{0,0,0}\makebox(0,0)[lt]{\lineheight{1.25}\smash{\begin{tabular}[t]{l}0.2\end{tabular}}}}%
    \put(0.56337833,0.20673444){\color[rgb]{0,0,0}\makebox(0,0)[lt]{\lineheight{1.25}\smash{\begin{tabular}[t]{l}0.4\end{tabular}}}}%
    \put(0.20706042,0.01680731){\color[rgb]{0,0,0}\makebox(0,0)[lt]{\lineheight{1.25}\smash{\begin{tabular}[t]{l}amplitude in mm\end{tabular}}}}%
    \put(0.71770519,0.0173825){\color[rgb]{0,0,0}\makebox(0,0)[lt]{\lineheight{1.25}\smash{\begin{tabular}[t]{l}amplitude in mm\end{tabular}}}}%
    \put(0.02507642,0.068137){\color[rgb]{0,0,0}\rotatebox{90}{\makebox(0,0)[lt]{\lineheight{1.25}\smash{\begin{tabular}[t]{l}frequency $\omega_1$/$\omega_{\mrm{lin},1}$\end{tabular}}}}}%
    \put(0.5412564,0.06004732){\color[rgb]{0,0,0}\rotatebox{90}{\makebox(0,0)[lt]{\lineheight{1.25}\smash{\begin{tabular}[t]{l}damping ratio $D_1$ in \%\end{tabular}}}}}%
    \put(0.00387324,0.27343477){\color[rgb]{0,0,0}\makebox(0,0)[lt]{\lineheight{1.25}\smash{\begin{tabular}[t]{l}(a)\end{tabular}}}}%
    \put(0.51533683,0.27659517){\color[rgb]{0,0,0}\makebox(0,0)[lt]{\lineheight{1.25}\smash{\begin{tabular}[t]{l}(b)\end{tabular}}}}%
    \put(0.15678938,0.09279285){\color[rgb]{0,0,0}\makebox(0,0)[lt]{\lineheight{1.25}\smash{\begin{tabular}[t]{l}\small{multi-scale, super-fine}\end{tabular}}}}%
    \put(0.15620133,0.11390376){\color[rgb]{0,0,0}\makebox(0,0)[lt]{\lineheight{1.25}\smash{\begin{tabular}[t]{l}\small{full-FE, super-fine}\end{tabular}}}}%
  \end{picture}%
\endgroup%
  \caption{Amplitude-dependent properties of the first in-phase bending mode for the contact topography in \fref{surfaceTopography_rough} (form, $\sigma=1~\mum$): (a) frequency, (b) damping ratio.
  }%
  \label{fig:roughValidation}
\end{figure}
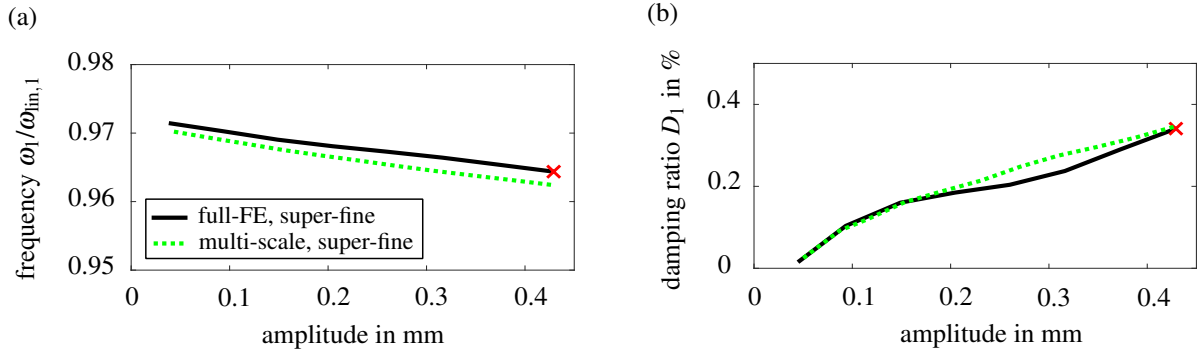
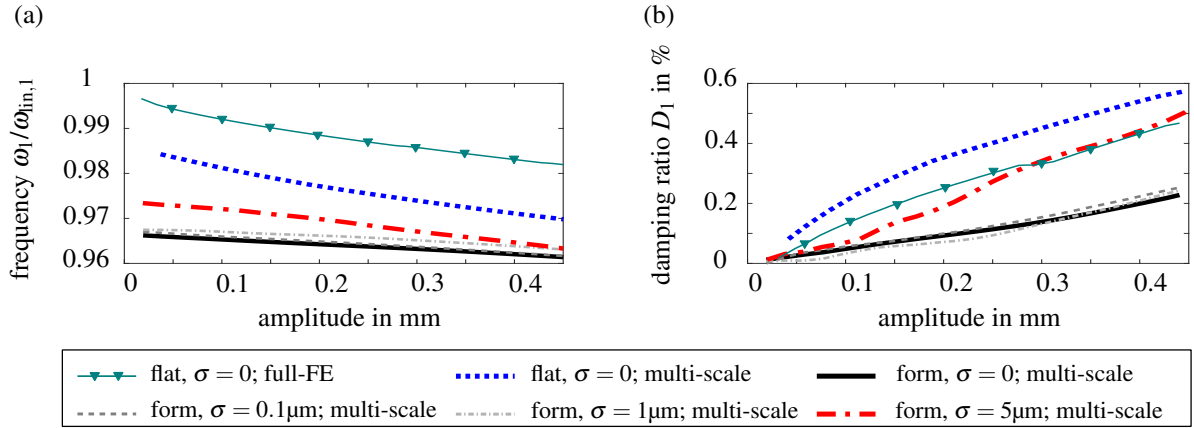
\begin{figure}[t]
 \centering
 \def\svgwidth{1.0\textwidth}
\begingroup%
  \makeatletter%
  \providecommand\color[2][]{%
    \errmessage{(Inkscape) Color is used for the text in Inkscape, but the package 'color.sty' is not loaded}%
    \renewcommand\color[2][]{}%
  }%
  \providecommand\transparent[1]{%
    \errmessage{(Inkscape) Transparency is used (non-zero) for the text in Inkscape, but the package 'transparent.sty' is not loaded}%
    \renewcommand\transparent[1]{}%
  }%
  \providecommand\rotatebox[2]{#2}%
  \newcommand*\fsize{\dimexpr\f@size pt\relax}%
  \newcommand*\lineheight[1]{\fontsize{\fsize}{#1\fsize}\selectfont}%
  \ifx\svgwidth\undefined%
    \setlength{\unitlength}{866.48433981bp}%
    \ifx\svgscale\undefined%
      \relax%
    \else%
      \setlength{\unitlength}{\unitlength * \real{\svgscale}}%
    \fi%
  \else%
    \setlength{\unitlength}{\svgwidth}%
  \fi%
  \global\let\svgwidth\undefined%
  \global\let\svgscale\undefined%
  \makeatother%
  \begin{picture}(1,0.35236681)%
    \lineheight{1}%
    \setlength\tabcolsep{0pt}%
    \put(0,0){\includegraphics[width=\unitlength]{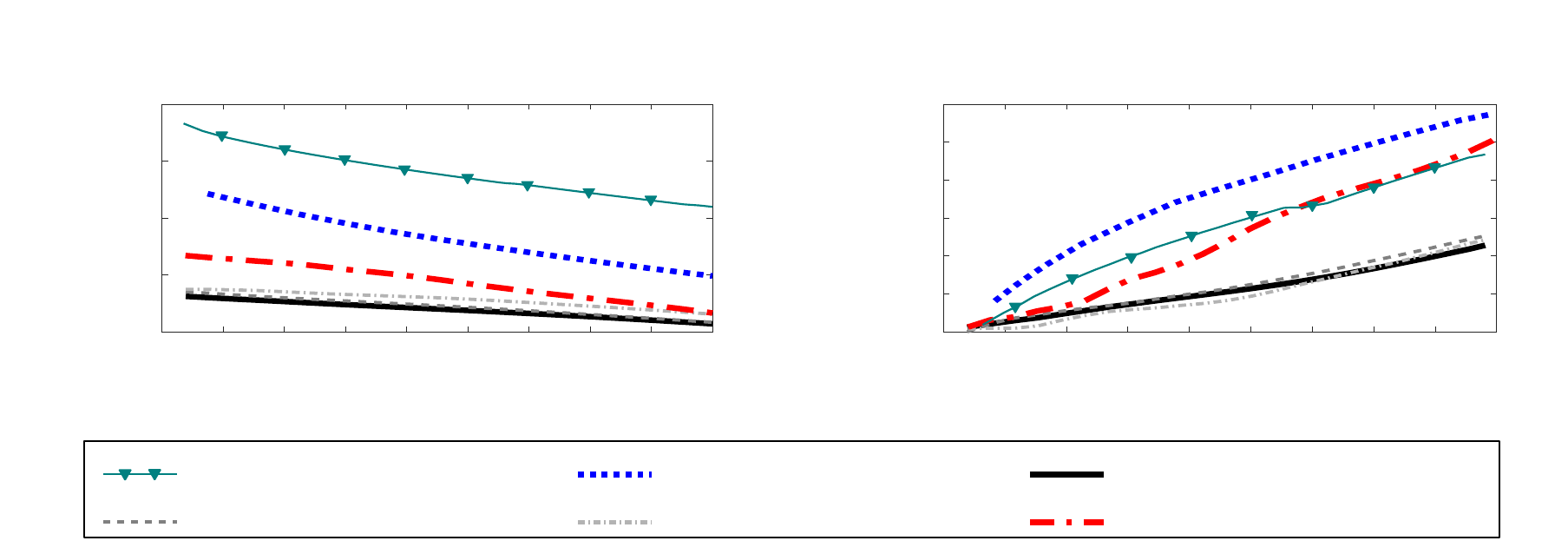}}%
    \put(0.10327293,0.11879903){\color[rgb]{0,0,0}\makebox(0,0)[lt]{\lineheight{1.25}\smash{\begin{tabular}[t]{l}0\end{tabular}}}}%
    \put(0.17729261,0.11869085){\color[rgb]{0,0,0}\makebox(0,0)[lt]{\lineheight{1.25}\smash{\begin{tabular}[t]{l}0.1\end{tabular}}}}%
    \put(0.25602864,0.11865136){\color[rgb]{0,0,0}\makebox(0,0)[lt]{\lineheight{1.25}\smash{\begin{tabular}[t]{l}0.2\end{tabular}}}}%
    \put(0.33329967,0.11880437){\color[rgb]{0,0,0}\makebox(0,0)[lt]{\lineheight{1.25}\smash{\begin{tabular}[t]{l}0.3\end{tabular}}}}%
    \put(0.41133581,0.11865134){\color[rgb]{0,0,0}\makebox(0,0)[lt]{\lineheight{1.25}\smash{\begin{tabular}[t]{l}0.4\end{tabular}}}}%
    \put(0.60310875,0.11897674){\color[rgb]{0,0,0}\makebox(0,0)[lt]{\lineheight{1.25}\smash{\begin{tabular}[t]{l}0\end{tabular}}}}%
    \put(0.67712842,0.11886855){\color[rgb]{0,0,0}\makebox(0,0)[lt]{\lineheight{1.25}\smash{\begin{tabular}[t]{l}0.1\end{tabular}}}}%
    \put(0.75586446,0.11882907){\color[rgb]{0,0,0}\makebox(0,0)[lt]{\lineheight{1.25}\smash{\begin{tabular}[t]{l}0.2\end{tabular}}}}%
    \put(0.8331355,0.11898208){\color[rgb]{0,0,0}\makebox(0,0)[lt]{\lineheight{1.25}\smash{\begin{tabular}[t]{l}0.3\end{tabular}}}}%
    \put(0.91117168,0.11882904){\color[rgb]{0,0,0}\makebox(0,0)[lt]{\lineheight{1.25}\smash{\begin{tabular}[t]{l}0.4\end{tabular}}}}%
    \put(0.21066807,0.09061116){\color[rgb]{0,0,0}\makebox(0,0)[lt]{\lineheight{1.25}\smash{\begin{tabular}[t]{l}amplitude in mm\end{tabular}}}}%
    \put(0.70654277,0.09063574){\color[rgb]{0,0,0}\makebox(0,0)[lt]{\lineheight{1.25}\smash{\begin{tabular}[t]{l}amplitude in mm\end{tabular}}}}%
    \put(0.56721683,0.18436572){\color[rgb]{0,0,0}\makebox(0,0)[lt]{\lineheight{1.25}\smash{\begin{tabular}[t]{l}0.2\end{tabular}}}}%
    \put(0.57468239,0.13773332){\color[rgb]{0,0,0}\makebox(0,0)[lt]{\lineheight{1.25}\smash{\begin{tabular}[t]{l}0\end{tabular}}}}%
    \put(0.56786602,0.23456859){\color[rgb]{0,0,0}\makebox(0,0)[lt]{\lineheight{1.25}\smash{\begin{tabular}[t]{l}0.4\end{tabular}}}}%
    \put(0.56710866,0.28455506){\color[rgb]{0,0,0}\makebox(0,0)[lt]{\lineheight{1.25}\smash{\begin{tabular}[t]{l}0.6\end{tabular}}}}%
    \put(0.05481,0.21063525){\color[rgb]{0,0,0}\makebox(0,0)[lt]{\lineheight{1.25}\smash{\begin{tabular}[t]{l}0.98\end{tabular}}}}%
    \put(0.05449232,0.17286596){\color[rgb]{0,0,0}\makebox(0,0)[lt]{\lineheight{1.25}\smash{\begin{tabular}[t]{l}0.97\end{tabular}}}}%
    \put(0.05473013,0.24628618){\color[rgb]{0,0,0}\makebox(0,0)[lt]{\lineheight{1.25}\smash{\begin{tabular}[t]{l}0.99\end{tabular}}}}%
    \put(0.06821746,0.28412228){\color[rgb]{0,0,0}\makebox(0,0)[lt]{\lineheight{1.25}\smash{\begin{tabular}[t]{l}1\end{tabular}}}}%
    \put(0.02647661,0.12780553){\color[rgb]{0,0,0}\rotatebox{90}{\makebox(0,0)[lt]{\lineheight{1.25}\smash{\begin{tabular}[t]{l}frequency $\omega_1$/$\omega_{\mrm{lin,1}}$\end{tabular}}}}}%
    \put(0.53969668,0.12424185){\color[rgb]{0,0,0}\rotatebox{90}{\makebox(0,0)[lt]{\lineheight{1.25}\smash{\begin{tabular}[t]{l}damping ratio $D_1$ in \%\end{tabular}}}}}%
    \put(0.12435086,0.04666655){\color[rgb]{0,0,0}\makebox(0,0)[lt]{\lineheight{1.25}\smash{\begin{tabular}[t]{l}\small{flat, $\sigma=0$; full-FE}\end{tabular}}}}%
    \put(0.42854882,0.04666615){\color[rgb]{0,0,0}\makebox(0,0)[lt]{\lineheight{1.25}\smash{\begin{tabular}[t]{l}\small{flat, $\sigma=0$; multi-scale}\end{tabular}}}}%
    \put(0.72128394,0.0462277){\color[rgb]{0,0,0}\makebox(0,0)[lt]{\lineheight{1.25}\smash{\begin{tabular}[t]{l}\small{form, $\sigma=0$; multi-scale}\end{tabular}}}}%
    \put(0.12469299,0.01726648){\color[rgb]{0,0,0}\makebox(0,0)[lt]{\lineheight{1.25}\smash{\begin{tabular}[t]{l}\small{form, $\sigma = 0.1 \mum$; multi-scale}\end{tabular}}}}%
    \put(0.42895979,0.01681243){\color[rgb]{0,0,0}\makebox(0,0)[lt]{\lineheight{1.25}\smash{\begin{tabular}[t]{l}\small{form, $\sigma = 1 \mum$; multi-scale}\end{tabular}}}}%
    \put(0.72028702,0.01681245){\color[rgb]{0,0,0}\makebox(0,0)[lt]{\lineheight{1.25}\smash{\begin{tabular}[t]{l}\small{form, $\sigma = 5 \mum$; multi-scale}\end{tabular}}}}%
    \put(0.01541792,0.33497432){\color[rgb]{0,0,0}\makebox(0,0)[lt]{\lineheight{1.25}\smash{\begin{tabular}[t]{l}(a)\end{tabular}}}}%
    \put(0.51881464,0.33516777){\color[rgb]{0,0,0}\makebox(0,0)[lt]{\lineheight{1.25}\smash{\begin{tabular}[t]{l}(b)\end{tabular}}}}%
    \put(0.05425323,0.13867733){\color[rgb]{0,0,0}\makebox(0,0)[lt]{\lineheight{1.25}\smash{\begin{tabular}[t]{l}0.96\end{tabular}}}}%
  \end{picture}%
\endgroup%
 \caption{Amplitude-dependent properties of the first in-phase bending mode for different contact topographies: (a) frequency, (b) damping ratio. Simulations for smooth ($\sigma=0$) topographies were run on the respective fine mesh/grid, for rough ($\sigma\neq 0$ on the super-fine one.
  }%
  \label{fig:sigmaVariation}
\end{figure}
\\
Remarkably, the edge of the bore hole is no longer in contact, already for $\sigma=0.1~\mum$.
Hence, accounting for smaller length scales of the contact topography mitigates the undesired edge effects, and is hence expected to increase the accuracy of the proposed multi-scale method.
This is supported by the agreement with the full-FE reference, which is excellent in the case with synthetic roughness (\fref{roughValidation}), and even slightly better than in the case where only the form deviation is considered (\fref{MeshConvergenceModal}).
\\
In \fref{sigmaVariation}, frequency and damping ratio of the first in-phase bending mode are shown for quite different topographies.
Besides the case of a pure form deviation, and the cases of a superimposed synthetic height profile with different magnitudes $\sigma$, the case of a smooth, flat topography is also shown.
The effect of the form deviation is clearly a very prominent one.
Considering the nominal (flat) form instead of the real one leads to a severe over-prediction of the damping here.
In the flat case, edge effects are most emphasized, leading to a severe error of the proposed multi-scale method, when using full-FE model as reference.
Whether or not the form deviation prevails over the smaller length scales of the topography depends on their relative magnitude.
Remarkably, even though the contact is highly localized already for $\sigma = 0.1~\mum$, and is quite different for $\sigma=1~\mum$, the modal frequency and damping ratio are only mildly affected.
This is explained by the integral character of the modal parameters; \ie, not the precise distribution of the contact stresses but their projection onto the vibration mode is of relevance, which corresponds to a spatial averaging (provided that the wave length of the vibration mode is sufficiently large compared to the dimensions of the apparent contact area).

\subsection{Computational effort\label{sec:effort}}
The computation time required to obtain the results shown in \frefs{MeshConvergenceModal} and \frefo{roughValidation} is listed in \tref{compTime}.
The first is deemed representative for the scenario where only the form deviation is considered, and the second for the scenario where also smaller length scales of the topography are resolved.
Accordingly, the fine respectively super fine mesh / grid was used for the contact simulation.
The reported time includes the computation of the preload step as well as the quasi-static modal analysis step up to the amplitude where the full-FE solver breaks down.
Depending on the scenario, and whether wall time or the CPU time is regarded, the computation time is reduced by 2 to 3 orders of magnitude.
Simulations were run using an Intel i9-13900K with 24 CPUs.
It is useful to remark that \ABAQUS makes use of parallel computing on all 24 CPUs.
No effort was made to exploit parallel computing in the present implementation of the proposed multi-scale method; yet  some of the \MATLAB built-in functions exploit this internally.
\begin{table}[h!]
\caption{Computation time.}
\begin{center}\label{tab:compTime}
\begin{tabular}{l l l l l l l}
\toprule
\rule[-1ex]{0pt}{2.5ex}
scenario & topography & contact & wall time & wall time & CPU time & CPU time\\
& & mesh/grid & (multi-scale) & (full-FE) & (multi-scale) & (full-FE)\\
\midrule
\midrule
\rule[-1ex]{0pt}{2.5ex} \fref{MeshConvergenceModal} & form deviation & fine & 5.46 min & 3.32 h & 27.54 min & 64.06 h \\
& $\sigma=0~\mum$ (smooth) &&&&&\\
\hline
\rule[-1ex]{0pt}{2.5ex} \fref{roughValidation} & form deviation & super fine & 3.06 min & 85.37 h & 9.23 min & 626.38 h \\
& $\sigma=1~\mum$ &&&&&\\
\hline
\end{tabular}
\end{center}
\end{table}
\\
When the resolution of the contact topography is refined, the computation effort required for the full-FE analysis grows dramatically.
Remarkably, the opposite holds for the proposed multi-scale method.
This is explained by the fact that the actual contact area is smaller in the second scenario.
Consequently, a smaller number of integration points is in fact considered thanks to the proposed geometric restriction strategy.
In the first scenario, where only the form deviation is considered, the geometric restriction factor is $65\%$, whereas it is $21\%$ in the scenario where also smaller length scales were resolved.
Also, the number of active points is by an average factor of $4$ smaller in the second scenario.
This observation gives rise to the idea that the consideration of smaller length scales of the contact topography may be viewed as a physics-motivated path towards contact (hyper) reduction, and may in this sense be an alternative to the approaches presented in \cite{Witteveen.2022,Morsy.2023}.

\section{Conclusions\label{sec:conclusions}}
The key idea of the proposed multi-scale method is to describe the mechanics of the contact region with a BE model, and the structural dynamics of the underlying bodies with an FE model, and to couple both models via the far field of the BE domain.
The most important benefit is that the actual contact topography has to be resolved with a sufficiently fine mesh only in the BE model, while a coarse mesh of the nominal geometry is sufficient in the FE model.
Compared to the full-FE reference obtained with \ABAQUS, the proposed method showed higher numerical robustness, even when treating rigid contact laws (without regularization) on very fine grids.
Also, the computational effort was reduced by 2-3 orders of magnitude for considered scenarios of the S4 beam benchmark
The agreement was good to excellent in terms of amplitude-dependent modal properties, depending on whether form deviations alone or also some smaller length scales of the contact topography were considered.
As the BE model relies on Boussinesq-Cerruti theory, it may suffer from edge effects.
The finer the actual contact topography is resolved, the less pronounced are those errors.
The finer the resolution, the smaller the real contact area.
The proposed method restricts the problem to the active contact area.
Thus, the computational effort scales favorably with increased resolution.
At some point, however, the fact that the compliance matrix associated with the BE model is fully-populated may become an obstacle, and ideas to overcome this may have to be pursued (\eg neglecting the interaction of remote asperities, implement means of parallel computing).
In any case, the method does not seem useful for smooth, complete contacts. 
\\
Efforts to develop and validate fully-dynamic implementations based on time step integration or Harmonic Balance are ongoing.
Also, friction-clamped structures undergoing large deflections, or contacts undergoing finite sliding seem particularly interesting perspectives of future work.
The proposed multi-scale method opens up new opportunities to address the question what length scales of the contact topography are important for an accurate damping prediction.
The proposed multi-scale method seems particularly attractive for probabilistic analysis of uncertainties related to the contact topography.


\appendix

\section{Closed-form expression of the elements of the BE compliance matrix\label{asec:C}}
If elastically identical materials are paired, one can show that the sub-matrices ${\mm C}_{j\ell}$ in \eref{C} are diagonal, which means that the normal and tangential directions are elastically decoupled, see \eg \cite{Willner.2008b,popov10},
\ea{
{\mm C}_{j\ell} = \matrix{ccc}{C_{{zz}} & 0 & 0 \\ 0 & C_{{xx}} & 0 \\ 0 & 0 & C_{{yy}}}_{j\ell}\fp \label{eq:Cjk}
}
In the isotropic case, the influence coefficients $C_{xx}$, $C_{yy}$ and $C_{zz}$ in \eref{Cjk} are (\cf, \eg, \cite{Willner.2008b}):
\ea{
C_{xx} =  \frac{2 (1-\nu^2)}{\Delta A \pi E} \Biggl[
& (\bar{x} + \Delta x) \ln \left( \frac{\bar{y} + \Delta y + \sqrt{(\bar{x} + \Delta x)^2 + (\bar{y} + \Delta y)^2}}{\bar{y} - \Delta y + \sqrt{(\bar{x} + \Delta x)^2 + (\bar{y} - \Delta y)^2}} \right) \\
& + (\bar{x} - \Delta x) \ln \left( \frac{\bar{y} - \Delta y + \sqrt{(\bar{x} - \Delta x)^2 + (\bar{y} - \Delta y)^2}}{\bar{y} + \Delta y + \sqrt{(\bar{x} - \Delta x)^2 + (\bar{y} + \Delta y)^2}} \right)  \nonumber \\
& + \frac{(\bar{y} + \Delta y)}{1 - \nu} \ln \left( \frac{\bar{x} + \Delta x + \sqrt{(\bar{x} + \Delta x)^2 + (\bar{y} + \Delta y)^2}}{\bar{x} - \Delta x + \sqrt{(\bar{x} - \Delta x)^2 + (\bar{y} + \Delta y)^2}} \right) \nonumber \\
& + \frac{(\bar{y} - \Delta y)}{1 - \nu} \ln \left( \frac{\bar{x} - \Delta x + \sqrt{(\bar{x} - \Delta x)^2 + (\bar{y} - \Delta y)^2}}{\bar{x} + \Delta x + \sqrt{(\bar{x} + \Delta x)^2 + (\bar{y} - \Delta y)^2}} \right) \Biggr]\fk  \label{eq:Cxx}
}
\ea{
C_{yy} = & \frac{2 (1-\nu^2)}{\Delta A \pi E} \Biggl[ \frac{(\bar{x} + \Delta x)}{1 - \nu} \ln \left( \frac{\bar{y} + \Delta y + \sqrt{(\bar{x} + \Delta x)^2 + (\bar{y} + \Delta y)^2}}{\bar{y} - \Delta y + \sqrt{(\bar{x} + \Delta x)^2 + (\bar{y} - \Delta y)^2}} \right) \\
& + \frac{(\bar{x} - \Delta x)}{1 - \nu} \ln \left( \frac{\bar{y} - \Delta y + \sqrt{(\bar{x} - \Delta x)^2 + (\bar{y} - \Delta y)^2}}{\bar{y} + \Delta y + \sqrt{(\bar{x} - \Delta x)^2 + (\bar{y} + \Delta y)^2}} \right) \nonumber \\
& + (\bar{y} + \Delta y) \ln \left( \frac{\bar{x} + \Delta x + \sqrt{(\bar{x} + \Delta x)^2 + (\bar{y} + \Delta y)^2}}{\bar{x} - \Delta x + \sqrt{(\bar{x} - \Delta x)^2 + (\bar{y} + \Delta y)^2}} \right) \nonumber  \\
& + (\bar{y} - \Delta y) \ln \left( \frac{\bar{x} - \Delta x + \sqrt{(\bar{x} - \Delta x)^2 + (\bar{y} - \Delta y)^2}}{\bar{x} + \Delta x + \sqrt{(\bar{x} + \Delta x)^2 + (\bar{y} - \Delta y)^2}} \right) \Biggr]\fk  \label{eq:Cyy}
}
\ea{
C_{zz} = &  \frac{2 (1-\nu^2)}{\Delta A \pi E}  \Biggl[ (\bar{x} + \Delta x) \ln \left( \frac{\bar{y} + \Delta y + \sqrt{(\bar{x} + \Delta x)^2 + (\bar{y} + \Delta y)^2}}{\bar{y} - \Delta y + \sqrt{(\bar{x} + \Delta x)^2 + (\bar{y} - \Delta y)^2}} \right) \\
& + (\bar{x} - \Delta x) \ln \left( \frac{\bar{y} - \Delta y + \sqrt{(\bar{x} - \Delta x)^2 + (\bar{y} - \Delta y)^2}}{\bar{y} + \Delta y + \sqrt{(\bar{x} - \Delta x)^2 + (\bar{y} + \Delta y)^2}} \right) \nonumber \\
& + (\bar{y} + \Delta y) \ln \left( \frac{\bar{x} + \Delta x + \sqrt{(\bar{x} + \Delta x)^2 + (\bar{y} + \Delta y)^2}}{\bar{x} - \Delta x + \sqrt{(\bar{x} - \Delta x)^2 + (\bar{y} + \Delta y)^2}} \right) \nonumber \\
& + (\bar{y} - \Delta y) \ln \left( \frac{\bar{x} - \Delta x + \sqrt{(\bar{x} - \Delta x)^2 + (\bar{y} - \Delta y)^2}}{\bar{x} + \Delta x + \sqrt{(\bar{x} + \Delta x)^2 + (\bar{y} - \Delta y)^2}} \right) \Biggr]\fp  \label{eq:Czz}
}
Herein, $E$ is the Young's modulus and $\nu$ is the Poisson ratio, and $\bar{x}=x_\ell-x_j$, $\bar{y}=y_\ell-y_j$ are the distances between points $\ell$ and $j$ in the respective direction (\fref{BE_grid}).
\\
Note that even if there is no elastic coupling of the $x$-, $y$- and $z$-direction via the influence coefficient matrices of the half spaces, such a coupling will generally be due to geometry and boundary conditions of the underlying solid bodies, and that this is accounted for via the compliance of the FE model (term $\tilde{\mm K}_{\mrm{bb}}$ in \eref{Cstar}).
Even if this is beyond the focus of the present work, the extension to dissimilar and/or non-isotropic materials seems feasible, leading to fully populated sub-matrices ${\mm C}_{j\ell}$ in general \cite{barber2007,bagault2012}.

\section{Solution of implicit algebraic inclusions\label{asec:iai}}
The implicit algebraic inclusion derived in \ssref{algorithm} takes the form $-(\mm G\mm x + \mm c) \in \mathcal N_{\mathcal C}(\mm x)$.
This problem has a unique solution if $\mm G$ is symmetric and positive definite.
One can easily verify from \erefs{G} and \erefo{Cstar} that $\mm G$ is indeed symmetric.
A sufficient condition for a positive definite $\mm G$ is that the considered static equilibrium is stable, which should be the case for any reasonable scenario involving jointed structures.
\\
The implicit algebraic inclusion can be rewritten as a non-smooth equation (see \eg \cite{Acary.2008,Studer.2009}),
\ea{
	\mm x = \operatorname{proj}_{\mathcal C}\left[~\mm x - \epsAL\left(\mm G\mm x+\mm c\right)~\right]\fk \label{eq:prox}
}
using the projection onto the admissible set $\mathcal C$.
We use the projected Jacobi over-relaxation algorithm to solve \eref{prox}.
This is known for its good convergence in the case of symmetric, positive definite matrices $\mm G$ \cite{Studer.2009}.
$\mm G$ is called \emph{Delassus matrix} in this context.
Compared to the common projected Gauss-Seidel algorithm, we found that the projected Jacobi algorithm is more easy to vectorize in order to make use of fast multi-threaded linear algebra operations.
The algorithm involves projections on the subsets of $\mathcal C$, \ie, on $\mathbb R_0^+$ and $\mathcal D\left(r\right)$.
These projections can be explicitly written as
\ea{
	\operatorname{proj}_{\mathbb R_0^+}\left(\xi\right) = \begin{cases} \xi & \xi \geq 0\\ 0 & \xi <0\end{cases}\fk \label{eq:projR}\\
	\operatorname{proj}_{\mathcal D(r)}\left(\mm\xi\right) = \begin{cases} r\frac{\mm\xi}{\sqrt{\mm\xi\tra\mm\xi}} & \sqrt{\mm\xi\tra\mm\xi}>r\\ \mm \xi & \sqrt{\mm\xi\tra\mm\xi}\leq r\end{cases}\fp \label{eq:projD}
}
The parameter $\epsAL$ was set as proposed in \cite{Studer.2009} based on the properties of the matrix $\mm G$.
To define reasonable tolerances for the solver, we found it useful to re-define the displacements for the dynamic load step, to count from the static equilibrium obtained after the preload step.

\section{Quasi-static modal analysis\label{asec:QSMA}}
To study the nonlinear vibration behavior in the present work, a quasi-static modal analysis was carried out.
In that method, a load of the following form is applied
\ea{
\tilde{\mm f}_{\mrm{ex}} = \tilde{\mm M}\mm\varphi_{\mrm{lin}}\alpha \fk\label{eq:QSMAload}
}
where $\mm\varphi_{\mrm{lin}}$ is the mass-normalized deflection shape of the considered mode.
The load scale $\alpha$ was step-wise increased (decreased), from zero, until a maximum value $\alpha_{\max}$ (minimum value $-\alpha_{\max}$).
This way, one obtains the displacement response $\tilde{\mm q}$ for each load scale, and the associated modal displacement defined as $\qmod = \mm\varphi^{\mrm T}_{\mrm{lin}}\tilde{\mm M}\tilde{\mm q}$.
\\
From the results of the quasi-static analysis, one can estimate the modal properties as function of the modal amplitude. 
The modal frequency $\omega$ and damping ratio $\Dmod$ are as follows \cite{Allen.2016d,Lacayo.2019b}:
\ea{
\omega = \sqrt{\frac{2\hat\alpha}{\left|\qmod(\hat\alpha) - \qmod(-\hat\alpha)\right|}}\fk \quad \Dmod = \frac{E_{\mrm{diss}}(\hat\alpha)}{2\pi \left(\,\omega\qmod\left(\hat\alpha\right)\,\right)^2}\fp \label{eq:QSMAmprop}
}
Herein, the dissipated energy is obtained as integral $E_{\mrm{diss}} = \oint \alpha \dd\qmod$ over one full hysteresis cycle in the range from $-\hat\alpha$ to $+\hat\alpha$.
Note that by increasing / decreasing the load scale monotonously from the equilibrium, one obtains an initial loading curve.
To estimate the full hysteresis cycle from the initial loading curve, for each considered amplitude $\hat\alpha$, the Masing rules were employed, as described \eg in \cite{Mathis.2020}.
\\
Finally, it is important to understand that the modal deflection shape, $\mm\varphi_{\mrm{lin}}$, is to be obtained from a linearization around the static equilibrium configuration obtained after the preload step.
The corresponding eigenvalue problem is:
\ea{
\left(\tilde{\mm K} - \Wb\frac{\partial \mm\lambda}{\partial \qb}- \omega^2_{\mrm{lin}}\tilde{\mm M}\right)\mm\varphi_{\mrm{lin}} = \mm 0\fk \label{eq:EVP}\\
\frac{\partial \mm\lambda^{(\mrm{sep})}}{\partial \qb} = \mm 0\fk \quad \frac{\partial \mm\lambda^{(\mrm{cl})}}{\partial \qb} = - \left(\mm C^{(\mrm{cl,cl})}\right)^{-1}\Wbtra\fp \label{eq:dlamdqb}
}
Herein, the derivatives $\partial X/\partial Y$ are to be evaluated at the static equilibrium, and the second term in the parenthesis of \eref{EVP} denotes the tangent contact stiffness.
The latter is defined separately for open and closed contact points in \eref{dlamdqb}.
The second of \eref{dlamdqb} follows from the condition that the gap at the closed contacts, $\mm g^{(\mrm{cl})}$, must not change under infinitesimal displacement.
Thus, $\partial \mm g^{(\mrm{cl})}/ \partial \qb = \mm 0$, which together with the first of \erefs{dlamdqb}, \erefo{BEM} and $\Wtra\mm q = \Wbtra\qb$ yields the above stated result.

\section{Contact solver verification\label{asec:verification}} 
It is important to ensure that the potentially different contact treatment within our implementation versus \ABAQUS have no significant effect on the observed results.
To assess this, a simulation was run directly on the same FE model.
To this end, the compliance of the BE model was neglected, $\mm C=\mm 0$, and the matrix of contact force directions was set to the identity matrix, $\Wb=\mm I$.
The FE nodes were directly used as integration points of the contact stress.
Note that linear shape functions were used at the interface, so that a node-based integration scheme appears reasonable.
As the mesh is not a regular grid, the areas associated with the respective nodes need to be accounted for.
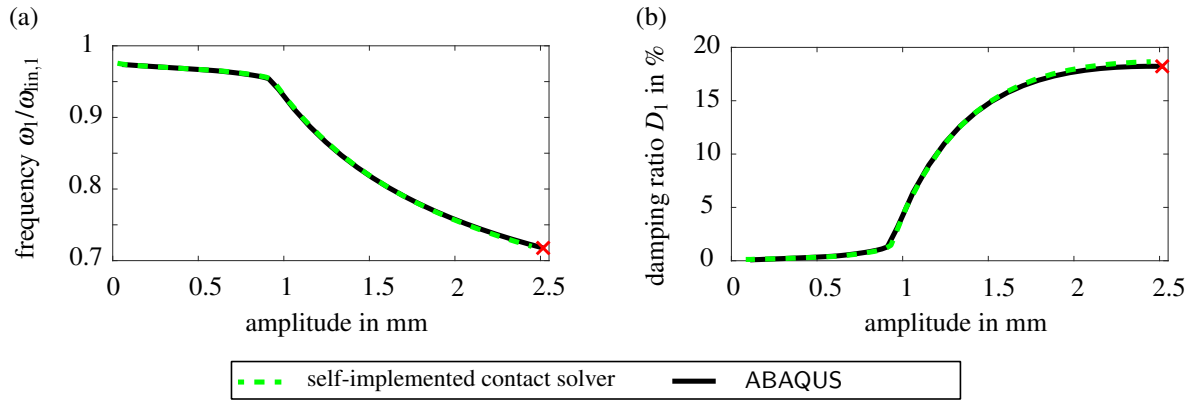
\begin{figure}[h]
 \centering
 \def\svgwidth{1.0\textwidth}
\begingroup%
  \makeatletter%
  \providecommand\color[2][]{%
    \errmessage{(Inkscape) Color is used for the text in Inkscape, but the package 'color.sty' is not loaded}%
    \renewcommand\color[2][]{}%
  }%
  \providecommand\transparent[1]{%
    \errmessage{(Inkscape) Transparency is used (non-zero) for the text in Inkscape, but the package 'transparent.sty' is not loaded}%
    \renewcommand\transparent[1]{}%
  }%
  \providecommand\rotatebox[2]{#2}%
  \newcommand*\fsize{\dimexpr\f@size pt\relax}%
  \newcommand*\lineheight[1]{\fontsize{\fsize}{#1\fsize}\selectfont}%
  \ifx\svgwidth\undefined%
    \setlength{\unitlength}{832.50000283bp}%
    \ifx\svgscale\undefined%
      \relax%
    \else%
      \setlength{\unitlength}{\unitlength * \real{\svgscale}}%
    \fi%
  \else%
    \setlength{\unitlength}{\svgwidth}%
  \fi%
  \global\let\svgwidth\undefined%
  \global\let\svgscale\undefined%
  \makeatother%
  \begin{picture}(1,0.3421988)%
    \lineheight{1}%
    \setlength\tabcolsep{0pt}%
    \put(0,0){\includegraphics[width=\unitlength]{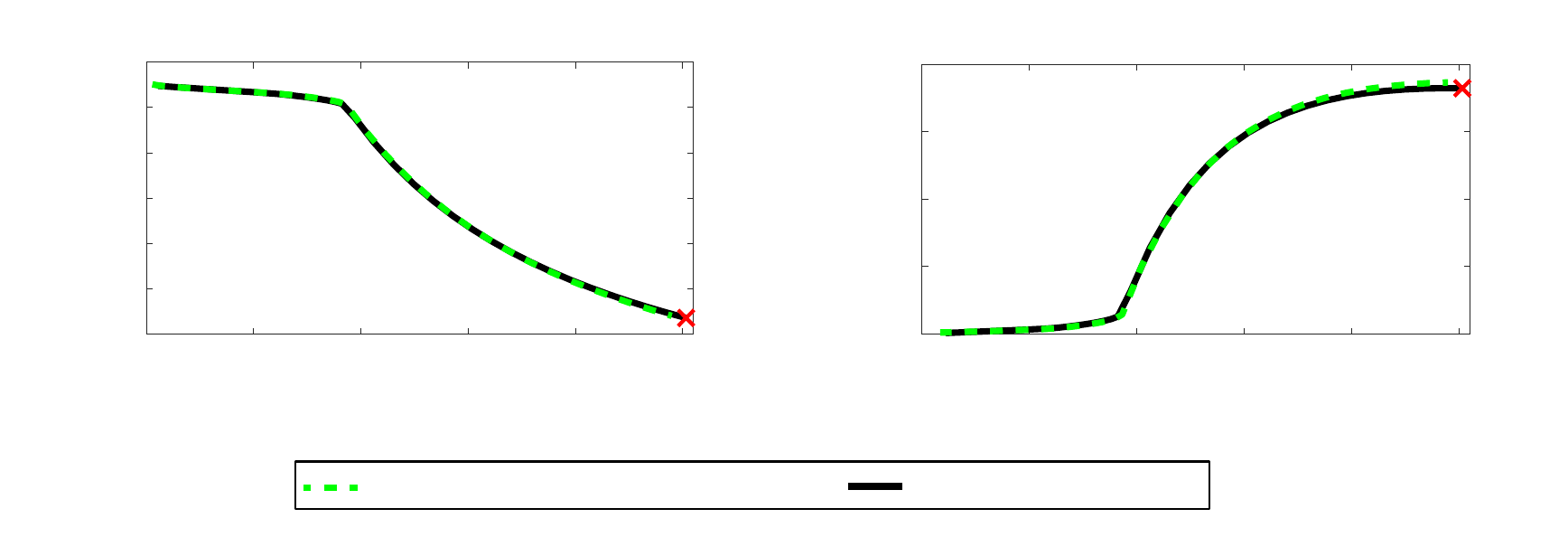}}%
    \put(0.08873874,0.10188349){\color[rgb]{0,0,0}\makebox(0,0)[lt]{\lineheight{1.25}\smash{\begin{tabular}[t]{l}0\end{tabular}}}}%
    \put(0.1545045,0.10132042){\color[rgb]{0,0,0}\makebox(0,0)[lt]{\lineheight{1.25}\smash{\begin{tabular}[t]{l}0.5\end{tabular}}}}%
    \put(0.22736488,0.10132042){\color[rgb]{0,0,0}\makebox(0,0)[lt]{\lineheight{1.25}\smash{\begin{tabular}[t]{l}1\end{tabular}}}}%
    \put(0.2927928,0.10132042){\color[rgb]{0,0,0}\makebox(0,0)[lt]{\lineheight{1.25}\smash{\begin{tabular}[t]{l}1.5\end{tabular}}}}%
    \put(0.3641892,0.10109522){\color[rgb]{0,0,0}\makebox(0,0)[lt]{\lineheight{1.25}\smash{\begin{tabular}[t]{l}2\end{tabular}}}}%
    \put(0.42916664,0.10120783){\color[rgb]{0,0,0}\makebox(0,0)[lt]{\lineheight{1.25}\smash{\begin{tabular}[t]{l}2.5\end{tabular}}}}%
    \put(0.58472931,0.10168053){\color[rgb]{0,0,0}\makebox(0,0)[lt]{\lineheight{1.25}\smash{\begin{tabular}[t]{l}0\end{tabular}}}}%
    \put(0.65049505,0.10111746){\color[rgb]{0,0,0}\makebox(0,0)[lt]{\lineheight{1.25}\smash{\begin{tabular}[t]{l}0.5\end{tabular}}}}%
    \put(0.72335542,0.10111746){\color[rgb]{0,0,0}\makebox(0,0)[lt]{\lineheight{1.25}\smash{\begin{tabular}[t]{l}1\end{tabular}}}}%
    \put(0.78878334,0.10111746){\color[rgb]{0,0,0}\makebox(0,0)[lt]{\lineheight{1.25}\smash{\begin{tabular}[t]{l}1.5\end{tabular}}}}%
    \put(0.86017984,0.10089225){\color[rgb]{0,0,0}\makebox(0,0)[lt]{\lineheight{1.25}\smash{\begin{tabular}[t]{l}2\end{tabular}}}}%
    \put(0.9251572,0.10100487){\color[rgb]{0,0,0}\makebox(0,0)[lt]{\lineheight{1.25}\smash{\begin{tabular}[t]{l}2.5\end{tabular}}}}%
    \put(0.56326998,0.12533149){\color[rgb]{0,0,0}\makebox(0,0)[lt]{\lineheight{1.25}\smash{\begin{tabular}[t]{l}0\end{tabular}}}}%
    \put(0.56357288,0.16779251){\color[rgb]{0,0,0}\makebox(0,0)[lt]{\lineheight{1.25}\smash{\begin{tabular}[t]{l}5\end{tabular}}}}%
    \put(0.55841959,0.21015514){\color[rgb]{0,0,0}\makebox(0,0)[lt]{\lineheight{1.25}\smash{\begin{tabular}[t]{l}10\end{tabular}}}}%
    \put(0.55818782,0.25323913){\color[rgb]{0,0,0}\makebox(0,0)[lt]{\lineheight{1.25}\smash{\begin{tabular}[t]{l}15\end{tabular}}}}%
    \put(0.55799885,0.29648237){\color[rgb]{0,0,0}\makebox(0,0)[lt]{\lineheight{1.25}\smash{\begin{tabular}[t]{l}20\end{tabular}}}}%
    \put(0.0587008,0.12546156){\color[rgb]{0,0,0}\makebox(0,0)[lt]{\lineheight{1.25}\smash{\begin{tabular}[t]{l}0.7\end{tabular}}}}%
    \put(0.05838229,0.18405921){\color[rgb]{0,0,0}\makebox(0,0)[lt]{\lineheight{1.25}\smash{\begin{tabular}[t]{l}0.8\end{tabular}}}}%
    \put(0.0587008,0.2400588){\color[rgb]{0,0,0}\makebox(0,0)[lt]{\lineheight{1.25}\smash{\begin{tabular}[t]{l}0.9\end{tabular}}}}%
    \put(0.06777852,0.29760099){\color[rgb]{0,0,0}\makebox(0,0)[lt]{\lineheight{1.25}\smash{\begin{tabular}[t]{l}1\end{tabular}}}}%
    \put(0.69490366,0.07256555){\color[rgb]{0,0,0}\makebox(0,0)[lt]{\lineheight{1.25}\smash{\begin{tabular}[t]{l}amplitude in mm\end{tabular}}}}%
    \put(0.19948671,0.0725728){\color[rgb]{0,0,0}\makebox(0,0)[lt]{\lineheight{1.25}\smash{\begin{tabular}[t]{l}amplitude in mm\end{tabular}}}}%
    \put(0.53394121,0.10884974){\color[rgb]{0,0,0}\rotatebox{90}{\makebox(0,0)[lt]{\lineheight{1.25}\smash{\begin{tabular}[t]{l}damping ratio $D_1$ in \%\end{tabular}}}}}%
    \put(0.02997861,0.13049323){\color[rgb]{0,0,0}\rotatebox{90}{\makebox(0,0)[lt]{\lineheight{1.25}\smash{\begin{tabular}[t]{l}frequency $\omega_1$/$\omega_{\mrm{lin,1}}$\end{tabular}}}}}%
    \put(0.2489009,0.02847481){\color[rgb]{0,0,0}\makebox(0,0)[lt]{\lineheight{1.25}\smash{\begin{tabular}[t]{l}\small{self-implemented contact solver}\end{tabular}}}}%
    \put(0.59899333,0.02622705){\color[rgb]{0,0,0}\makebox(0,0)[lt]{\lineheight{1.25}\smash{\begin{tabular}[t]{l}\small{\ABAQUS}\end{tabular}}}}%
    \put(0.01048759,0.31959899){\color[rgb]{0,0,0}\makebox(0,0)[lt]{\lineheight{1.25}\smash{\begin{tabular}[t]{l}(a)\end{tabular}}}}%
    \put(0.51182068,0.31940979){\color[rgb]{0,0,0}\makebox(0,0)[lt]{\lineheight{1.25}\smash{\begin{tabular}[t]{l}(b)\end{tabular}}}}%
  \end{picture}%
\endgroup%
  \caption{Amplitude-dependent properties of the first in-phase bending mode for the contact topography in \fref{surfaceTopography} (form, smooth): (a) frequency, (b) damping ratio. The fine FE model is used. Crosses indicate where \ABAQUS failed to converge.
  }%
  \label{fig:SolverVerification}
\end{figure}
\\
Results obtained for the medium mesh are shown in \fref{SolverVerification} in terms of the amplitude-dependent frequency and damping ratio of the first in-phase bending mode.
Apparently, the self-implemented contact solver agrees perfectly with the \ABAQUS reference.
Recall that finite sliding was switched on in \ABAQUS, while our implementation exploits a small sliding assumption.
The excellent agreement implies that the small sliding assumption is valid in the considered amplitude range.
The maximum slip distance is $0.2~\mrm{mm}$ in depicted range.


\end{document}